\documentclass{emulateapj}
\usepackage{natbib}
\usepackage{graphicx}
\singlespace
\shorttitle{\WMAP\ 5-year Cosmological Interpretation}
\shortauthors{Komatsu et al.}
\newcommand{\map}    {{\sl WMAP}}
\newcommand{\WMAP}    {{\sl WMAP}}
\newcommand{\cobe}   {{\sl COBE} }
\newcommand{\fnlKS}  {f_{NL}^{\rm local}}
\newcommand{\fnleq}  {f_{NL}^{\rm equil}} 
\newcommand{\bsrc}   {b_{src}}
\slugcomment{Accepted for Publication in the Astrophysical Journal Supplement Series}
\begin{document}
\title{Five-Year Wilkinson Microwave Anisotropy Probe 
(WMAP\altaffilmark{1}) Observations:\\
Cosmological Interpretation}
\author{{E. Komatsu} \altaffilmark{1}, 
{J. Dunkley} \altaffilmark{2,3,4},
{M. R. Nolta}   \altaffilmark{5},
{C. L. Bennett} \altaffilmark{6},
{B. Gold} \altaffilmark{6},
{G. Hinshaw} \altaffilmark{7},
{N. Jarosik} \altaffilmark{2},
{D. Larson} \altaffilmark{6},
{M. Limon}  \altaffilmark{8}
{L. Page}    \altaffilmark{2},
{D. N. Spergel} \altaffilmark{3,9},
{M. Halpern} \altaffilmark{10},
{R. S. Hill}    \altaffilmark{11},
{A. Kogut}   \altaffilmark{7}, 
{S. S. Meyer}   \altaffilmark{12},
{G. S. Tucker}  \altaffilmark{13},
{J. L. Weiland} \altaffilmark{11},
{E. Wollack} \altaffilmark{7},
and {E. L. Wright}  \altaffilmark{14}
}
\altaffiltext{1}{\map\ is the result of a partnership between Princeton 
                 University and NASA's Goddard Space Flight Center. Scientific 
		 guidance is provided by the \map\ Science Team.}
\altaffiltext{1}{{Univ. of Texas, Austin, Dept. of Astronomy,  2511 Speedway, RLM 15.306, Austin, TX 78712}}
\altaffiltext{2}{{Dept. of Physics, Jadwin Hall,  Princeton University, Princeton, NJ 08544-0708}}
\altaffiltext{3}{{Dept. of Astrophysical Sciences,  Peyton Hall, Princeton University, Princeton, NJ 08544-1001}}
\altaffiltext{4}{{Astrophysics, University of Oxford,  Keble Road, Oxford, OX1 3RH, UK}}
\altaffiltext{5}{{Canadian Institute for Theoretical Astrophysics,  60 St. George St, University of Toronto,  Toronto, ON  Canada M5S 3H8}}
\altaffiltext{6}{{Dept. of Physics \& Astronomy,  The Johns Hopkins University, 3400 N. Charles St.,  Baltimore, MD  21218-2686}}
\altaffiltext{7}{{Code 665, NASA/Goddard Space Flight Center,  Greenbelt, MD 20771}}
\altaffiltext{8}{{Columbia Astrophysics Laboratory,  550 W. 120th St., Mail Code 5247, New York, NY  10027-6902}}
\altaffiltext{9}{{Princeton Center for Theoretical Physics,  Princeton University, Princeton, NJ 08544}}
\altaffiltext{10}{{Dept. of Physics and Astronomy, University of  British Columbia, Vancouver, BC  Canada V6T 1Z1}}
\altaffiltext{11}{{Adnet Systems, Inc.,  7515 Mission Dr., Suite A100, Lanham, Maryland 20706}}
\altaffiltext{12}{{Depts. of Astrophysics and Physics, KICP and EFI,  University of Chicago, Chicago, IL 60637}}
\altaffiltext{13}{{Dept. of Physics, Brown University, 182 Hope St., Providence, RI 02912-1843}}
\altaffiltext{14}{{UCLA Physics \& Astronomy, PO Box 951547,  Los Angeles, CA 90095-1547}}
\email{komatsu@astro.as.utexas.edu}
\begin{abstract}
The \map\ 5-year data provide stringent limits on deviations from the
minimal, 6-parameter $\Lambda$CDM model. We report these limits and
use them to constrain the physics of cosmic inflation via
Gaussianity,  adiabaticity, the power spectrum of primordial fluctuations,
gravitational waves, and spatial curvature. We also constrain
models of dark energy via its equation of state, parity-violating
interaction, and neutrino properties such as mass and the number of species.
We detect no convincing deviations from the minimal model. The 6
 parameters and the corresponding 68\% uncertainties, derived 
 from the \map\ 
data combined with the distance measurements from the Type Ia
 supernovae (SN) and the Baryon Acoustic
Oscillations (BAO) in the distribution of galaxies, are: 
\ensuremath{\Omega_bh^2 = 0.02267^{+ 0.00058}_{- 0.00059}},
\ensuremath{\Omega_ch^2 = 0.1131\pm 0.0034},
\ensuremath{\Omega_\Lambda = 0.726\pm 0.015},
\ensuremath{n_s = 0.960\pm 0.013},
\ensuremath{\tau = 0.084\pm 0.016}, and
\ensuremath{\Delta_{\cal R}^2 = (2.445\pm 0.096)\times 10^{-9}} at
$k=0.002~{\rm Mpc^{-1}}$. From these we derive
\ensuremath{\sigma_8 = 0.812\pm 0.026},
\ensuremath{H_0 = 70.5\pm 1.3}~${\rm
km~s^{-1}~Mpc^{-1}}$,
\ensuremath{\Omega_b = 0.0456\pm 0.0015},
\ensuremath{\Omega_c = 0.228\pm 0.013},
\ensuremath{\Omega_mh^2 = 0.1358^{+ 0.0037}_{- 0.0036}},
\ensuremath{z_{\rm reion} = 10.9\pm 1.4},
and 
\ensuremath{t_0 = 13.72\pm 0.12\ \mbox{Gyr}}.
With the \WMAP\ data combined with BAO and SN, we find the limit on 
the tensor-to-scalar ratio of
\ensuremath{r < 0.22\ \mbox{(95\% CL)}}, and that
$n_s>1$ is disfavored even when gravitational waves are
included, which constrains the models of inflation that can
produce significant gravitational waves, such as chaotic or
power-law inflation models, or a blue spectrum, such as hybrid
inflation models. We obtain tight, simultaneous limits on the (constant)
equation of state of dark energy and the spatial curvature of the
universe: \ensuremath{-0.14<1+w<0.12\ \mbox{(95\% CL)}}
and
\ensuremath{-0.0179<\Omega_k<0.0081\ \mbox{(95\% CL)}}. 
We provide a set of ``\map\
distance priors,'' to test a variety of dark energy models with
spatial curvature. We test a time-dependent $w$ with a present
value constrained as $-0.33<1+w_0<0.21$ (95\% CL).
Temperature and dark matter fluctuations are found to obey the
adiabatic relation to within 8.9\% and 2.1\% for the axion-type and
curvaton-type dark matter, respectively. The power spectra of 
TB and EB correlations constrain a parity-violating interaction, which
rotates the polarization angle and converts E to B. The
polarization angle could not be rotated more than
$-5.9^\circ<\Delta\alpha<2.4^\circ$ (95\% CL) between the decoupling
and the present epoch. We find the limit on the total mass of massive
 neutrinos of 
\ensuremath{\sum m_\nu < 0.67\ \mbox{eV}\ \mbox{(95\% CL)}}, which
is free from the uncertainty in the normalization of the large-scale
structure data. The number of relativistic degrees of freedom,
expressed in units of the effective number of neutrino species, is
constrained as
\ensuremath{N_{\rm eff} = 4.4\pm 1.5}
(68\%), consistent with the standard value of 3.04. Finally, 
quantitative limits on physically motivated primordial
non-Gaussianity parameters are $-9 < \fnlKS <111$ (95\% CL)
and $-151 < \fnleq < 253$ (95\% CL) for the local and equilateral
 models, respectively.
\end{abstract}
\keywords{cosmic microwave background, cosmology: observations, early
universe, dark matter, space vehicles, space vehicles: instruments, 
instrumentation: detectors, telescopes}
\section{Introduction}\label{sec:intro}
Measurements of microwave background fluctuations by the Cosmic
Background Explorer
\citep[\cobe;][]{smoot/etal:1992,bennett/etal:1994,bennett/etal:1996},
the Wilkinson Microwave Anisotropy Probe
\citep[\map;][]{bennett/etal:2003,bennett/etal:2003b}, and ground
and balloon-borne experiments
\citep{miller/etal:1999,miller/etal:2002,debernardis/etal:2000,hanany/etal:2000,netterfield/etal:2002,ruhl/etal:2003,mason/etal:2003,sievers/etal:2003,sievers/etal:2007,pearson/etal:2003,readhead/etal:2004,dickinson/etal:2004,kuo/etal:2004,kuo/etal:2007,reichardt/etal:prep,jones/etal:2006,montroy/etal:2006,piacentini/etal:2006}
have addressed many of the questions that were the focus
of cosmology for the past 50 years: How old is the universe? How fast is
it expanding? What is the size and shape of the universe? What is the
composition of the universe? What seeded the formation of galaxies and
large-scale structure?

By accurately measuring the statistical properties of the microwave
background fluctuations, \map\ has helped establish a standard
cosmology: a flat $\Lambda$CDM model composed of atoms, dark matter and dark
energy, with nearly scale-invariant adiabatic Gaussian fluctuations.
With our most recent measurements, \map\ has measured the basic
parameters of this cosmology to high precision: with the \map\ 5-year
data alone, we find the density of dark matter (21.4\%), the density of
atoms  (4.4\%), the expansion rate of the universe, the amplitude of
density fluctuations, and their scale dependence, as well as the optical
depth due to reionization \citep{dunkley/etal:prep}.  

Cosmologists are now focused on a new set of questions: What is the
nature of the dark energy?  What is the dark matter? Did inflation seed
the primordial fluctuations?  If so, what is the class of the
inflationary model? 
How did the first stars form?  Microwave background observations from
\map, {\sl Planck}, and from the upcoming generation of CMB ground and
balloon-borne experiments  will play an important role in addressing
these questions.  

This paper will discuss how the \map\ results, particularly when
combined with other astronomical observations (mainly the distance
measurements), are now providing new insights into these questions
through constraints on gravitational waves and non-adiabatic (entropic)
fluctuations, measurements of primordial non-Gaussianity, accurate
determination of the primordial spectral index and the geometry of the
universe, and limits on parity-violating interactions.

This paper is one of 7 papers on the analysis of the \WMAP\ 5-year data:
 \citet{hinshaw/etal:prep}
report on the data processing, map-making, and systematic error limits,
\citet{hill/etal:prep} on the physical optics modeling of beams and the
5-year window functions (beam transfer functions),
\citet{gold/etal:prep} on the modeling, understanding, and subtraction of the
temperature and polarized foreground emission, \citet{wright/etal:prep}
 on the catalogue of point sources detected in the 5-year temperature
 data, \citet{nolta/etal:prep} on the measurements of the temperature
 and polarization power spectra, and \citet{dunkley/etal:prep} on the
 parameter estimation methodology, the cosmological parameters inferred
 from the \map\ data alone, and comparison between different
 cosmological data sets.

This paper is organized as follows. In \S~\ref{sec:analysis} we briefly
summarize new aspects of our analysis of the \map\ 5-year temperature
and polarization data. In \S~\ref{sec:inflation} we constrain the
spatial curvature of the observable universe,
Gaussianity/adiabaticity/scale-invariance of the primordial
fluctuations, and the amplitude of primordial gravitational waves.  We
discuss their implications for the physics of the early, primordial
universe. In \S~\ref{sec:TB} we demonstrate that the power spectra of
TB and EB correlations\footnote{Here, ``TB'' refers to the power
spectrum of a cross-correlation between the temperature and B-mode
polarization, while ``EB'' refers to a correlation between the E-mode
and B-mode polarization.}
which are usually ignored in the cosmological analysis, can be used to
constrain a certain parity-violating interaction that couples to 
photons. In \S~\ref{sec:DE} we explore the nature of dark energy, and in
\S~\ref{sec:neutrino} we study the properties of neutrinos in cosmology.  
We conclude in \S~\ref{sec:conclusion}.

\begin{deluxetable*}{llcccc}
\tablecolumns{6}
\small
\tablewidth{0pt}
\tablecaption{%
Summary of the cosmological parameters of $\Lambda$CDM
 model and the corresponding 68\% intervals
}
\tablehead{\colhead{Class} &
\colhead{Parameter}
&\colhead{\map\ 5-year ML\footnote{\citet{dunkley/etal:prep}. ``ML''
 refers to the Maximum Likelihood parameters}}
&\colhead{\map+BAO+SN ML}
&\colhead{\map\ 5-year Mean\footnote{\citet{dunkley/etal:prep}. ``Mean''
 refers to the mean of the posterior distribution of each parameter}}
&\colhead{\map+BAO+SN Mean}
}
\startdata
Primary &
$100\Omega_bh^2$
&2.268
&2.262
&\ensuremath{2.273\pm 0.062} 
&\ensuremath{2.267^{+ 0.058}_{- 0.059}} \nl
&
$\Omega_ch^2$
&0.1081
&0.1138
&\ensuremath{0.1099\pm 0.0062} 
&\ensuremath{0.1131\pm 0.0034} \nl
&
$\Omega_\Lambda$
&0.751
&0.723
&\ensuremath{0.742\pm 0.030} 
&\ensuremath{0.726\pm 0.015} \nl
&
$n_s$
&0.961
&0.962
&\ensuremath{0.963^{+ 0.014}_{- 0.015}} 
&\ensuremath{0.960\pm 0.013} \nl
&
$\tau$
&0.089
&0.088
&\ensuremath{0.087\pm 0.017} 
&\ensuremath{0.084\pm 0.016} \nl
&
$\Delta^2_{\cal R}(k_0\footnote{$k_0=0.002~{\rm
 Mpc}^{-1}$. $\Delta^2_{\cal R}(k)=k^3P_{\cal R}(k)/(2\pi^2)$ (Eq.~[\ref{eq:pR}])})$
&$2.41\times 10^{-9}$
&$2.46\times 10^{-9}$
&\ensuremath{(2.41\pm 0.11)\times 10^{-9}} 
&\ensuremath{(2.445\pm 0.096)\times 10^{-9}} \nl
\hline
Derived &
$\sigma_8$
&0.787
&0.817
&\ensuremath{0.796\pm 0.036} 
&\ensuremath{0.812\pm 0.026} \nl
&
$H_0$
&$72.4~{\rm km/s/Mpc}$
&$70.2~{\rm km/s/Mpc}$
&\ensuremath{71.9^{+ 2.6}_{- 2.7}\ \mbox{km/s/Mpc}} 
&\ensuremath{70.5\pm 1.3\ \mbox{km/s/Mpc}} \nl
&
$\Omega_b$
&0.0432
&0.0459
&\ensuremath{0.0441\pm 0.0030} 
&\ensuremath{0.0456\pm 0.0015} \nl
&
$\Omega_c$
&0.206
&0.231
&\ensuremath{0.214\pm 0.027} 
&\ensuremath{0.228\pm 0.013} \nl
&
$\Omega_mh^2$
&0.1308
&0.1364
&\ensuremath{0.1326\pm 0.0063} 
&\ensuremath{0.1358^{+ 0.0037}_{- 0.0036}} \nl
&
$z_{\rm reion}$\footnote{``Redshift of reionization,'' if the universe was
 reionized instantaneously from the neutral state to the fully ionized
 state at $z_{\rm reion}$}
&11.2
&11.3
&\ensuremath{11.0\pm 1.4} 
&\ensuremath{10.9\pm 1.4}\nl 
&
$t_0$\footnote{The present-day age of the universe}
&13.69~{\rm Gyr}
&13.72~{\rm Gyr}
&\ensuremath{13.69\pm 0.13\ \mbox{Gyr}} 
&\ensuremath{13.72\pm 0.12\ \mbox{Gyr}} 
\enddata
\label{tab:summary}
\end{deluxetable*}

\begin{deluxetable*}{lllcc}
\tablecolumns{5}
\small
\tablewidth{0pt}
\tablecaption{%
Summary of the 95\% confidence limits on deviations from the simple 
(flat, Gaussian, adiabatic, power-law) $\Lambda$CDM model
}
\tablehead{\colhead{Section}&
\colhead{Name}&\colhead{Type}&
\colhead{\map\ 5-year}&\colhead{\map+BAO+SN}}
\startdata
\S~\ref{sec:GW}&
Gravitational Wave\footnote{In the form of the tensor-to-scalar ratio, $r$, at
 $k=0.002~{\rm Mpc}^{-1}$}
& No Running Ind.
& $r<0.43$\footnote{\citet{dunkley/etal:prep}
}
& $r<0.22$
\nl
\S~\ref{sec:PK_results_running}&
Running Index
& No Grav. Wave
& $-0.090<dn_s/d\ln k<0.019$\footnote{\citet{dunkley/etal:prep}
}
& $-0.068<dn_s/d\ln k<0.012$
\nl
\S~\ref{sec:OK}&
Curvature\footnote{(Constant) dark energy equation of state allowed to
 vary ($w\ne -1$)}
&
&$-0.063<\Omega_k<0.017$\footnote{With the HST prior, $H_0=72\pm 8~{\rm
 km/s/Mpc}$. For $w=-1$, 
\ensuremath{-0.052<\Omega_k<0.013\ \mbox{(95\% CL)}}
}
&$-0.0179<\Omega_k<0.0081$\footnote{
For $w=-1$,
\ensuremath{-0.0178<\Omega_k<0.0066\ \mbox{(95\% CL)}}
}\nl
&
Curvature Radius\footnote{$R_{\rm curv}=(c/H_0)/\sqrt{|\Omega_k|}=3/\sqrt{|\Omega_k|}~h^{-1}$Gpc}
& Positive Curv.
& $R_{\rm curv}>12~h^{-1}$Gpc
& $R_{\rm curv}>22~h^{-1}$Gpc\nl
& 
& Negative Curv.
& $R_{\rm curv}>23~h^{-1}$Gpc
& $R_{\rm curv}>33~h^{-1}$Gpc\nl
\S~\ref{sec:NG} &
Gaussianity
&Local
&$-9<\fnlKS<111$\footnote{Cleaned V$+$W map with $l_{\rm max}=500$ and
 the {\it KQ75} mask, after the point  source correction}
&N/A\nl
&
&Equilateral
&$-151 < \fnleq < 253$\footnote{Cleaned V$+$W map with $l_{\rm max}=700$
 and the {\it KQ75} mask, after the point  source correction}
&N/A\nl
\S~\ref{sec:AD} &
Adiabaticity
&Axion
&$\alpha_0<0.16$\footnote{\citet{dunkley/etal:prep}
}
&$\alpha_0<0.072$\footnote{In terms of the adiabaticity deviation
 parameter, $\delta_{adi}^{(c,\gamma)}=\sqrt{\alpha}/3$
 (Eq.~[\ref{eq:deltaadi}]), the axion-like dark matter and photons are found to
 obey the adiabatic relation (Eq.~[\ref{eq:adi}]) to 8.9\%. 
}\nl 
&
&Curvaton
&$\alpha_{-1}<0.011$\footnote{\citet{dunkley/etal:prep}
}
&$\alpha_{-1}<0.0041$\footnote{In terms of the adiabaticity deviation
 parameter, $\delta_{adi}^{(c,\gamma)}=\sqrt{\alpha}/3$
 (Eq.~[\ref{eq:deltaadi}]), the curvaton-like dark matter and photons
 are found to obey the adiabatic relation (Eq.~[\ref{eq:adi}]) to 2.1\%.
}\nl
\S~\ref{sec:TB} &
Parity Violation
& Chern-Simons\footnote{For an interaction of the form given by
$[\phi(t)/M]F_{\alpha\beta}\tilde{F}^{\alpha\beta}$, the polarization
 rotation angle is $\Delta\alpha=M^{-1}\int \frac{dt}{a} \dot{\phi}$}
& $-5.9^\circ<\Delta\alpha<2.4^\circ$
&N/A\nl
\S~\ref{sec:DE} &
Dark Energy
& Constant $w$\footnote{For spatially curved universes ($\Omega_k\ne 0$)}
& $-1.37<1+w<0.32$\footnote{With the HST prior, $H_0=72\pm 8~{\rm
 km/s/Mpc}$
}
& $-0.14<1+w<0.12$
\nl
&
& Evolving $w(z)$\footnote{For a flat universe ($\Omega_k=0$)}
& N/A
& $-0.33<1+w_0<0.21$\footnote{$w_0\equiv w(z=0)$}\nl 
\S~\ref{sec:massnu}&
Neutrino Mass\footnote{$\sum m_\nu=94(\Omega_\nu h^2)~{\rm eV}$}
& 
& $\sum m_\nu<1.3~{\rm eV}$\footnote{\citet{dunkley/etal:prep}
}
& $\sum m_\nu<0.67~{\rm eV}$\footnote{For $w=-1$. For $w\ne-1 $, 
\ensuremath{\sum m_\nu < 0.80\ \mbox{eV}\ \mbox{(95\% CL)}}
}\nl 
\S~\ref{sec:neff}&
Neutrino Species
&
& $N_{\rm eff}>2.3$\footnote{\citet{dunkley/etal:prep}
}
&
\ensuremath{N_{\rm eff} = 4.4\pm 1.5}\footnote{With
 the HST prior, $H_0=72\pm 8~{\rm km/s/Mpc}$. The 95\% limit is
 $1.8<N_{\rm eff}<7.6$} (68\%)
\enddata
\label{tab:deviation}
\end{deluxetable*}

\section{Summary of 5-year analysis}
\label{sec:analysis}
\subsection{WMAP 5-year data: temperature and polarization}
\label{sec:analysis_wmap}
With 5 years of observations of Jupiter and an extensive physical optics
modeling of beams \citep{hill/etal:prep}, our understanding of the beam
transfer  function, $b_l$, has improved significantly: the fractional
beam errors, $\Delta b_l/b_l$, have been nearly halved in most
Differencing Assemblies (DAs). In some cases, e.g., W4, the errors have
been reduced by as much as a factor of 4. 

Many of the small-scale CMB experiments have been calibrated to the
\map\ 3-year data at high multipoles. Since the new beam model raises
the 5-year power spectrum almost uniformly by $\sim 2.5$\% relative to the 
3-year power spectrum over
$l\gtrsim 200$ \citep{hill/etal:prep}, those small-scale CMB
experiments have been 
under-calibrated by the same amount, i.e., $\sim 2.5$\% in power, and
$1.2$\% in temperature. For example, the latest ACBAR data
\citep{reichardt/etal:prep} report on the calibration error of 2.23\% in
temperature (4.5\% in power), which is twice as large as the magnitude of
mis-calibration; thus, we expect the effect of mis-calibration to be
sub-dominant in the error budget.
Note that the change in the beam is fully consistent with the 1-$\sigma$
error of the previous \map\  beam reported in \citet{page/etal:2003b}. Since the ACBAR calibration error includes the
previous \map\ beam error, the change in the beam should have a minimal
impact on the current ACBAR calibration.
 
While we use only V and W
bands for the cosmological analysis of the temperature data, the
treatment of $b_l$ in Q band affects our determination of the point
source contamination to the angular power spectrum,
$A_{ps}$\footnote{This quantity, $A_{ps}$, is the value of the power
spectrum, $C_l$, from unresolved point sources in Q band, {\it in units of the
antenna temperature}. To convert this value to the thermodynamic units,
use $C_{ps}=1.089A_{ps}$ \citep{nolta/etal:prep}.}. 
The 5-year estimate of the point source correction, $A_{ps}=0.011\pm
0.001~\mu{\rm K}^2~{\rm sr}$ \citep{nolta/etal:prep}, is slightly lower than the
3-year estimate, $A_{ps}=0.014\pm 0.003~\mu{\rm K}^2~{\rm sr}$
\citep{hinshaw/etal:2007}, partly because more sources have
been detected and  masked by the 5-year source mask (390 sources have
been detected in the 5-year temperature data, whereas 323 sources were
detected in the 3-year data \citep{wright/etal:prep}). 

Note that the uncertainty in
$A_{ps}$ has been reduced by a factor of 3. The
uncertainty in the previous estimate was inflated to include the lower
value found by \citet{huffenberger/eriksen/hansen:2006} (0.011) and
higher value from our original estimate (0.017). Much of the discrepancy
between these estimates is due to the multipole range over which $A_{ps}$
is fit. With the improved beam model from the 5-year analysis, the dependence
on the multipole range has disappeared, and thus we no longer need
to inflate our uncertainty. See \citet{nolta/etal:prep} for more details. 

The method for cleaning foreground emission in both temperature and
polarization data is the same as we used for the
3-year data, i.e., the template-based cleaning method described in
\S~5.3 of \citet{hinshaw/etal:2007} for temperature and \S~4.3 of
\citet{page/etal:2007} for polarization. \citet{gold/etal:prep}
describe the results from the template cleaning of the 5-year data with
the new coefficients. In addition, \citet{gold/etal:prep} and
\citet{dunkley/etal:prep} explore alternative modelings of the foreground
emission. All of these methods give consistent results.
\citet{gold/etal:prep} also describe definitions of the new masks, {\it
KQ75} and {\it KQ85}, that replace the previous masks, {\it Kp0} and
{\it Kp2}, that are recommended for the analysis of Gaussianity tests
and the power spectrum, respectively.

The method for measuring the TT and TE spectra at higher multipoles,
i.e., $l\ge 33$ for TT and $l\ge 24$ for TE, is also the same as
we used for the 3-year data \citep{hinshaw/etal:2007}. As for
the estimation of the cosmological parameters from these spectra,
we now include the 
weak gravitational lensing effect of CMB due to the intervening matter
fluctuations \citep[see][for a review]{lewis/challinor:2006}, which was not
included in the 3-year analysis. We continue to marginalize over a
potential contribution from the Sunyaev--Zel'dovich effect (SZE), using
exactly the same template SZE power spectrum that we used for the
3-year analysis: $C_l^{\rm SZE}$ from \citet{komatsu/seljak:2002} with
$\Omega_m=0.26$, $\Omega_b=0.044$, $h=0.72$, $n_s=0.97$, and
$\sigma_8=0.80$ \citep[see also \S~2.1 of][]{spergel/etal:2007}.
We continue to use the V and W band data for estimating the high-$l$
temperature power spectrum, and the Q and V band data for the high-$l$
polarization power spectra.

We have improved our treatment of the temperature and polarization power
spectra at lower multipoles, as described below. 

{\it Low-$l$ temperature} -- We use the Gibbs sampling technique and the
Blackwell-Rao (BR) estimator to evaluate the likelihood of the
temperature power spectrum at $l\leq 32$
\citep{jewell/levin/anderson:2004,wandelt:2003,wandelt/larson/lakshminarayanan:2004,odwyer/etal:2004,eriksen/etal:2004e,eriksen/etal:2007,eriksen/etal:2007b,chu/etal:2005,larson/etal:2007}. For the
3-year analysis we used the resolution 4 Internal Linear Combination
(ILC) temperature map ($N_{\rm side}=16$) with a Gaussian smoothing of
$9.183^\circ$ (FWHM). Since the ILC map has an intrinsic Gaussian
smoothing of $1^\circ$, we have added an extra smoothing of
$9.1285^\circ$. We then evaluated the likelihood directly in the pixel
space for a given $C_l$. For the 5-year analysis we use a higher
resolution map, the resolution 5 ILC map ($N_{\rm side}=32$) with a
smaller Gaussian smoothing of $5^\circ$ (FWHM). 
The potential foreground leakage due to smoothing is therefore reduced.
The BR estimator has an advantage of being much
faster to compute, which is why we have adopted the Gibbs sampling and
the BR estimator for the 5-year data release. We have confirmed that
both the resolution 4 pixel-based likelihood and the resolution 5
Gibbs-based likelihood yield consistent results \citep[see][for
details]{dunkley/etal:prep}. Both options are made publicly available in
the released likelihood code. 

{\it Low-$l$ polarization} --
While we continue to use the direct evaluation of the likelihood of
polarization power spectra in pixel space from coadded resolution 3
($N_{\rm side}=8$) polarization maps (Stokes Q and U maps), we now add
the Ka band data to the coadded maps; we used only Q and V band data for
the 3-year analysis. We believe that we understand the polarized
foreground emission (dominated by synchrotron, traced well by the K-band
data) in the Ka band data well enough to justify the inclusion of the Ka
band \citep{gold/etal:prep}. This, with 2 years more integration, has
led to a significant reduction of the noise power spectra (averaged over
$l=2-7$) in the polarization EE and BB power spectra by a factor of as
much as 2.3 compared to the 3-year analysis. As a result, the EE power
spectrum averaged over $l=2-7$ exceeds the noise by a factor of 10,
i.e., our measurement of the EE power spectrum averaged over $l=2-7$ is
now limited 
by cosmic variance and the possibility of residual foreground emission
and/or systematic errors\footnote{For our limits on the residual
polarized foreground contamination, see \citet{dunkley/etal:prep}.},
rather than by noise. In addition, we have added a 
capability of computing the likelihood of TB and EB power spectra to the
released likelihood code. This allows us to test models in which
non-zero TB and EB correlations can be generated. We discuss this
further in \S~\ref{sec:TB}. 

We continue to use the Markov Chain Monte Carlo (MCMC) technique to
explore the posterior distribution of cosmological parameters
given the measured temperature and polarization power spectra. For
details on the implementation and convergence criteria, see
\citet{dunkley/etal:prep}.   

\subsection{Comments on systematic errors in the  cosmological
  parameters derived from the \map\ 5-year data}

\citet{hinshaw/etal:prep} give extensive descriptions of our limits on
the systematic errors in the \map\ 5-year temperature and polarization
data. With the improved treatment of beams and gain calibrations, we are
confident that the instrumental systematic errors in the 
cosmological results derived from the temperature data are negligible
compared to the statistical errors. As for the polarization data, we find
that the W band polarization data still contain the systematic errors
that we do not 
understand fully, and thus we do not use the W band polarization for the
cosmological analysis. We do not find any evidence for the unaccounted
instrumental systematic errors in the other bands that we use for cosmology.

The most dominant systematic errors in the cosmological analysis are the
foreground emission. Since CMB dominates over the foreground emission in
the temperature data in V and W bands outside the galaxy mask, and we
also reduce the 
sub-dominant foreground
contamination at high Galactic latitudes further by using the K- and
Ka-band data 
(for synchrotron emission), 
the external H$\alpha$ map (for free-free emission), and the external
dust map, the systematic errors from the foreground emission are
unimportant for the temperature data, even at the lowest multipoles where
the foreground is most important \citep{gold/etal:prep}.

We, however, find that the uncertainty in our modeling of the {\it
polarized} foreground is not negligible compared to the statistical
errors. For the 5-year polarization analysis we have used two
independent foreground-cleaning algorithms: one based upon the template
fitting \citep[as developed for the 3-year analysis;
see][]{page/etal:2007}, and the other based upon the Gibbs sampling
\citep{dunkley/etal:prep}. The optical depth, $\tau$, is the parameter
that is most affected by the uncertainty in the polarized foreground model.
The template fitting method gives $\tau=0.087\pm 0.017$.
The Gibbs sampling method gives a range of values from $\tau=0.085\pm 0.025$ to 
$\tau=0.103\pm 0.018$, depending upon various assumptions made about the
properties of the polarized synchrotron and dust emission. Therefore,
the systematic error in $\tau$ is comparable to the statistical error.

This has an implication for the determination of the primordial tilt,
$n_s$, as there is a weak correlation between $n_s$ and $\tau$ (see
Fig.~\ref{fig:ns}): for $\tau=0.087$ we find $n_s=0.963$, while for
$\tau=0.105$ we find $n_s=0.98$. Since the statistical error of $n_s$ is
0.015, the systematic error in $n_s$ (from the polarized foreground) is
comparable to the statistical one. The other parameters that are
correlated with $n_s$, i.e., the baryon density (Fig.~\ref{fig:ns}),
the tensor-to-scalar ratio (Fig.~\ref{fig:tens}), and the amplitude of
non-adiabatic fluctuations (Fig.~\ref{fig:axion} and
\ref{fig:curvaton}), would be similarly affected.
For the parameters that are not correlated with $n_s$ or $\tau$ the
systematic errors are insignificant.

\subsection{External data sets: Hubble constant, luminosity and angular
  diameter distances}
\label{sec:analysis_ex}

\begin{deluxetable}{lccc}
\tablecolumns{4}
\small
\tablewidth{0pt}
\tablecaption{%
Sound horizon scales determined by the \WMAP\ 5-year data.
CMB: the sound horizon scale at the photon decoupling epoch, $z_*$, 
imprinted on the CMB power spectrum;
Matter: the sound horizon scale at the baryon drag epoch, $z_d$,
 imprinted on the matter (galaxy) power spectrum.    
}
\tablehead{ & \colhead{Quantity} & \colhead{Eq.} & \colhead{5-year WMAP}}
\startdata
CMB & $z_*$ & (\ref{eq:zstar}) &
\ensuremath{1090.51\pm 0.95} \nl
CMB & $r_s(z_*)$ & (\ref{eq:rs}) &
\ensuremath{146.8\pm 1.8\ \mbox{Mpc}} \nl
Matter & $z_d$ & (\ref{eq:zd}) &
\ensuremath{1020.5\pm 1.6} \nl
Matter & $r_s(z_d)$ & (\ref{eq:rs}) &
\ensuremath{153.3\pm 2.0\ \mbox{Mpc}} 
\enddata
\label{tab:ruler}
\end{deluxetable}

Aside from the CMB data (including the small-scale CMB measurements),
the main external astrophysical results that we shall use in this
paper for the joint cosmological analysis are the following distance-scale
indicators: 
\begin{itemize}
 \item A Gaussian prior on the present-day Hubble's constant from the
       Hubble Key Project final results, $H_0=72\pm 8~{\rm
 km~s^{-1}~Mpc^{-1}}$ \citep{freedman/etal:2001}. While the uncertainty
       is larger than the \WMAP's determination of $H_0$ for the minimal
       $\Lambda$CDM model (see Table~\ref{tab:summary}), this
       information improves upon limits on the other models, such as
       models with non-zero spatial curvature.
 \item The luminosity distances out to Type Ia supernovae with their
       absolute magnitudes marginalized over uniform priors. 
       We use the ``union'' supernova samples compiled by
       \citet{kowalski/etal:prep}. The union compilation contains 57
       nearby ($0.015<z\lesssim 0.15$) Type Ia supernovae, and 250 high-$z$
       Type Ia supernovae, after selection cuts. The high-$z$ samples
       contain the Type Ia supernovae from the Hubble Space Telescope
       \citep{knop/etal:2003,riess/etal:2004,riess/etal:2007}, the
       SuperNova Legacy 
       Survey (SNLS) \citep{astier/etal:2006}, the Equation of
       State: SupErNovae trace Cosmic Expansion (ESSENCE) survey
       \citep{wood-vasey/etal:2007}, as well as those used in the
       original papers of the discovery of the acceleration of the universe
       \citep{riess/etal:1998,perlmutter/etal:1999}, and the samples
       from \citet{barris/etal:2004,tonry/etal:2003}.
       The nearby Type Ia supernovae are taken from
       \citet{hamuy/etal:1996,riess/etal:1999,jha/etal:2006,krisciunas/etal:2001,krisciunas/etal:2004,krisciunas/etal:2004b}. \citet{kowalski/etal:prep}
       have processed all of these samples using the same light curve
       fitter called {\sf SALT} \citep{guy/etal:2005}, which allowed
       them to combine all the data in a self-consistent fashion.
       The union compilation is the largest to date. The previous
       compilation by \citet{davis/etal:2007} used a smaller number of
       Type Ia supernovae, and did not use the same light curve fitter
       for all the samples. We examine the difference in the derived
       $\Lambda$CDM cosmological parameters between the union
       compilation and Davis et al.'s 
       compilation in Appendix~\ref{sec:sn}.
       While we ignore the systematic errors when we fit the Type
       Ia supernova data, we examine the effects of systematic errors on
       the $\Lambda$CDM  parameters and the dark energy parameters in
       Appendix~\ref{sec:sn}. 
 \item Gaussian priors on the distance ratios, $r_s/D_V(z)$, at $z=0.2$
       and 0.35 measured from the Baryon Acoustic Oscillations (BAO)
       in the distribution of galaxies \citep{percival/etal:2007c}. 
       The CMB observations measure the acoustic oscillations
       in the photon-baryon plasma, which can be used to measure the
       angular diameter distance to the photon decoupling epoch.  The
       same oscillations are  
       imprinted on the distribution of matter, traced by the distribution of
       galaxies, which can be used to measure the angular distances to
       the galaxies that one observes from galaxy surveys. While both
       CMB and galaxies measure the same oscillations, in this paper we
       shall use the term,
       BAO, to refer only to the oscillations in the distribution of
       matter, for definiteness.
\end{itemize}

Here, we describe how we use the BAO data in more detail.
The BAO can be used to measure not only the angular diameter distance,
$D_A(z)$, through the clustering perpendicular to the line of sight, but
also the expansion rate of the universe, $H(z)$, through the clustering
along the line of sight. This is a powerful probe of dark energy;
however, the accuracy of the current data do not allow us to extract
$D_A(z)$ and $H(z)$ separately, as one can barely measure BAO in the
spherically averaged correlation function \citep{okumura/etal:2008}. 

The spherical average gives us the following effective distance measure
\citep{eisenstein/etal:2005}: 
\begin{equation}
 D_V(z) \equiv \left[(1+z)^2D_A^2(z)\frac{cz}{H(z)}\right]^{1/3},
\end{equation}
where $D_A(z)$ is the proper (not comoving) angular diameter distance:
\begin{equation}
 D_A(z) = \frac{c}{H_0}\frac{f_k\left[H_0\sqrt{\left|\Omega_{
				k}\right|}\int_0^z\frac{dz'}{H(z')}\right]}{(1+z)\sqrt{\left|\Omega_{
 k}\right|}},
\label{eq:da}
\end{equation}
where $f_k[x]=\sin x$, $x$, and $\sinh x$ for $\Omega_{k}<0$
($k=1$), $\Omega_{k}=0$ ($k=0$), and $\Omega_{k}>0$ ($k=-1$), 
respectively.

There is an additional subtlety. The peak positions of the (spherically
averaged) BAO depend actually on the {\it ratio} of $D_V(z)$ to the
sound horizon size at the so-called drag epoch, $z_d$, at which baryons
were released from photons. Note that there is no reason why the
decoupling epoch of photons, $z_*$, needs to be the same as the drag
epoch, $z_d$. They would be equal only when the energy density of
photons and that of baryons were equal at the decoupling epoch -- more
precisely, they would be equal only when $R(z)\equiv
3\rho_b/(4\rho_\gamma)=(3\Omega_b/4\Omega_\gamma)/(1+z)\simeq
0.67(\Omega_bh^2/0.023)[1090/(1+z)]$ was unity at $z=z_*$. Since we
happen to live in the universe in which $\Omega_bh^2\simeq 0.023$, this
ratio is less than unity, and thus the drag epoch is slightly later than
the photon decoupling epoch, $z_d<z_*$. As a result, the sound
horizon size at the drag epoch happens to be slightly larger than  that
at the photon decoupling epoch. 
In Table~\ref{tab:ruler} we give the CMB decoupling epoch, BAO drag
epoch, as well as the corresponding sound horizon radii that are
determined from the \WMAP\ 5-year data.

We use a fitting formula for $z_d$ proposed by \citet{eisenstein/hu:1998}:
\begin{equation}
z_d  =
 \frac{1291(\Omega_mh^2)^{0.251}}{1+0.659(\Omega_mh^2)^{0.828}}
\left[1+b_1(\Omega_bh^2)^{b2}\right],
\label{eq:zd}
\end{equation}
where
\begin{eqnarray}
  b_1&=&0.313(\Omega_mh^2)^{-0.419}\left[1+0.607(\Omega_mh^2)^{0.674}\right],\\
  b_2&=&0.238(\Omega_mh^2)^{0.223}.
\end{eqnarray}

In this paper we frequently combine the \map\ data with
$r_s(z_d)/D_V(z)$ extracted from 
the Sloan Digital Sky Survey (SDSS) and the Two Degree Field Galaxy Redshift
Survey (2dFGRS)
\citep{percival/etal:2007c}, where $r_s(z)$ is the comoving sound horizon
size given by  
\begin{equation}
 r_s(z)
=\frac{c}{\sqrt{3}}\int_0^{1/(1+z)}
\frac{da}{a^2H(a)\sqrt{1+(3\Omega_b/4\Omega_\gamma)a}},
\label{eq:rs}
\end{equation} 
where $\Omega_\gamma=2.469\times 10^{-5}h^{-2}$ for $T_{\rm
cmb}=2.725$~K, and
\begin{equation}
 H(a) = H_0\left[\frac{\Omega_m}{a^3}+\frac{\Omega_r}{a^4}+\frac{\Omega_k}{a^2}+\frac{\Omega_\Lambda}{a^{3(1+w_{\rm
	    eff}(a))}}\right]^{1/2}.
\label{eq:hubble}
\end{equation}
The radiation density parameter, $\Omega_r$, is the sum of photons and
relativistic neutrinos,
\begin{equation}
 \Omega_r = \Omega_\gamma \left(1+0.2271N_{\rm eff}\right),
\end{equation}
where $N_{\rm eff}$ is the effective number of neutrino species (the
standard value is 3.04). For more details on $N_{\rm eff}$, see
\S~\ref{sec:neff}. When neutrinos are non-relativistic at $a$, one needs
to reduce the value of $N_{\rm eff}$ accordingly. Also, the matter
density must contain the neutrino contribution when they are non-relativistic,
\begin{equation}
 \Omega_m = \Omega_c + \Omega_b + \Omega_\nu,
\end{equation}
where $\Omega_\nu$ is related to the sum of neutrino masses as
\begin{equation}
 \Omega_\nu = \frac{\sum m_\nu}{94h^2~{\rm eV}}.
\end{equation}
For more details on the neutrino mass, see \S~\ref{sec:massnu}.

All the density parameters refer to the values at the present epoch, and
add up to unity: 
\begin{equation}
 \Omega_m + \Omega_r + \Omega_k + \Omega_\Lambda = 1.
\end{equation}
Throughout this paper, we shall use $\Omega_\Lambda$ to denote the dark energy
density parameter at present:
\begin{equation}
 \Omega_\Lambda \equiv \Omega_{de}(z=0).
\end{equation}

Here, $w_{\rm eff}(a)$ is the effective equation of state of dark energy
given by 
\begin{equation}
 w_{\rm eff}(a)\equiv \frac1{\ln a}\int_0^{\ln a} d\ln a'~w(a'),
\end{equation}
and $w(a)$ is the usual dark energy equation of state, i.e., the dark
energy pressure divided by the dark energy density:
\begin{equation}
  w(a)\equiv \frac{P_{de}(a)}{\rho_{de}(a)}.
\end{equation}
For vacuum energy (cosmological constant), $w$ does not depend on time,
and $w=-1$.

\citet{percival/etal:2007c} have determined $r_s(z_d)/D_V(z)$ out to two
redshifts, $z=0.2$ and 0.35, as $r_s(z_d)/D_V(z=0.2)=0.1980\pm 0.0058$
and $r_s(z_d)/D_V(z=0.35)=0.1094 \pm 0.0033$, with a correlation
coefficient of 0.39. We follow the description given in Appendix~A of
\citet{percival/etal:2007c} to implement these constraints in the
likelihood code. We have checked that our calculation of $r_s(z_d)$
using the formulae above (including $z_d$) matches the value that they
quote\footnote{In \citet{percival/etal:2007c} the authors use a different
notation for the drag redshift, $z_*$, instead of $z_d$. We have
confirmed that they have used equation~(6) of \citet{eisenstein/hu:1998}
for $r_s$, which makes an explicit use of the drag redshift (Will
Percival, private communication).}, $111.426~h^{-1}$~Mpc, to within
$0.2~h^{-1}$~Mpc, for their quoted cosmological parameters, $\Omega_m=0.25$,
$\Omega_bh^2=0.0223$, and $h=0.72$.

We have decided to use these results, as they measure the distances
only, and are not sensitive to the growth of structure. This properly
enables us to identify the information added by the external
astrophysical results more 
clearly. 
In addition to these, we shall also use the BAO
measurement by \citet{eisenstein/etal:2005}\footnote{We use a Gaussian
prior on
$A=D_V(z=0.35)\sqrt{\Omega_mH_0^2}/(0.35c)=0.469(n_s/0.98)^{-0.35}\pm
0.017$.}, and the flux power spectrum
of Ly$\alpha$ forest from \citet{seljak/slosar/mcdonald:2006} in the
appropriate context.  

For the 3-year data analysis in \citet{spergel/etal:2007} we also
used the shape of the galaxy power spectra measured from the SDSS main
sample as well as the Luminous Red Galaxies
\citep{tegmark/etal:2004a,tegmark/etal:2006}, and 2dFGRS
\citep{cole/etal:2005}. We have found some tension between these data
sets, which could be indicative of the degree by which  our
understanding of non-linearities, such as the non-linear matter
clustering, non-linear bias, and non-linear redshift space distortion,
is still limited at low redshifts, i.e., $z\lesssim 1$.
See \citet{dunkley/etal:prep} for more
detailed study on this issue. See also \citet{sanchez/cole:2008} on the
related subject. The galaxy power spectra should provide us with
important information on the growth of structure (which helps constrain
the dark energy and neutrino masses) as our understanding of
non-linearities improves in the future. 
 In this paper we do not combine these data sets because of the limited
 understanding of the consequences of non-linearities. 

\section{Flat, Gaussian, Adiabatic, Power-law $\Lambda$CDM Model and Its
 Alternatives}
\label{sec:inflation}
The theory of inflation, the idea that the universe underwent a brief
period of rapid accelerated expansion
\citep{starobinsky:1979,starobinsky:1982,kazanas:1980,guth:1981,sato:1981,linde:1982,albrecht/steinhardt:1982},
has become an indispensable building block of the standard model of our
universe. 

Models of the early universe must explain the following
observations: the universe is nearly flat and the fluctuations observed
by \WMAP\ appear to be nearly Gaussian \citep{komatsu/etal:2003},
scale-invariant, super-horizon, and adiabatic 
\citep{zaldarriaga/spergel:1997,spergel/etal:2003,peiris/etal:2003}. Inflation
models have been able to explain these 
properties successfully
\citep{mukhanov/chibisov:1981,hawking:1982,starobinsky:1982,guth/pi:1982,bardeen/steinhardt/turner:1983}. 

Although many models have been ruled out observationally \citep[see][for
recent surveys]{kinney/kolb/melchiorri:2006,alabidi/lyth:2006,martin/ringeval:2006}, there
are more than one hundred
candidate inflation models available \citep[see][for
reviews]{liddle/lyth:CIALSS,bassett/tsujikawa/wands:2006,linde:2008}. Therefore,
we now focus on the question, ``{\it which model is the
correct inflation model?}'' This must be answered by the observational data.

On the other hand, an inflationary expansion may not be the only way to
solve cosmological puzzles and create primordial
fluctuations. Contraction of the primordial universe followed by a
bounce to expansion can, in principle, make a set of predictions that
are qualitatively similar to those of inflation models
\citep{khoury/etal:2001,khoury/etal:2002,khoury/etal:2002b,khoury/steinhardt/turok:2003,buchbinder/khoury/ovrut:2007,buchbinder/khoury/ovrut:2008,koyama/wands:2007,koyama/etal:2007,creminelli/senatore:2007}, although building concrete models and making robust predictions have been challenging \citep{kallosh/kofman/linde:2001,kallosh/etal:2001,linde:prep,kallosh/etal:2008}.

There is also a fascinating possibility that one can learn something about
the fundamental physics from cosmological observations. For example, a recent
progress in implementing de Sitter vacua and inflation in the context of
String Theory \citep[see][for a review]{mcallister/silverstein:2008},
makes it possible to connect the cosmological observations to the
ingredients of the fundamental physics via their predicted properties of
inflation such as the shape of the power spectrum, spatial curvature of
the universe, and non-Gaussianity of primordial fluctuations.  

\subsection{Power spectrum of primordial fluctuations}
\label{sec:PK}

\begin{deluxetable*}{llccccc}
\tablecolumns{7}
\small
\tablewidth{0pt}
\tablecaption{Primordial tilt $n_s$, running index $dn_s/d\ln k$, and
 tensor-to-scalar ratio $r$}
\tablehead{\colhead{Sec.}&\colhead{Model} 
& \colhead{Parameter\footnote{Defined at $k_0=0.002~{\rm Mpc}^{-1}$}} 
& \colhead{5-year WMAP\footnote{\citet{dunkley/etal:prep}}} 
& \colhead{5-year WMAP+CMB\footnote{``CMB'' includes the small-scale CMB
 measurements from CBI
 \citep{mason/etal:2003,sievers/etal:2003,sievers/etal:2007,pearson/etal:2003,readhead/etal:2004},
 VSA \citep{dickinson/etal:2004}, ACBAR
 \citep{kuo/etal:2004,kuo/etal:2007}, and BOOMERanG
 \citep{ruhl/etal:2003,montroy/etal:2006,piacentini/etal:2006}}}
& \colhead{5-year WMAP+ACBAR08\footnote{``ACBAR08'' is the complete
 ACBAR data set presented in \citet{reichardt/etal:prep}. We used the
 ACBAR data in the multipole range of $900<l<2000$}}
&\colhead{5-year WMAP+BAO+SN}}
 \startdata
 \S~\ref{sec:PK_results_tilt} & 
 Power-law & $n_s$ & 
\ensuremath{0.963^{+ 0.014}_{- 0.015}} &
\ensuremath{0.960\pm 0.014} &
\ensuremath{0.964\pm 0.014} &
\ensuremath{0.960\pm 0.013} \nl
\hline
 \S~\ref{sec:PK_results_running} & 
Running & $n_s$ &
\ensuremath{1.031^{+ 0.054}_{- 0.055}}\footnote{At the
 pivot point for WMAP only, $k_{pivot}=0.080~{\rm Mpc}^{-1}$, where $n_s$ and $dn_s/d\ln k$
 are uncorrelated, $n_s(k_{pivot})=0.963\pm 0.014$} &
\ensuremath{1.059^{+ 0.051}_{- 0.049}} &
\ensuremath{1.031\pm 0.049} &
\ensuremath{1.017^{+ 0.042}_{- 0.043}}\footnote{At
 the pivot point for WMAP+BAO+SN, $k_{pivot}=0.106~{\rm Mpc}^{-1}$, where $n_s$ and $dn_s/d\ln k$
 are uncorrelated, $n_s(k_{pivot})=0.961\pm 0.014$} \nl
& & $dn_s/d\ln k$ &
\ensuremath{-0.037\pm 0.028} &
\ensuremath{-0.053\pm 0.025} &
\ensuremath{-0.035^{+ 0.024}_{- 0.025}} &
\ensuremath{-0.028\pm 0.020}\footnote{With
 the Ly$\alpha$ forest data \citep{seljak/slosar/mcdonald:2006}, $dn_s/d\ln k=-0.012\pm 0.012$} \nl
\hline
  \S~\ref{sec:GW_results} &
 Tensor & $n_s$ & 
\ensuremath{0.986\pm 0.022} &
\ensuremath{0.979\pm 0.020}&
\ensuremath{0.985^{+ 0.019}_{- 0.020}} &
\ensuremath{0.970\pm 0.015} \nl
& & $r$ & 
\ensuremath{< 0.43\ \mbox{(95\% CL)}} &
\ensuremath{< 0.36\ \mbox{(95\% CL)}}  &
\ensuremath{< 0.40\ \mbox{(95\% CL)}} &
\ensuremath{< 0.22\ \mbox{(95\% CL)}} \nl
\hline
\S~\ref{sec:GW_results} &
Running & $n_s$ & 
\ensuremath{1.087^{+ 0.072}_{- 0.073}} &
\ensuremath{1.127^{+ 0.075}_{- 0.071}}  &
\ensuremath{1.083^{+ 0.063}_{- 0.062}} &
\ensuremath{1.089^{+ 0.070}_{- 0.068}} \nl
 &+Tensor& $r$ & 
\ensuremath{< 0.58\ \mbox{(95\% CL)}} &
\ensuremath{< 0.64\ \mbox{(95\% CL)}}  &
\ensuremath{< 0.54\ \mbox{(95\% CL)}} &
\ensuremath{< 0.55\ \mbox{(95\% CL)}}\footnote{With
 the Ly$\alpha$ forest data \citep{seljak/slosar/mcdonald:2006},
 $r<0.28$ (95\% CL)} \nl
 & & $dn_s/d\ln k$ & 
\ensuremath{-0.050\pm 0.034} &
\ensuremath{-0.072^{+ 0.031}_{- 0.030}}  &
\ensuremath{-0.048\pm 0.027} &
\ensuremath{-0.053\pm 0.028}\footnote{With the Ly$\alpha$ forest data \citep{seljak/slosar/mcdonald:2006},
 $dn_s/d\ln k=-0.017^{+0.014}_{-0.013}$} 
\enddata
\label{tab:ns}
\end{deluxetable*}

\begin{figure*}[ht]
\centering \noindent
\includegraphics[width=18cm]{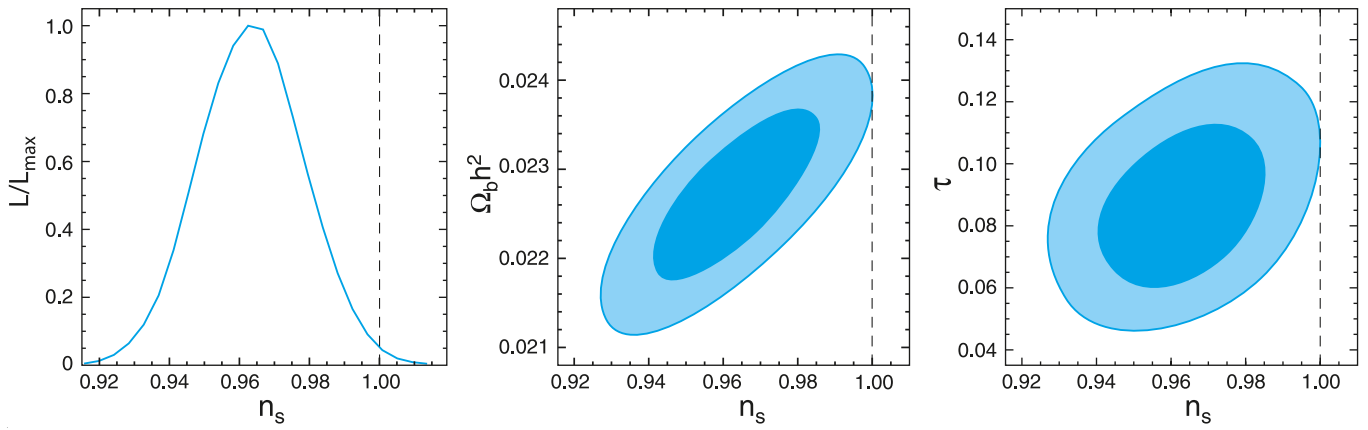}
\caption{%
Constraint on the primordial tilt, $n_s$ (\S~\ref{sec:PK_results_tilt}).
No running index or gravitational
 waves are included in the analysis.
({\it Left}) One-dimensional marginalized constraint on $n_s$ from the
 WMAP-only analysis.
({\it Middle}) Two-dimensional joint marginalized constraint (68\% and
 95\% CL), showing
a strong correlation between $n_s$ and $\Omega_bh^2$.  
({\it Right}) A mild correlation with $\tau$. None of these correlations
 are reduced
 significantly by including  BAO or SN data, as these data sets are not
 sensitive to $\Omega_bh^2$ or $\tau$; however, the situation changes
   when the gravitational wave contribution is included (see
 Fig.~\ref{fig:tens}).} 
\label{fig:ns}
\end{figure*} 

\subsubsection{Motivation and Analysis}
\label{sec:PK_motivation}
The shape of the  power spectrum of primordial curvature perturbations,
$P_{\cal R}(k)$, is one of the most powerful and
practical tool for distinguishing among inflation models. Inflation models
with featureless scalar-field potentials usually predict
that $P_{\cal R}(k)$ is {\it nearly} a power-law
\citep{kosowsky/turner:1995} 
\begin{equation}
\Delta^2_{\cal R}(k) \equiv \frac{k^3P_{\cal R}(k)}{2\pi^2} 
= \Delta^2_{\cal
R}(k_0)\left(\frac{k}{k_0}\right)^{n_s(k_0)-1+\frac12dn_s/d\ln k}. 
\label{eq:pR}
\end{equation}
Here, $\Delta^2_{\cal R}(k)$ is a useful quantity, which gives an
approximate contribution of ${\cal R}$ at a given scale per logarithmic
interval in $k$ to the total variance of ${\cal R}$, as $\langle{\cal
R}^2({\mathbf x})\rangle = \int d\ln k~\Delta^2_{\cal R}(k)$. It is
clear that the special case with $n_s=1$ and $dn_s/d\ln k=0$ results in the
``scale-invariant spectrum,'' in which the contributions of ${\cal R}$
at any scales per logarithmic interval in $k$ to the total variance are
equal. (Hence the term, ``scale invariance.'') Following the usual
terminology, we shall call $n_s$ and $dn_s/d\ln k$ the tilt of the
spectrum and the running index, respectively. We shall take $k_0$ to be
$0.002~{\rm Mpc}^{-1}$. 

The significance of $n_s$ and $dn_s/d\ln k$ is that different inflation
models motivated by different physics make specific, testable
predictions for the values of  $n_s$ and $dn_s/d\ln k$.  For a given
shape of the scalar field potential, $V(\phi)$, of a single-field model,
for instance, one finds that $n_s$ is given by a combination of the
first derivative and second derivative of the potential,
$1-n_s=3M_{pl}^2(V'/V)^2 - 2M_{pl}^2(V''/V)$ (where $M_{pl}^2=1/(8\pi G)$ is
the reduced Planck mass), and $dn_s/d\ln k$ is given by a
combination of $V'/V$, $V''/V$, and $V'''/V$ \citep[see][for a
review]{liddle/lyth:CIALSS}. 

This means that one can reconstruct the shape of $V(\phi)$ up to the
first three derivatives in this way. As the expansion rate squared is
proportional to $V(\phi)$ via the Friedmann equation,
$H^2=V/(3M_{pl}^2)$, one can reconstruct the expansion history during
inflation from measuring the shape of the primordial power spectrum. 

How generic are $n_s$ and $dn_s/d\ln k$? They are motivated physically
by the fact that most inflation models satisfy the slow-roll conditions,
and thus deviations from a pure power-law, scale-invariant spectrum,
$n_s=1$ and $dn_s/d\ln k=0$, are expected to be small, and the
higher-order derivative terms such as $V''''$ and higher are
negligible. On the other hand, there is always a danger of missing some
important effects, such as sharp features, when one relies too much on a
simple parametrization like this. Therefore, a number of people have
investigated various, more general ways of reconstructing the shape of
$P_{\cal R}(k)$
\citep{matsumiya/sasaki/yokoyama:2002,matsumiya/sasaki/yokoyama:2003,mukherjee/wang:2003,mukherjee/wang:2003b,mukherjee/wang:2003c,bridle/etal:2003,kogo/etal:2004,kogo/sasaki/yokoyama:2005,hu/okamoto:2004,hannestad:2004,shafieloo/souradeep:2004,sealfon/verde/jimenez:2005,tocchini-valentini/doupis/silk:2005,tocchini-valentini/hoffman/silk:2006,spergel/etal:2007,shafieloo/souradeep:2007,verde/peiris:prep},  
and $V(\phi)$
\citep{lidsey/etal:1997,grivell/liddle:2000,kadota/etal:2005,covi/etal:2006,lesgourgues/valkenburg:2007,powell/kinney:2007}.  

These studies have indicated that the parametrized form
(Eq.~[\ref{eq:pR}]) is basically a good fit, and no significant features
were detected. Therefore, we do not repeat this type of analysis in this
paper, but focus on constraining the parametrized form given by
Eq.~(\ref{eq:pR}). 

Finally, let us comment on the choice of priors. We impose uniform priors
on $n_s$ and $dn_s/d\ln k$, but there are other possibilities for the
choice  of priors. For example, one may impose uniform priors on the
slow-roll parameters, $\epsilon=(M_{pl}^2/2)(V'/V)^2$,
$\eta=M_{pl}^2(V''/V)$, and  $\xi=M_{pl}^4(V'V'''/V^2)$, as well as on
the number of e-foldings, $N$, rather than on $n_s$ and $dn_s/d\ln k$
\citep{peiris/easther:2006,peiris/easther:2006b,easther/peiris:2006}. It
has been found that, as long as one imposes a reasonable lower bound on
$N$, $N>30$, both approaches yield similar results. 

To constrain $n_s$ and $dn_s/d\ln k$, we shall use the \map\ 5-year
temperature and polarization data, the small-scale CMB data, and/or BAO
and SN distance measurements.  
In Table~\ref{tab:ns} we summarize our results presented in
\S~\ref{sec:PK_results_tilt}, \ref{sec:PK_results_running}, and
\ref{sec:GW_results}.  

\subsubsection{Results: Tilt}
\label{sec:PK_results_tilt}
First, we test the inflation models with $dn_s/d\ln k=0$, and negligible
gravitational waves.  The \WMAP\ 5-year temperature and polarization data yield
\ensuremath{n_s = 0.963^{+ 0.014}_{- 0.015}}, which is slightly above
the 3-year value with a smaller uncertainty, $n_s(3{\rm yr})=0.958\pm
0.016$ \citep{spergel/etal:2007}. We shall provide the reason for this
small upward shift in \S~\ref{sec:PK_results_running}.

The scale-invariant,
Harrison--Zel'dovich--Peebles spectrum, $n_s=1$, is at 2.5 standard
deviations away from the mean of the likelihood for the WMAP-only analysis. The
significance increases to 3.1 standard deviations for WMAP+BAO+SN.
Looking at the two-dimensional constraints that include $n_s$, we find
that the most dominant correlation that is still left is the correlation
between $n_s$ and $\Omega_bh^2$ (see Fig.~\ref{fig:ns}).
The larger $\Omega_bh^2$ is, the
smaller the second peak becomes, and the larger $n_s$ is required to
compensate it. Also, the larger $\Omega_bh^2$ is, the larger the Silk
damping (diffusion damping) becomes, and the larger $n_s$ is required to
compensate it. 

This argument suggests that the constraint on $n_s$
should improve as we add more information from the small-scale CMB
measurements that probe the Silk damping scales; however, the current
data do not improve the 
constraint very much yet: \ensuremath{n_s = 0.960\pm 0.014}
from WMAP+CMB, where ``CMB'' includes the small-scale CMB measurements
from CBI
\citep{mason/etal:2003,sievers/etal:2003,sievers/etal:2007,pearson/etal:2003,readhead/etal:2004},
VSA \citep{dickinson/etal:2004}, ACBAR
\citep{kuo/etal:2004,kuo/etal:2007}, and BOOMERanG
\citep{ruhl/etal:2003,montroy/etal:2006,piacentini/etal:2006}, all of
which go well beyond the \WMAP\ angular resolution, so that their
small-scale data are statistically independent of the \WMAP\ data.

We find that the small-scale CMB data do not improve the determination
of $n_s$ because of their relatively large statistical errors.
We also find that the calibration and beam errors are important.
Let us examine this using the latest ACBAR data
\citep{reichardt/etal:prep}. The WMAP+ACBAR yields
\ensuremath{0.964\pm 0.014}.
When the beam error of ACBAR is ignored, we find 
$n_s=0.964\pm 0.013$.
When the calibration error is ignored, we find 
$n_s=0.962\pm 0.013$.
Therefore, both the beam and calibration error are important in the
error budget of the ACBAR data.

The Big Bang Nucleosynthesis (BBN), combined with measurements of the
deuterium-to-hydrogen ratio, D/H, from quasar absorption systems, has
been used extensively for determining $\Omega_bh^2$, independent of any other
cosmological parameters \citep[see][for a recent summary]{steigman:2007}.
The precision of the latest determination of
$\Omega_bh^2$ from BBN \citep{pettini/etal:prep} is comparable to that
of the \WMAP\ data-only analysis. More precise measurements of D/H will help
reduce the correlation between $n_s$ and $\Omega_bh^2$, yielding a better
determination of $n_s$.

There is still a bit of correlation left between $n_s$ and the
electron-scattering optical depth, $\tau$ (see Fig.~\ref{fig:ns}). While
a better measurement of $\tau$ from future \WMAP\ observations as
well as the {\sl Planck} satellite should help reduce the uncertainty in
$n_s$ via a better measurement of $\tau$, the effect of $\Omega_bh^2$ is
much larger. 

We find that the other data sets, such as BAO, SN, and the shape of
galaxy power spectrum from SDSS or 2dFGRS, do not improve our
constraints on $n_s$, as these data sets are not sensitive to
$\Omega_bh^2$ or $\tau$; however, this will change when we include the 
running index, $dn_s/d\ln k$ (\S~\ref{sec:PK_results_running}) and/or
the tensor-to-scalar ratio, $r$ (\S~\ref{sec:GW_results}).

\subsubsection{Results: Running index}
\label{sec:PK_results_running}
Next, we explore more general models in which a sizable running index may
be generated. (We still do not include gravitational waves; see
\S~\ref{sec:GW} for the analysis that includes gravitational waves.) We
find no evidence for $dn_s/d\ln k$ from WMAP only, 
\ensuremath{-0.090<dn_s/d\ln{k}<0.019\ \mbox{(95\% CL)}},
 or WMAP+BAO+SN 
\ensuremath{-0.068<dn_s/d\ln{k}<0.012\ \mbox{(95\% CL)}}.
The improvement from WMAP-only to WMAP+BAO+SN is only modest.

We find a slight upward shift from the 3-year result, 
$dn_s/d\ln k=-0.055^{+0.030}_{-0.031}$ \citep[68\% CL;][]{spergel/etal:2003},
to the 5-year result, 
\ensuremath{dn_s/d\ln{k} = -0.037\pm 0.028} (68\% CL; WMAP only).
This is caused by a combination of three effects:
\begin{itemize}
 \item[(i)] The 3-year number for $dn_s/d\ln k$ was derived from an older
	    analysis pipeline for the temperature data, namely, the
	    resolution 3 (instead of 4) pixel-based low-$l$ temperature
	    likelihood and a higher point-source amplitude,
	    $A_{ps}=0.017~{\rm \mu K^2~sr}$. 
 \item[(ii)] With two years more integration, we have a better
	     signal-to-noise near the third acoustic peak, whose
	     amplitude is slightly higher than the 3-year determination
	     \citep{nolta/etal:prep}. 
 \item[(iii)] With the improved beam model \citep{hill/etal:prep}, the
	      temperature power spectrum at 
	      $l\gtrsim 200$ has been raised nearly uniformly by $\sim 2.5$\%
	      \citep{hill/etal:prep,nolta/etal:prep}.
\end{itemize}
All of these effects go in the same direction, i.e., to increase the
power at high multipoles and reduce a negative running index. 
We find that these factors contribute to the upward shift in
$dn_s/d\ln k$ at comparable levels.

Note that an upward shift in $n_s$ for a power-law model,
0.958 to 0.963 (\S~\ref{sec:PK_results_tilt}), is not subject to (i)
because the 3-year number for $n_s$ 
was derived from an updated analysis pipeline
using the resolution 4 low-$l$ code and $A_{ps}=0.014~{\mu\rm K}^2~{\rm
sr}$. We find that (ii) and (iii) contribute to the shift in $n_s$ at comparable
levels. 
An upward shift in $\sigma_8$ from the 3-year value, 0.761, to the
5-year value, 0.796, can be explained similarly.

We do not find any significant evidence for the running index
when the \map\ data and small-scale CMB data (CBI, VSA, ACBAR07,
BOOMERanG) are combined, 
\ensuremath{-0.1002<dn_s/d\ln{k}<-0.0037\ \mbox{(95\% CL)}},
or 
the \map\ data and  the latest results from the analysis of the complete ACBAR data
\citep{reichardt/etal:prep} are combined,
\ensuremath{-0.082<dn_s/d\ln{k}<0.015\ \mbox{(95\% CL)}}.

Our best 68\% CL constraint from WMAP+BAO+SN shows no evidence for the
running index, 
\ensuremath{dn_s/d\ln{k} = -0.028\pm 0.020}.
In order to improve upon the limit on $dn_s/d\ln k$, one needs to
determine $n_s$ at small scales, as $dn_s/d\ln k$ is simply given by
the difference between $n_s$'s measured at two different scales, divided
by the logarithmic separation between two scales.
The Ly$\alpha$ forest provides such information
(see \S~\ref{sec:conclusion}; also Table~\ref{tab:ns}).

\subsection{Primordial gravitational waves}
\label{sec:GW}
\subsubsection{Motivation}
\label{sec:GW_motivation} 
The presence of primordial gravitational waves is a robust prediction of 
inflation models, as the same mechanism that generated primordial
density fluctuations should also generate primordial gravitational waves
\citep{grishchuk:1975,starobinsky:1979}. The amplitude of gravitational waves
relative to that of density fluctuations is model-dependent. 

The primordial gravitational waves leave their signatures imprinted on the CMB
temperature anisotropy
\citep{rubakov/sazhin/veryaskin:1982,fabbri/pollock:1983,abbott/wise:1984,starobinsky:1985,crittenden/etal:1993},
as well as on polarization
\citep{basko/polnarev:1980,bond/efstathiou:1984,polnarev:1985,crittenden/davis/steinhardt:1993,crittenden/coulson/turok:1995,coulson/crittenden/turok:1994}.\footnote{See,
e.g., \citet{watanabe/komatsu:2006} for the spectrum of the primordial
gravitational waves itself.} The
spin-2 nature of gravitational waves leads to two types of polarization
pattern on the sky
\citep{zaldarriaga/seljak:1997,kamionkowski/kosowsky/stebbins:1997}: (i)
the curl-free mode (E mode), in which the polarization directions are
either purely radial or purely tangential to hot/cold spots in
temperature, and (ii) the divergence-free mode (B mode), in which the
pattern formed by polarization directions around hot/cold spots possess
non-zero vorticity.  

In the usual gravitational instability picture, in the linear regime
(before shell crossing of fluid elements)  velocity perturbations can be
written in terms of solely a gradient of a scalar velocity potential
$u$, $\vec{v}=\vec{\nabla}u$. This means that no vorticity would arise,
and therefore no B mode polarization can be generated from density or
velocity perturbations in the linear regime. However, primordial gravitational
waves can generate both E and B mode polarization; thus, the B mode
polarization offers a smoking-gun signature for the presence of
primordial gravitational waves
\citep{seljak/zaldarriaga:1997,kamionkowski/kosowsky/stebbins:1997b}. This
gives us a strong motivation to look for signatures of the primordial
gravitational waves in CMB. 

In Table~\ref{tab:ns} we summarize our constraints on the amplitude of
gravitational waves, expressed in terms of the tensor-to-scalar ratio, $r$,
defined by equation~(\ref{eq:tsratio}).

\subsubsection{Analysis}
\label{sec:GW_analysis}
We quantify the amplitude and spectrum of primordial gravitational waves in
the following form:
\begin{equation}
 \Delta^2_h(k) \equiv \frac{k^3P_h(k)}{2\pi^2}=
\Delta^2_h(k_0) \left(\frac{k}{k_0}\right)^{n_t},
\label{eq:Ph}
\end{equation}
where we have ignored a possible scale dependence of $n_t(k)$, as the
current data cannot constrain it. Here, by $P_h(k)$ we mean
\begin{equation}
 \langle \tilde{h}_{ij}({\mathbf k})\tilde{h}^{ij}({\mathbf k}')\rangle
= (2\pi)^3P_h(k)\delta^3({\mathbf k}-{\mathbf k}'),
\end{equation}
where $\tilde{h}_{ij}({\mathbf k})$ is the Fourier transform of 
the tensor metric perturbations, $g_{ij}=a^2(\delta_{ij}+h_{ij})$, which
can be further decomposed into the two polarization states ($+$ and
$\times$) with the appropriate polarization tensor, $e^{(+,\times)}_{ij}$, as
\begin{equation}
 \tilde{h}_{ij}({\mathbf k}) 
= \tilde{h}_+({\mathbf k})e_{ij}^+({\mathbf k}) + \tilde{h}_\times({\mathbf
k})e_{ij}^{\times}({\mathbf k}),
\end{equation}
with the normalization that
$e^{+}_{ij}e^{+,ij}=e^{\times}_{ij}e^{\times,ij}=2$ and 
$e^{+}_{ij}e^{\times,ij}=0$.
Unless there was a parity-violating interaction term such as 
$f(\phi) R_{\mu\nu\rho\sigma}\tilde{R}^{\mu\nu\rho\sigma}$ where
$f(\phi)$ is an arbitrary function of a scalar field, 
$R_{\mu\nu\rho\sigma}$ is the Riemann tensor, and 
$\tilde{R}^{\mu\nu\rho\sigma}$ is a dual tensor
\citep{lue/wang/kamionkowski:1999}, 
both polarization states are statistically independent with the same
amplitude, meaning
\begin{equation}
 \langle |\tilde{h}_+|^2\rangle = \langle
  |\tilde{h}_\times|^2\rangle\equiv \langle|\tilde{h}^2|\rangle,\qquad \langle 
  \tilde{h}_\times\tilde{h}_+^*\rangle=0.
\end{equation}
This implies that parity-violating correlations such as the TB and EB
correlations must vanish. We shall explore such parity-violating
correlations in \S~\ref{sec:TB} in a slightly different context.
For limits on the difference between 
$\langle |\tilde{h}_+|^2\rangle$ and $\langle
  |\tilde{h}_\times|^2\rangle$
 from the TB and EB spectra of the \map\ 3-year data, see
\citet{saito/ichiki/taruya:2007}. 

In any case, this definition suggests that $P_h(k)$ is given by
$P_h(k) = 4\langle |\tilde{h}|^2\rangle$. Notice a factor of 4.
This is the definition of $P_h(k)$ that we have been using consistently
since the first year release
\citep{peiris/etal:2003,spergel/etal:2003,spergel/etal:2007,page/etal:2007}. We
continue to use this definition.

With this definition of $\Delta^2_h(k)$ (Eq.~[\ref{eq:Ph}]), we define the tensor-to-scalar
ratio, $r$, at $k=k_0$, as
\begin{equation}
 r\equiv \frac{\Delta_h^2(k_0)}{\Delta_{\cal R}^2(k_0)},
\label{eq:tsratio}
\end{equation}
where $\Delta_{\cal R}(k)$ is the curvature perturbation spectrum given
by  equation~(\ref{eq:pR}). We shall take $k_0$ to be $0.002~{\rm Mpc}^{-1}$.
In this paper we sometimes call this quantity loosely the ``amplitude of gravitational
waves.'' 

What about the tensor spectral tilt, $n_t$? 
In single-field inflation models, there exists the so-called consistency
relation between $r$ and $n_t$ \citep[see][for a review]{liddle/lyth:CIALSS}
\begin{equation}
 n_t = -\frac{r}{8}.
\end{equation}
In order to reduce the number of parameters, we continue to impose this
relation at $k=k_0=0.002~{\rm Mpc}^{-1}$. For a discussion on how to
impose this constraint in a more self-consistent way, see
\citet{peiris/easther:2006}. 

To constrain $r$, we shall use the \map\ 5-year
temperature and polarization data, the small-scale CMB data, and/or BAO
and SN distance measurements. 

\subsubsection{How \WMAP\ constrains the amplitude of gravitational waves}
\label{sec:GW_pedagogical}

\begin{figure*}[ht]
\centering \noindent
\includegraphics[width=18cm]{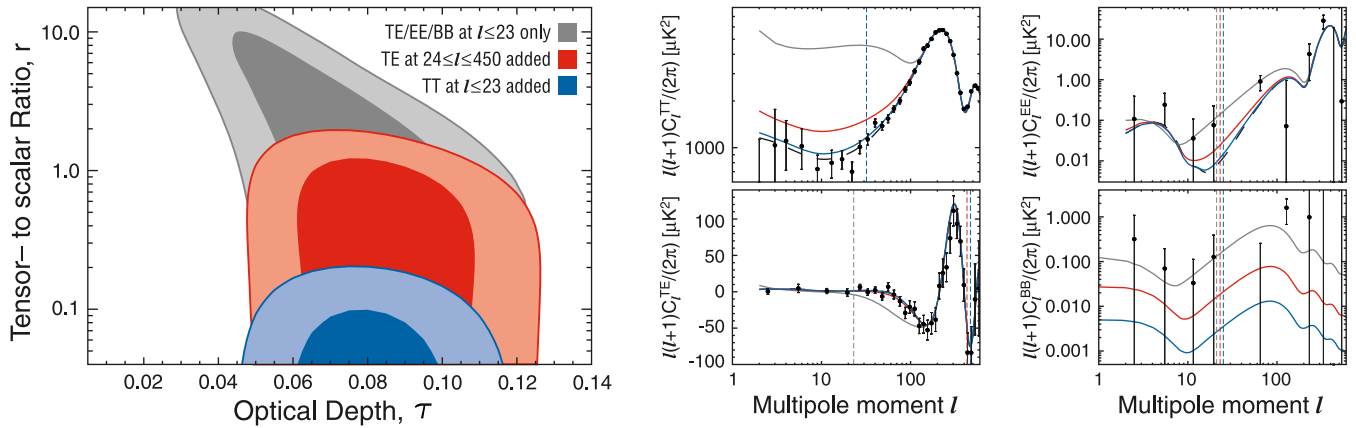}
\caption{%
 How the \map\ temperature and polarization data constrain the
 tensor-to-scalar ratio, $r$.
 ({\it Left}) The contours show 68\% and 95\% CL. The gray region is derived
 from the low-$l$ polarization data (TE/EE/BB at $l\le 23$) only, the
 red region from the low-$l$ polarization plus the high-$l$ TE data at
 $l\le 450$, and the blue region from the low-$l$ polarization, the
 high-l TE, and the low-$l$ temperature data at $l\le 32$. ({\it Right})
 The gray curves show $(r,\tau)=(10,0.050)$, the red curves
 $(r,\tau)=(1.2,0.075)$, and the blue curves $(r,\tau)=(0.20,0.080)$,
 which are combinations of $r$ and $\tau$ that give the upper edge of
 the 68\%  CL contours shown on the left panel. The vertical lines
 indicate the maximum multipoles below which the data are used for each
 color. The data points with 68\% CL errors are the  WMAP 5-year
 measurements \citep{nolta/etal:prep}.
 (Note that the BB power
 spectrum at $l\sim 130$ is consistent with zero within 95\% CL.)
}
\label{fig:pedagogical}
\end{figure*}

Let us show how the gravitational wave contribution is constrained by the
\map\ data (see Fig.~\ref{fig:pedagogical}). In this pedagogical
analysis, we vary only $r$ and $\tau$, while adjusting the overall amplitude of
fluctuations such that the height of the first peak of the temperature power
spectrum is always held fixed. All the other cosmological parameters are
fixed at $\Omega_k=0$, $\Omega_bh^2=0.02265$, $\Omega_ch^2=0.1143$,
$H_0=70.1~{\rm km~s^{-1}~Mpc^{-1}}$, and $n_s=0.960$.
Note that the limit on $r$ from this analysis should not be taken as our
best limit, as this analysis ignores the correlation between $r$ and the
other cosmological parameters, especially $n_s$. The limit on $r$ from
the full analysis will be given in \S~\ref{sec:GW_results}.

\begin{itemize}
 \item[(1)] (The gray contours in the left panel, and the upper right of the
	    right panel of Fig.~\ref{fig:pedagogical}.)
	    The low-l polarization data (TE/EE/BB) at $l\lesssim 10$ are
	    unable to place meaningful limits on $r$. A large $r$ can be
	    compensated by a small $\tau$, producing nearly the same EE
	    power spectrum at $l\lesssim 10$. (Recall that the gravitational
	    waves contribute also to EE.) As a result, $r$ that is
	    as large as 10 is still allowed within 68\% CL.\footnote{We have performed a similar, but 
 different, analysis in \S~6.2 of \citet{page/etal:2007}. In this paper
 we include both the scalar and tensor contributions to EE, whereas in
 \citet{page/etal:2007} we have ignored the tensor contribution to EE
 and found a somewhat tighter limit, $r<4.5$ (95\% CL), from the low-$l$
 polarization data. This is because, when the 
 tensor contribution was ignored, the EE polarization could still be
 used to fix $\tau$, whereas in our case $r$ and $\tau$ are fully
 degenerate when $r\gtrsim 1$ (see Fig.~\ref{fig:pedagogical}), as the
 EE is also dominated by the tensor contribution for such a high value of $r$.}
 \item[(2)] (The red contours in the left panel, and the lower left of the right panel of
	    Fig.~\ref{fig:pedagogical}.) Such a high value of $r$,
	    however, produces too negative a TE correlation between
	    $30\lesssim l\lesssim 150$. Therefore, we can improve the
	    limit on $r$ significantly -- by nearly an order of
	    magnitude -- by simply using the high-l TE data. The 95\%
	    upper bound at this point is still as large as  $r\sim
	    2$.\footnote{See
	    \citet{polnarev/miller/keating:2008,miller/keating/polnarev:prep}
	    for a way to constrain $r$ from the TE power spectrum alone.} 
 \item[(3)] (The blue contours in the left panel, and the upper left of
	    the right panel of 
	    Fig.~\ref{fig:pedagogical}.) Finally, the low-$l$
	    temperature data at $l\lesssim 30$ severely limit the excess
	    low-$l$ power due to gravitational waves, bringing the upper bound
	    down to $r\sim 0.2$. Note that this bound is about a half of
	    what we actually obtain from the full Markov Chain Monte
	    Carlo of the WMAP-only analysis,
	    \ensuremath{r < 0.43\ \mbox{(95\% CL)}}. This is
	    because we have fixed $n_s$, and thus ignored the correlation
	    between $r$ and $n_s$ shown by Fig.~\ref{fig:tens}. 
\end{itemize}

\subsubsection{Results}
\label{sec:GW_results}
\begin{figure*}[ht]
\centering \noindent
\includegraphics[width=18cm]{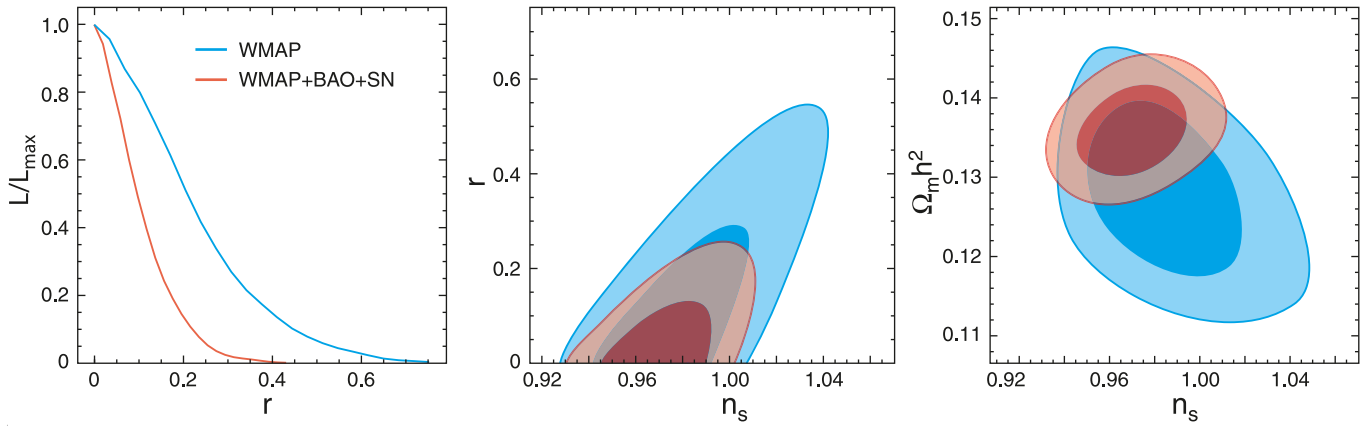}
\caption{%
Constraint on the tensor-to-scalar ratio, $r$, at $k=0.002~{\rm
 Mpc}^{-1}$ (\S~\ref{sec:GW_results}). No running index is assumed. See
 Fig.~\ref{fig:run+tens} for $r$ with the running index. 
 In all panels we show
 the WMAP-only results in blue and WMAP+BAO+SN in red.
 ({\it Left}) One-dimensional marginalized distribution of $r$, showing 
 the WMAP-only limit, \ensuremath{r < 0.43\ \mbox{(95\% CL)}},
 and WMAP+BAO+SN, 
\ensuremath{r < 0.22\ \mbox{(95\% CL)}}.
({\it Middle}) Joint two-dimensional marginalized distribution (68\% and
 95\% CL), showing  a strong correlation between $n_s$ and $r$.
({\it Right}) Correlation between $n_s$ and $\Omega_mh^2$. The BAO
 and SN data help to break this correlation which, in turn, reduces
 the correlation between $r$ and $n_s$, resulting in a factor of 2.2
 better limit on  $r$.
}
\label{fig:tens}
\end{figure*}
\begin{figure*}[ht]
\centering \noindent
\includegraphics[width=18cm]{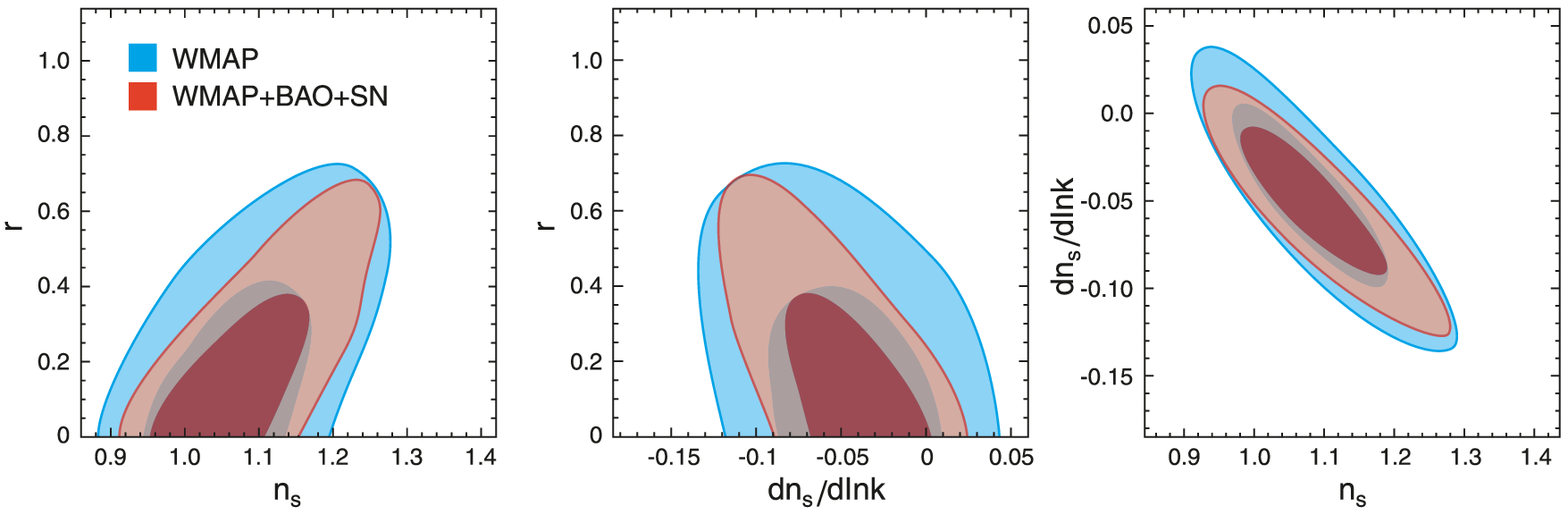}
\caption{%
Constraint on the tensor-to-scalar ratio, $r$,
the tilt, $n_s$, and the running index, $dn_s/d\ln k$, when all of them
 are allowed to vary (\S~\ref{sec:GW_results}). In all panels we show
 the WMAP-only results in blue and WMAP+BAO+SN in red.
({\it Left}) Joint two-dimensional marginalized distribution of 
$n_s$ and $r$ at $k=0.002~{\rm Mpc}^{-1}$ (68\% and
 95\% CL).
({\it Middle}) $n_s$ and $dn_s/d\ln k$.
({\it Right}) $dn_s/d\ln k$ and $r$.
We find no evidence for the running index. While the inclusion of the
 running index weakens our constraint on $n_s$ and $r$, the data do not
 support any need for treating the running index as a free parameter: changes
 in $\chi^2$ between the power-law model and the running model are
 $\chi^2({\rm running})-\chi^2({\rm power-law})\simeq -1.8$ with and without
 the tensor modes for WMAP5+BAO+SN, and 1.2 for WMAP5.
}
\label{fig:run+tens}
\end{figure*}

Having understood which parts of the temperature and polarization
spectra constrain $r$, we obtain the upper limit on $r$ from the full
exploration of the likelihood space using the Markov Chain Monte Carlo.
Figure~\ref{fig:tens} shows the 1-d constraint on $r$ as well as the
2-d constraint on $r$ and $n_s$, assuming a negligible running
index. With the \map\ 5-year data alone, we find  
\ensuremath{r < 0.43\ \mbox{(95\% CL)}}.
Since the B-mode contributes little here, and most of the
information essentially comes from TT and TE, our limit on $r$ is
highly degenerate with $n_s$, and thus we can obtain a better limit on
$r$ only when we have a better limit on $n_s$. 

When we add BAO and SN data, the limit improves significantly to 
\ensuremath{r < 0.22\ \mbox{(95\% CL)}}.
This is because the distance information from BAO and SN
reduces the uncertainty in the matter density, and thus
it helps to determine $n_s$ better because $n_s$ is also degenerate with
the matter density. This ``chain of correlations'' helped us improve our
limit on $r$ significantly from the previous results. This limit, 
\ensuremath{r < 0.22\ \mbox{(95\% CL)}}, is the best
limit on $r$ to date.\footnote{This is the one-dimensional
marginalized 95\% limit. From the joint two-dimensional marginalized
distribution of $n_s$ and $r$, we find $r<0.27$ (95\% CL) at
$n_s=0.99$. See Fig.~\ref{fig:tens}.}
With the new data, we are able to get more
stringent the limits than our earlier analyses \citep{spergel/etal:2007}
that combined the \WMAP\ data with measurements of the galaxy power
spectrum and found $r < 0.30$ (95\% CL). 

A dramatic reduction in the uncertainty in $r$ has an important
implication for $n_s$ as well. Previously $n_s>1$ was within the 95\% CL
when the gravitational wave contribution 
was allowed, owing to the correlation between $n_s$
and $r$. Now, we are beginning to disfavor
$n_s> 1$ even when $r$ is non-zero: with WMAP+BAO+SN we find
\ensuremath{-0.0022<1-n_s<0.0589\ \mbox{(95\% CL)}}.\footnote{This
is is the one-dimensional marginalized 95\% limit. From the joint
two-dimensional marginalized distribution of $n_s$ and $r$, we find
$n_s<1.007$ (95\% CL) at $r=0.2$. See Fig.~\ref{fig:tens}.} 

However, these stringent limits on $r$ and $n_s$ weaken to
\ensuremath{-0.246<1-n_s<0.034\ \mbox{(95\% CL)}}
and \ensuremath{r < 0.55\ \mbox{(95\% CL)}} when
a sizable running index is allowed. 
The BAO and SN data helped reduce the uncertainty in $dn_s/d\ln k$
(Fig.~\ref{fig:run+tens}), but not enough to improve on the other
parameters compared to the \map-only constraints. 
The Ly$\alpha$ forest data can improve the limit on $dn_s/d\ln k$ even
when $r$ is present (see \S~\ref{sec:conclusion}; also Table~\ref{tab:ns}).

\subsection{Implications for inflation models}
\label{sec:summary_inflation}
\begin{figure}[ht]
\centering \noindent
\includegraphics[width=7.3cm]{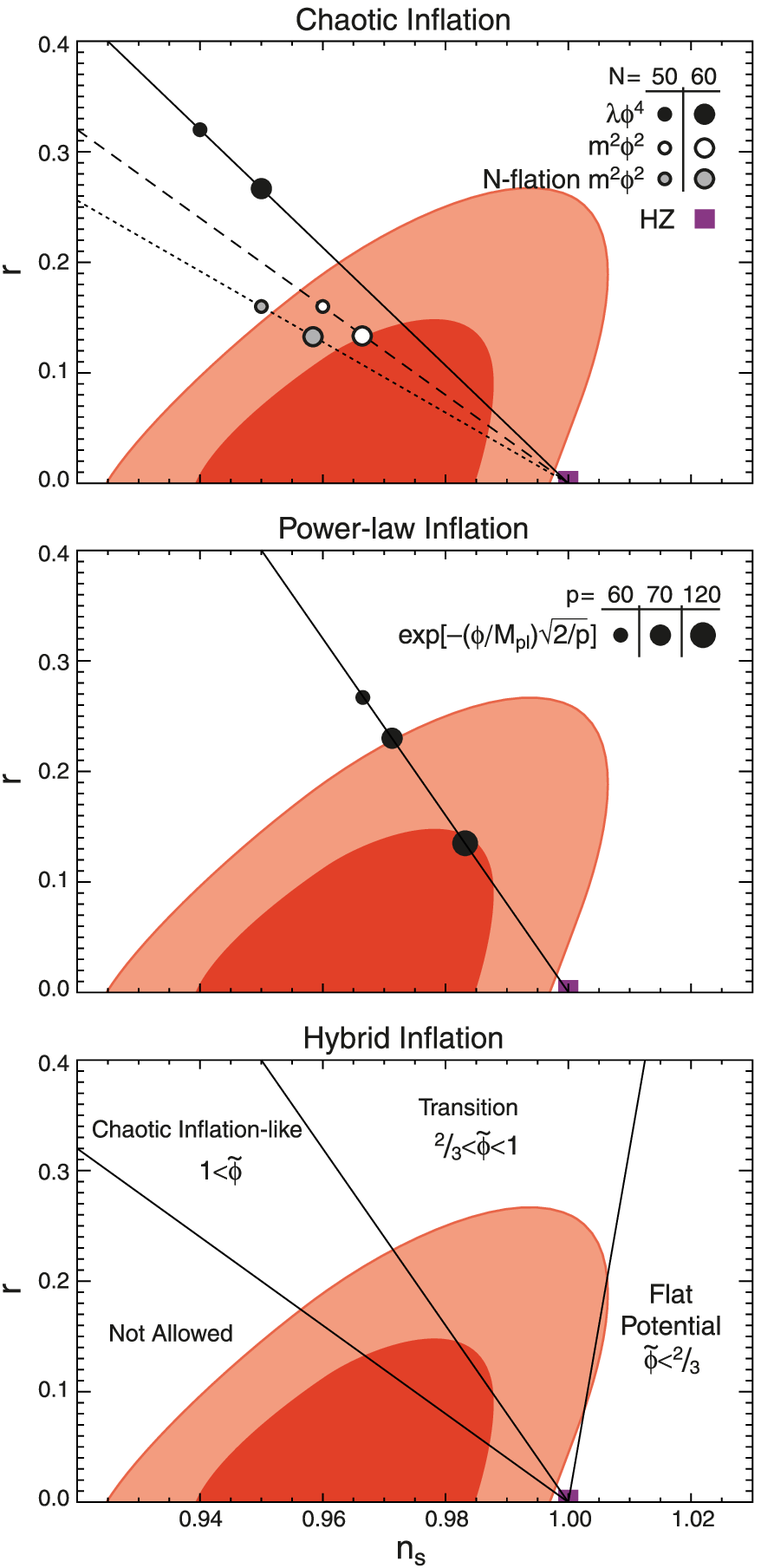}
\caption{%
 Constraint on three representative inflation models whose potential is
 positively curved, $V''>0$ (\S~\ref{sec:summary_inflation}). The
 contours show the 68\% and 95\% CL derived from WMAP+BAO+SN.
 ({\it Top}) The monomial, chaotic-type
 potential, $V(\phi)\propto \phi^\alpha$ \citep{linde:1983}, with
 $\alpha=4$ (solid) and 
 $\alpha=2$ (dashed) for single-field models, and $\alpha=2$ for
 multi-axion field models with $\beta=1/2$
 \citep{easther/mcallister:2006}  (dotted). The symbols show the
 predictions from 
 each of these models with the number of  $e$-folds of inflation equal
 to 50 and 60. The $\lambda\phi^4$ potential is excluded convincingly,
 the $m^2\phi^2$ single-field model lies outside of (at the boundary of)
 the 68\% region for $N=50$ (60), and the $m^2\phi^2$ multi-axion model
 with $N=50$ lies outside of the 95\% region. ({\it Middle}) The
 exponential potential, $V(\phi)\propto \exp[-(\phi/M_{pl})\sqrt{2/p}]$,
 which leads to a power-law inflation, $a(t)\propto t^p$
 \citep{abbott/wise:1984,lucchin/matarrese:1985}. All models but $p\sim
 120$ are outside of the 68\% region. The models with $p<60$ are
 excluded at more than 99\% CL, and those with $p<70$ are outside of the
 95\% region. For multi-field models these limits can be translated into
 the number of fields as $p\rightarrow np_i$, where $p_i$ is the
 $p$-parameter of each field \citep{liddle/mazumdar/schunck:1998}. The
 data favour $n\sim 120/p_i$ fields. 
 ({\it Bottom}) The hybrid-type potential,
 $V(\phi)=V_0+(1/2)m^2\phi^2=V_0(1+\tilde{\phi}^2)$, where
 $\tilde{\phi}\equiv m\phi/(2V_0)^{1/2}$ \citep{linde:1994}. The models with
 $\tilde{\phi}<2/3$ drive inflation by the vacuum energy term,
 $V_0$, and are disfavoured at  more than 95\% CL, while those with
 $\tilde{\phi}>1$ drive inflation by the quadratic term, and are similar
 to the chaotic type (the left panel with $\alpha=2$). The transition
 regime, $2/3<\tilde{\phi}<1$ are outside of the 68\% region, but still
 within the 95\% region.
}
\label{fig:inflation}
\end{figure}
How do the \WMAP\ 5-year limits on $n_s$ and $r$ constrain inflationary
models?\footnote{For recent surveys of inflation models in light of the
\WMAP\ 3-year data, see
\citet{alabidi/lyth:2006,kinney/kolb/melchiorri:2006,martin/ringeval:2006}.}   
In the context of single-field models, one can write down $n_s$ and $r$
in terms of the derivatives of potential, $V(\phi)$, as
\citep{liddle/lyth:CIALSS}:
\begin{eqnarray}
 1-n_s &=& 3M_{pl}^2\left(\frac{V'}{V}\right)^2-2M_{pl}^2\frac{V''}{V},\\
 r &=& 8M_{pl}^2\left(\frac{V'}{V}\right)^2,
\end{eqnarray}
where $M_{pl}=1/\sqrt{8\pi G}$ is the reduced Planck mass, 
and the derivatives are evaluated at the mean value of the scalar field at the
time that a given scale leaves the horizon.  These
equations may be combined to give a relation between $n_s$ and $r$: 
\begin{equation}
 r=\frac83(1-n_s)+\frac{16M_{pl}^2}3\frac{V''}{V}.
\label{eq:rns}
\end{equation}
This equation indicates that it is the {\it curvature} of the potential
that divides models on the $n_s$-$r$ plane; thus, it makes sense to classify
inflation models on the basis of the sign and magnitude of the curvature 
of the potential \citep{peiris/etal:2003}.\footnote{This classification scheme
 is similar to, but different from, the most
widely used one, which is based upon the field value (small-field,
large-field, hybrid) \citep{dodelson/kinney/kolb:1997,kinney:1998}.}

What is the implication of our bound on $r$ for inflation models? 
Equation~(\ref{eq:rns}) suggests that a large $r$ can be generated when
the curvature of the potential is positive, i.e., $V''>0$, 
at the field value that corresponds to the scales probed by the \WMAP\ data.
Therefore, it
is a set of {\it positive curvature models} that we can constrain from
the limit on $r$. On the other hand, negative curvature models
are more difficult to constrain from $r$, as they tend to predict 
small $r$ \citep{peiris/etal:2003}. We shall not discuss negative
curvature models in this paper: many of these models, including those
based upon the Coleman-Weinberg potential, fit the \WMAP\ data 
\citep[see, e.g.,][]{dvali/shafi/schaefer:1994,shafi/senoguz:2006}.

Here we shall pick three, simple but representative, forms of $V(\phi)$
that can produce $V''>0$\footnote{These choices are used to sample the space of positive curvature models.
Realistic potentials may be much more complicated: see, for example,
\citet{destri/devega/sanchez:2008} for the WMAP 3-year limits on
trinomial potentials. Also, the classification scheme based upon 
derivatives of potentials sheds little light on the models with
non-canonical kinetic terms such as $k$-inflation
\citep{armendariz-picon/damour/mukhanov:1999,garriga/mukhanov:1999}, 
ghost inflation \citep{arkani-hamed/etal:2004}, 
DBI inflation
\citep{silverstein/tong:2004,alishahiha/silverstein/tong:2004}, 
or infrared-DBI (IR-DBI) inflation \citep{chen:2005,chen:2005b},
as the tilt, $n_s$,
depends also on the derivative of the effective speed of sound of a
scalar field \citep[for recent constraints on this class of models from
the \WMAP\ 3-year data,
see][]{bean/etal:2007,bean/etal:2008,lorenz/martin/ringeval:2008}.}:
\begin{itemize}
 \item[(a)] Monomial (chaotic-type) potential, $V(\phi)\propto
	    \phi^\alpha$. This form of the potential was proposed by,
	    and is best known for, Linde's chaotic inflation models
	    \citep{linde:1983}. 
	    This model also approximates a pseudo Nambu-Goldstone boson
	    potential \citep[natural
	    inflation;][]{freese/frieman/olinto:1990,adams/etal:1993} with
	    the negative sign, $V(\phi)\propto 1-\cos(\phi/f)$, when
	    $\phi/f\ll 1$, or with the positive sign, $V(\phi)\propto
	    1+\cos(\phi/f)$, when $\phi/f\sim 1$.\footnote{The positive
	    sign case, $V(\phi)\propto 1+\cos(\phi/f)$, belongs to a
	    negative curvature model when $\phi/f\ll 1$. See 
	    \citet{savage/freese/kinney:2006} for constraints on this
	    class of models from the \WMAP\ 3-year data.} This model can
	    also approximate the Landau-Ginzburg type of spontaneous
	    symmetry breaking potential, $V(\phi)\propto
	    (\phi^2-v^2)^2$, in the appropriate limits.
 \item[(b)] Exponential potential, $V(\phi)\propto
	    \exp[-(\phi/M_{pl})\sqrt{2/p}]$. A unique feature of this
	    potential is that the dynamics of inflation is exactly
	    solvable, and the solution is a power-law expansion,
	    $a(t)\propto t^p$, rather than an exponential one. For this
	    reason this type of model is called power-law inflation
	    \citep{abbott/wise:1984,lucchin/matarrese:1985}. They often
	    appear in models of scalar-tensor theories of gravity
	    \citep{accetta/zoller/turner:1985,la/steinhardt:1989,futamase/maeda:1989,steinhardt/accetta:1990,kalara/kaloper:1990}.    
 \item[(c)] $\phi^2$ plus vacuum energy, $V(\phi)=
	    V_0+m^2\phi^2/2$. These models are known as Linde's hybrid
	    inflation \citep{linde:1994}. This model is a ``hybrid''
	    because the potential combines the chaotic-type (with $\alpha=2$)
	    with a Higgs-like potential for the second field (which is
	    not shown here). This model behaves as if it were a
	    single-field model until the second field terminates
	    inflation when $\phi$ reaches some critical value. When
	    $\phi\gg (2V_0)^{1/2}/m$ this model is the same as the model (a)
	    with $\alpha=2$, although one of Linde's motivation was to avoid
	    having such a large field value that exceeds $M_{pl}$.
\end{itemize}

These potentials\footnote{In the language of \S~3.4 in \citet{peiris/etal:2003},
the models (a) and (b) belong to ``small positive curvature models,'' and the
model (c) to ``large positive curvature models'' for $\tilde{\phi}\ll
1$, ``small positive curvature models'' for $\tilde{\phi}\gg 1$, and 
``intermediate positive curvature models'' for $\tilde{\phi}\sim 1$.}
make the following 
predictions for $r$ and $n_s$ as a function of their parameters:
\begin{itemize}
 \item[(a)] $r=8(1-n_s)\frac{\alpha}{\alpha+2}$
 \item[(b)] $r=8(1-n_s)$
 \item[(c)] $r=8(1-n_s)\frac{\tilde{\phi}^2}{2\tilde{\phi}^2-1}$
\end{itemize}
Here, for (c) we have defined a  dimensionless variable,
$\tilde{\phi}\equiv m\phi/(2V_0)^{1/2}$. 
This model approaches to the model (a) with $\alpha=2$ for $\tilde{\phi}\gg
1$, and yields the scale-invariant spectrum, $n_s=1$, when
$\tilde{\phi}=1/\sqrt{2}$. 

We summarize our findings below, and in Fig.~\ref{fig:inflation}:
\begin{itemize}
 \item[(a)] Assuming that the monomial potentials are valid to the end
	    of inflation including the reheating of the universe, one
	    can relate $n_s$ and $r$ to the number of $e$-folds of
	    inflation, $N\equiv \ln(a_{\rm end}/a_{WMAP})$, between the
	    expansion factors at the end of inflation, $a_{\rm end}$, and 
	    the epoch when the wavelength of fluctuations that we probe
	    with \WMAP\ left the horizon during inflation,
	    $a_{WMAP}$. The relations are 
	    \citep{liddle/lyth:CIALSS}:
	    \begin{equation}
	     r=\frac{4\alpha}{N},\qquad
	      1-n_s=\frac{\alpha+2}{2N}.
	    \end{equation}
	    We take $N=50$ and 60 as a reasonable range
	    \citep{liddle/leach:2003}. For $\alpha=4$, i.e., inflation
	    by a massless self-interacting scalar field
	    $V(\phi)=(\lambda/4)\phi^4$, we find that both $N=50$ and
	    60 are far away from the 95\% region, and they are excluded
	    convincingly at more than 99\% CL. For $\alpha=2$, i.e.,
	    inflation by a massive free scalar field
	    $V(\phi)=(1/2)m^2\phi^2$, the model with $N=50$ lies outside
	    of the 68\% region, whereas the model with $N=60$ is at the
	    boundary of the 68\% region. Therefore, both of these models
	    are consistent 
	    with the data within the 95\% CL. 
	    While this limit applies to a single massive free field,
	    \citet{easther/mcallister:2006} show that a model with many
	    massive axion fields \citep[$N$-flation
	    model;][]{dimopoulos/etal:2005} can shift the predicted
	    $n_s$ further away from unity,   
	    \begin{equation}
	     1-n_s^{\rm N.f.}=(1-n_s^{\rm  s.f.})\left(1+\frac{\beta}2\right), 
	    \end{equation}
	    where ``N.f'' refers to ``$N$ fields,'' and ``s.f.'' to ``single
	    field,'' and $\beta$ is a free parameter of the
	    model. \citet{easther/mcallister:2006} argue that $\beta\sim 1/2$ is
	    favoured, for which $1-n_s$ is larger than the single-field
	    prediction by as much as
	    25\%. The prediction for the tensor-to-scalar ratio, $r$,
	    is the same as the single-field case \citep{alabidi/lyth:2006b}.
	    Therefore, this model lies outside of the 95\% region for
	    $N=50$. As usual, however, these monomial potentials can be made
	    a better fit to the data by invoking a non-minimal coupling
	    between the inflaton and gravity, as the non-minimal
	    coupling can reduce $r$ to negligible levels
	    \citep{komatsu/futamase:1999,hwang/noh:1998,tsujikawa/gumjudpai:2004}.  \citet{piao:2006}
 	    has shown that $N$-flation models with monomial potentials,
 	    $V(\phi)\propto \phi^\alpha$, generically predict $n_s$ that
 	    is smaller than the corresponding single-field predictions.
 \item[(b)] For an exponential potential, $r$ and $n_s$ are uniquely
	    determined by a single parameter, $p$, that determines a
	    power-law index of the expansion factor, $a(t)\propto t^p$, as 
	    \begin{equation}
	     r=\frac{16}{p},\qquad 1-n_s=\frac{2}{p}.
	    \end{equation}
	    We find that $p<60$ is excluded at more than 99\% CL,
	    $60<p<70$ is within the 99\% region but outside of the 95\%
	    region, and $p>70$ is within the 95\% region. The models with
	    $p\sim 120$ lie on the boundary of the 68\% region, but other
	    parameters are not within the 68\% CL. This model can be
	    thought of as a single-field inflation with $p\gg 1$, or
	    multi-field inflation with $n$ fields, each having $p_i\sim
	    1$ or even $p_i<1$ \citep[assisted
	    inflation;][]{liddle/mazumdar/schunck:1998}. In this
	    context, therefore, one can translate the above limits on
	    $p$ into the limits on the number of fields. The data favour
	    $n\sim 120/p_i$ fields. 
 \item[(c)] For this model we can divide the parameter space into 3
	    regions, depending upon the value of $\tilde{\phi}$ that
	    corresponds to the field value when the wavelength of
	    fluctuations that we probe with \WMAP\ left the
	    horizon. When $\tilde{\phi}\ll 1$, the potential is
	    dominated by a constant term, which we call ``Flat Potential
	    Regime.'' When $\tilde{\phi}\gg 1$, the potential is
	    indistinguishable from the chaotic-type (model (a)) with
	    $\alpha=2$. We call this region ``Chaotic Inflation-like
	    Regime.'' When $\tilde{\phi}\sim 1$, the model shows a
	    transitional behaviour, and thus we call it ``Transition
	    Regime.'' We find that the flat potential regime with
	    $\tilde{\phi}\lesssim 2/3$ lies outside of the 95\%
	    region. The transition regime with $2/3\lesssim
	    \tilde{\phi}\lesssim 1$ is within the 95\% region, but
	    outside of the 68\% region. Finally, the chaotic-like regime
	    contains the 68\% region. Since inflation in this model ends
	    by the second field whose dynamics depends on other
	    parameters, there is no constraint from the number of $e$-folds.
\end{itemize}

These examples show that the \WMAP\ 5-year data, combined with the distance
information from BAO and SN, begin to disfavour a number of popular
inflation models. 

\subsection{Curvature of the observable universe}
\label{sec:OK}
\subsubsection{Motivation}
\label{sec:OK_motivation}
The flatness of the observable universe is one of the predictions
of conventional inflation models. How much curvature can we expect from
inflation? The common view is that inflation naturally produces the
spatial curvature parameter, $\Omega_k$, on the order of the magnitude
of quantum fluctuations, i.e., $\Omega_k\sim 10^{-5}$. On the other
hand, the current limit on $\Omega_k$ is of order $10^{-2}$; thus, the
current data are not capable of reaching the level of $\Omega_k$ that is
predicted by the common view. 

Would a detection of $\Omega_k$ rule out inflation? It is 
possible that the value of $\Omega_k$ is just below our current
detection limit, even within the context of inflation: inflation may not
have lasted for so long, and the curvature radius of our universe may
just be large enough for us not to see the evidence for curvature within
our measurement accuracy, yet. While this sounds like 
fine-tuning, it is a possibility.

This is something we
can test by constraining $\Omega_k$ better. There is also a revived (and
growing) interest in measurements of $\Omega_k$, as $\Omega_k$ is
degenerate with the equation of state of dark energy, $w$. Therefore, a
better determination of $\Omega_k$ has an important implication for our
ability to constrain the nature of dark energy.  

\subsubsection{Analysis}
\label{sec:OK_analysis}
Measurements of the CMB power spectrum alone do not strongly constrain 
$\Omega_k$. More precisely, any 
experiments that measure the angular diameter or luminosity distance to
a {\it single} redshift are not able to constrain $\Omega_k$ uniquely,
as the distance depends not only on $\Omega_k$, but also on the
expansion history of the universe. For a universe containing matter and
vacuum energy, it is essential to combine {\it at least two} absolute
distance indicators, or the expansion rates, out to different redshifts,
in order to constrain 
the spatial curvature well.
Note that CMB is also sensitive to $\Omega_\Lambda$ via the late-time
integrated Sachs-Wolfe (ISW) effect, as well as to $\Omega_m$ via
the signatures of gravitational
lensing in the CMB power spectrum.   
These properties can be used to reduce the correlation between $\Omega_k$
and $\Omega_m$  \citep{stompor/efstathiou:1999} or $\Omega_\Lambda$
\citep{ho/etal:prep,giannantonio/etal:prep}.

It has been pointed out by a number of people
\citep[e.g.,][]{eisenstein/etal:2005} that a combination of distance
measurements from BAO and CMB is a powerful way to constrain
$\Omega_k$. One needs more distances, if dark energy is not a constant
but dynamical.  

In this section, we shall make one important assumption that  the dark
energy component is vacuum energy, i.e., a cosmological constant. We
shall study the case in which the equation of state, $w$, and $\Omega_k$
are varied simultaneously, in \S~\ref{sec:DE}. 

\subsubsection{Results}
\label{sec:OK_results}
\begin{figure*}[ht]
\centering \noindent
\includegraphics[width=18cm]{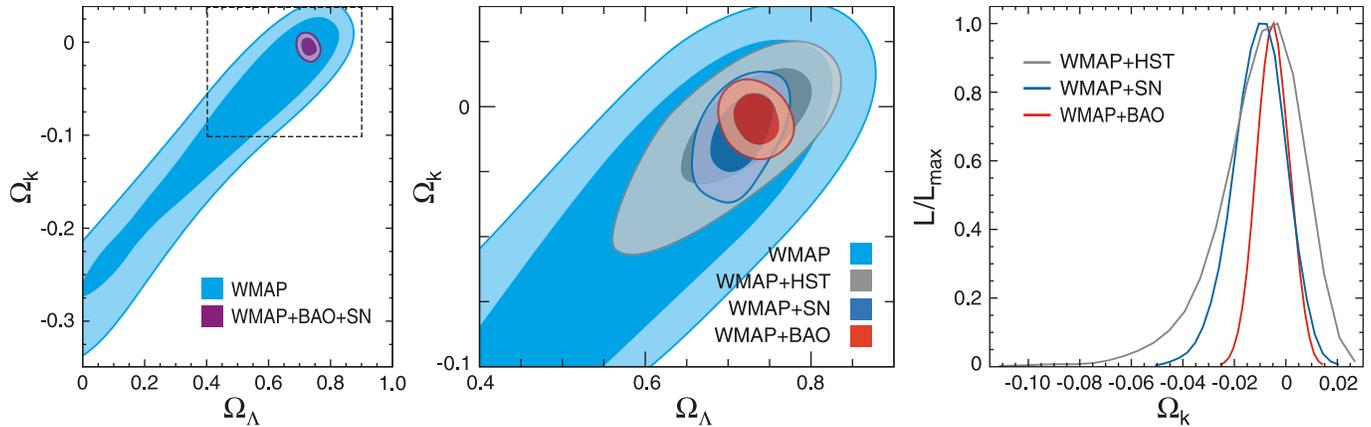}
\caption{%
 Joint two-dimensional marginalized constraint on the vacuum energy density,
 $\Omega_\Lambda$, and the spatial
 curvature parameter, $\Omega_{k}$ (\S~\ref{sec:OK_results}). 
 The contours show the 68\% and 95\% CL.
 ({\it
 Left}) The WMAP-only 
 constraint (light blue) compared with WMAP+BAO+SN (purple). Note that we have
 a prior on  $\Omega_\Lambda$, $\Omega_\Lambda>0$. This figure shows how
 powerful the extra distance information is for constraining $\Omega_k$.
 ({\it Middle}) A blow-up of the region within the dashed lines in the
 left panel, showing WMAP-only (light blue), 
 WMAP+HST (gray), WMAP+SN (dark blue), and WMAP+BAO (red). The BAO
 provides the most stringent constraint on $\Omega_k$.
 ({\it Right}) One-dimensional marginalized constraint on $\Omega_{k}$
from WMAP+HST, WMAP+SN, and WMAP+BAO. We find the best limit, 
\ensuremath{-0.0178<\Omega_k<0.0066\ \mbox{(95\% CL)}}, from
 WMAP+BAO+SN, which is essentially the same as WMAP+BAO.
See Fig.~\ref{fig:okw} for the constraints on $\Omega_k$ when dark
 energy is dynamical, i.e., $w\neq -1$, with time-independent $w$.
 Note that neither BAO nor SN alone is able to constrain $\Omega_k$:
 they need the \map\ data for lifting the degeneracy. Note also that
 BAO+SN is unable to lift the degeneracy either, as BAO needs the
 sound horizon size measured by the \map\ data.
}
\label{fig:ok}
\end{figure*}

Figure~\ref{fig:ok} shows the limits on $\Omega_\Lambda$ and
$\Omega_{k}$. While the \WMAP\ data alone cannot constrain $\Omega_{k}$
(see the left panel), the \WMAP\ data combined with the HST's constraint on
$H_0$ tighten the constraint significantly, to
\ensuremath{-0.052<\Omega_k<0.013\ \mbox{(95\% CL)}}. The \WMAP\
data combined with SN yield $\sim 50$\%
better limits,
\ensuremath{-0.0316<\Omega_k<0.0078\ \mbox{(95\% CL)}}, compared to
WMAP+HST. Finally, the WMAP+BAO yields the smallest statistical
uncertainty, \ensuremath{-0.0170<\Omega_k<0.0068\ \mbox{(95\% CL)}},
which is  a factor of 2.6 and 1.7 better than WMAP+HST and WMAP+SN,
respectively. This shows how powerful the BAO is in terms of
constraining the spatial curvature of the universe; however, this
statement needs to be re-evaluated when dynamical dark energy is
considered, e.g., $w\neq -1$. We shall come back to this point in
\S~\ref{sec:DE}. 

Finally, when \WMAP, BAO, and SN are combined, we find 
\ensuremath{-0.0178<\Omega_k<0.0066\ \mbox{(95\% CL)}}. As one
can see from the  right panel of Fig.~\ref{fig:ok}, the constraint
on $\Omega_k$ is totally dominated by that from WMAP+BAO; thus, the size
of the uncertainty does not change very much from WMAP+BAO to
WMAP+BAO+SN. Note that the above result indicates that we have reached
1.3\% accuracy (95\% CL) in determining $\Omega_k$, which is rather
good. The future BAO surveys at $z\sim 3$ are expected to yield an order
of magnitude better determination, i.e., 0.1\% level, of $\Omega_k$
\citep{knox:2006}. 

It is instructive to convert our limit on $\Omega_k$ to the limits on
the curvature radius of the universe.
As $\Omega_k$ is defined as $\Omega_k=-kc^2/(H_0^2R_{\rm curv}^2)$,
where $R_{\rm curv}$ is the present-day curvature radius, one can
convert the upper bounds on $\Omega_k$ into the lower bounds on $R_{\rm
curv}$, as $R_{\rm
curv}=(c/H_0)/\sqrt{\left|\Omega_k\right|}=3/\sqrt{\left|\Omega_k\right|}~h^{-1}{\rm
Gpc}$.  For negatively curved universes, we find $R_{\rm
curv}>37~h^{-1}$Gpc, whereas for positively curved universes, $R_{\rm
curv}>22~h^{-1}$Gpc. Incidentally these values are greater than the
particle horizon at present, $9.7~h^{-1}$Gpc (computed for the same
model).  

The 68\% limits from the 3-year data \citep{spergel/etal:2007} were
$\Omega_k=-0.012\pm 0.010$  from WMAP-3yr+BAO 
(where BAO is from the SDSS LRG of \citet{eisenstein/etal:2005}),
and $\Omega_k=-0.011\pm 0.011$ from WMAP-3yr+SN
(where SN is from the SNLS data of
\citet{astier/etal:2006}).
The 68\% 
limit from WMAP-5yr+BAO+SN (where both BAO and SN have more
data than for the 3-year analysis) is
\ensuremath{\Omega_k = -0.0050^{+ 0.0061}_{- 0.0060}}. 
A significant
improvement in the constraint is due to a combination of the better
\WMAP, BAO, and SN data.

We conclude that, {\it if} dark energy is vacuum energy (cosmological
constant) with $w=-1$, we do not find any deviation from a spatially
flat universe. 

\subsubsection{Implications for the duration of inflation}
What does this imply for inflation? Since we do not detect any finite
curvature radius, inflation had to last for a long enough period in
order to make the observable universe sufficiently flat within the
observational limits. This argument allows us to find a lower bound on
the {\it total} number of $e$-foldings of the expansion factor during
inflation, from the beginning to the end, $N_{\rm tot}\equiv \ln(a_{\rm 
end}/a_{\rm begin})$ \citep[see also \S~4.1 of][]{weinberg:COS}.

When the curvature parameter, $\Omega_k$, is much smaller than unity, it
evolves with the expansion factor, 
$a$, as $\Omega_k\propto a^{-2}$, $a^2$, and $a$ during inflation,
radiation, and matter era, respectively. Therefore, the observed
$\Omega_k$ is related to $\Omega_k$ at the beginning of inflation
as\footnote{To simplify our discussion, we ignore the dark energy
contribution, and assume that the universe is dominated matter at the
present epoch. This leads to a small error in the estimated lower bound
on $N_{\rm tot}$.}
\begin{eqnarray}
 \frac{\Omega_k^{\rm obs}}{\Omega_k^{\rm begin}} &=&
  \left(\frac{a_{\rm today}}{a_{\rm eq}}\right)\left(\frac{a_{\rm
   eq}}{a_{\rm end}}\right)^2\left(\frac{a_{\rm begin}}{a_{\rm end}}\right)^2\\
&=&
(1+z_{\rm eq})\left(\frac{T_{\rm end}g_{*,\rm end}^{1/3}}{T_{\rm
	       eq}g_{*,\rm eq}^{1/3}}\right)^2e^{-2N_{\rm tot}},
\end{eqnarray}
where $g_*$ is the effective number of relativistic degrees of freedom
contributing to entropy,
$z_{\rm eq}$ is the matter-radiation equality redshift, and $T_{\rm end}$
and $T_{\rm eq}$ are the reheating temperature of the universe at the end of
inflation\footnote{For simplicity we assume that reheating occurred as
soon as inflation ended.} and the temperature 
at the equality epoch, respectively. 
To within 10\% accuracy, we take $1+z_{\rm eq}= 3200$,
$T_{\rm eq}=0.75$~eV, and $g_{*,eq}=3.9$. We find
\begin{equation}
 N_{\rm tot} = 47
-\frac12\ln\frac{\Omega_k^{\rm obs}/0.01}{\Omega_k^{\rm begin}}
+\ln\frac{T_{\rm end}}{10^8~{\rm GeV}}+\frac13\ln\frac{g_{*,\rm end}}{200}.
\end{equation}
Here, it is plausible that $g_{*,\rm end}\sim 100$ in the Standard Model of
elementary particles, and $\sim 200$ when the supersymmetric partners
are included. The difference between these two cases gives the error of
only $\Delta N_{\rm tot}=-0.2$, and thus can be ignored.

The curvature parameter at the beginning of inflation must be 
below of order unity, as inflation would not begin otherwise. 
However, it is plausible that $\Omega_k^{\rm begin}$ was not too much
smaller than 1; otherwise, we have to explain why it was so small before
inflation, and probably we would have to explain it by inflation before
inflation. In that case $N_{\rm tot}$ would refer to the sum of the
number of $e$-foldings from two periods of inflation.
From this argument we shall take $\Omega_k^{\rm begin}\sim 1$.

The reheating temperature can be anywhere between 1~MeV and
$10^{16}$~GeV. It is more likely that it is between 1~TeV and $10^8$~GeV
for various reasons, but the allowed region is still large.
If we scale the result to 
a reasonably conservative lower limit on the reheating
temperature, $T_{\rm end}\sim 1$~TeV, then we find, from our limit on
the curvature of the universe, 
\begin{equation}
 N_{\rm tot}>36 + \ln\frac{T_{\rm end}}{1~{\rm TeV}}.
\end{equation}
A factor of 10 improvement in the upper limit on 
$|\Omega_k^{\rm begin}|$ will raise this limit by $\Delta N_{\rm
tot}=1.2$.  

Again, $N_{\rm tot}$ here refers to the total number of $e$-foldings of
inflation. In \S~\ref{sec:summary_inflation} we use $N\equiv \ln(a_{\rm
end}/a_{WMAP})$, which is the number of
$e$-foldings  between the end of inflation and
	    the epoch when the wavelength of fluctuations that we probe
	    with \WMAP\ left the horizon during inflation. Therefore, by
	    definition $N$ is less than $N_{\rm tot}$.

\subsection{Primordial non-Gaussianity}
\label{sec:NG}
\subsubsection{Motivation and Background}
\label{sec:NG_motivation}
In the simplest model of inflation, the distribution of
primordial fluctuations is close to a Gaussian with random phases.
The level of deviation from a Gaussian distribution and random phases,
called {\it non-Gaussianity}, predicted by the simplest model of
inflation is well below the current limit of measurement. Thus, any
detection of non-Gaussianity would be a significant challenge to the
currently favored models of the early universe.

The assumption of Gaussianity is motivated by the following view:
the probability distribution of quantum fluctuations, $P(\varphi)$, of free
scalar fields in the ground state of the Bunch-Davies vacuum, $\varphi$,
is a Gaussian distribution; thus, the probability distribution of
primordial curvature perturbations (in the comoving gauge), ${\cal R}$,
generated from $\varphi$ (in the flat gauge) as ${\cal
R}=-[H(\phi)/\dot{\phi_0}]\varphi$ 
\citep{mukhanov/chibisov:1981,hawking:1982,starobinsky:1982,guth/pi:1982,bardeen/steinhardt/turner:1983},
would also be a Gaussian distribution. Here, $H(\phi)$ is the expansion
rate during inflation, and $\phi_0$ is the mean field, i.e.,
$\phi=\phi_0+\varphi$. 

This argument suggests that non-Gaussianity can be generated when (a)
scalar fields are not free, but have some interactions,  (b) there are
non-linear corrections to the relation between ${\cal R}$ and $\varphi$,
and (c) the initial state is not in the Bunch-Davies vacuum. 

For (a) one can think of expanding a general scalar field potential
$V(\phi)$ to the cubic order or higher,
$V(\phi)=\bar{V}+V'\varphi+(1/2)V''\varphi^2+(1/6)V'''\varphi^3+\dots$. The
cubic (or higher-order) interaction terms can yield non-Gaussianity in
$\varphi$ \citep{falk/rangarajan/srendnicki:1993}. When perturbations in
gravitational fields are included, there are many more interaction terms
that arise from expanding the Ricci scalar to the cubic order, with
coefficients containing derivatives of $V$ and $\phi_0$, such as
$\dot{\phi_0}V''$, $\dot{\phi_0}^3/H$, etc. \citep{maldacena:2003}.

For (b) one can think of this relation,  ${\cal
R}=-[H(\phi)/\dot{\phi_0}]\varphi$,  as the leading-order term of a
Taylor series expansion of the underlying non-linear (gauge)
transformation law between ${\cal R}$ and
$\varphi$. \citet{salopek/bond:1990} show that, in the single-field
models, ${\cal R}=4\pi
G\int_{\phi_0}^{\phi_0+\varphi}d\phi\left(\partial\ln
H/\partial\phi\right)^{-1}$. Therefore, even if $\varphi$ is precisely
Gaussian, ${\cal R}$ can be non-Gaussian due to non-linear terms such as
$\varphi^2$ in a Taylor series expansion of this relation. One can write
this relation in the following form, up to second order in ${\cal R}$,
\begin{equation}
 {\cal R} = {\cal R}_{\rm L} 
- \frac1{8\pi G}\left(\frac{\partial^2\ln H}{\partial
		 \phi^2}\right){\cal R}_{\rm L}^2,
\label{eq:R}
\end{equation}
where ${\cal R}_{\rm L}$ is a linear  part of the curvature
perturbation. We thus find that the second term makes ${\cal R}$
non-Gaussian, even when ${\cal R}_{\rm L}$ is precisely Gaussian. This
formula has also been found independently by other researchers, extended
to multi-field cases,  and often referred to as the ``$\delta N$ formalism''
\citep{sasaki/stewart:1996,lyth/malik/sasaki:2004,lyth/rodriguez:2005}.

The observers like us, however, do not measure the primordial curvature
perturbations, ${\cal R}$, directly. A more observationally relevant
quantity is the curvature perturbation during the matter era, $\Phi$. At
the linear order these quantities are related by $\Phi=(3/5){\cal
R}_{\rm L}$ \citep[e.g.,][]{kodama/sasaki:1984}, but the actual relation
is more complicated at the non-linear order \citep[see][for a
review]{bartolo/etal:2004}. In any case, this argument has motivated our
defining the ``{\it local} non-linear coupling parameter,'' $\fnlKS$, as
\citep{komatsu/spergel:2001}\footnote{Note that $\fnlKS$ can be related
to the quantities discussed in earlier, pioneering work: $-\Phi_3/2$
\citep{gangui/etal:1994}, $-A_{\rm infl}/2$
\citep{wang/kamionkowski:2000}, and $-\alpha$ \citep{verde/etal:2000}.}
\begin{equation}
 \Phi = \Phi_{\rm L} + \fnlKS\Phi^2_{\rm L}.
\label{eq:fnllocal}
\end{equation}
If we take equation~(\ref{eq:R}), for example, we find $\fnlKS=-(5/24\pi
G)(\partial^2\ln H/\partial \phi^2)$ \citep{komatsu:prep}. Here, we have
followed the terminology proposed by
\citet{Babich/creminelli/zaldarriaga:2004} and called $\fnlKS$ the ``local'' parameter, as both sides of
equation~(\ref{eq:fnllocal}) are evaluated at the same location in
space. (Hence the term, ``local''.) 

Let us comment on the magnitude of the second term in
equation~(\ref{eq:fnllocal}). Since $\Phi\sim 10^{-5}$, the second term
is smaller than the first term by $10^{-5}\fnlKS$; thus, the second term
is only 0.1\% of the first term for $\fnlKS\sim
10^2$. As we shall see below, the existing limits on  $\fnlKS$ have reached this level of Gaussianity already, and thus it is
clear that we are already talking about a tiny deviation from Gaussian
fluctuations. This limit is actually better than the current limit on
the spatial curvature, which is only on the order of 1\%. Therefore,
Gaussianity tests offer a stringent test of early universe models.

In the context of single-field inflation in which the scalar field is
rolling down the potential  slowly, the 
quantities, $H$, $V$, and $\phi$, are changing slowly. Therefore,
one generically expects that $\fnlKS$ is small, on the order of the
so-called slow-roll parameters $\epsilon$ and $\eta$, which are
typically of order $10^{-2}$ or smaller. In this sense, the
single-field, slow-roll inflation models are expected to result in a
tiny amount of non-Gaussianity
\citep{salopek/bond:1990,falk/rangarajan/srendnicki:1993,gangui/etal:1994,maldacena:2003,acquaviva/etal:2003}. These
contributions from the epoch of inflation are much smaller than those
from the ubiquitous, 
second-order cosmological perturbations, i.e., the non-linear
corrections to the relation between $\Phi$ and ${\cal R}$, which result
in $\fnlKS$ of order unity 
\citep{liguori/etal:2006,smith/zaldarriaga:prep}.
See also \citet{bartolo/etal:2004} for a review on this subject.

One can use the cosmological observations, such as the CMB data, to
constrain $\fnlKS$ \citep{verde/etal:2000,komatsu/spergel:2001}. While
the temperature anisotropy, $\Delta T/T$, is related to $\Phi$ via the
Sachs--Wolfe formula as $\Delta T/T=-\Phi/3$ at the linear order on very
large angular scales \citep{sachs/wolfe:1967}, there are non-linear
corrections (non-linear Sachs--Wolfe effect, non-linear Integrated
Sachs--Wolfe effect, gravitational lensing, etc) to this  relation,
which add terms of order unity to $\fnlKS$ by the time we  observe it in
CMB \citep{pyne/carroll:1996,mollerach/matarrese:1997}. On smaller
angular scales one must include the effects of acoustic oscillations of
photon-baryon plasma by solving the Boltzmann equations. The
second-order corrections to the Boltzmann equations can also yield
$\fnlKS$ of order unity 
\citep{bartolo/matarrese/riotto:2006,bartolo/matarrese/riotto:2007}. 

Any detection of $\fnlKS$ at the level that is currently accessible 
would have a profound 
implication for the physics of inflation. How can a large $\fnlKS$ be
generated? We essentially need to break either (a) single field, or (b)
slow-roll. For example, a multi-field model known as the curvaton
scenario can result in much larger values of $\fnlKS$
\citep{linde/mukhanov:1997,lyth/ungarelli/wands:2003}, so can the models
with field-dependent (variable) decay widths for reheating of the
universe after inflation
\citep{dvali/gruzinov/zaldarriaga:2004a,dvali/gruzinov/zaldarriaga:2004b}. A
more violent, non-linear reheating process called ``preheating'' can
give rise to a large $\fnlKS$
\citep{enqvist/etal:2005,jokinen/mazumdar:2006,chambers/rajantie:2007}.  

Although breaking of slow-roll usually results in a premature
termination of inflation, it is possible to break it temporarily for a
brief period, without terminating inflation, by some features (steps,
dips, etc) in the shape of the potential. In such a scenario, a large
non-Gaussianity may be generated at a certain limited scale at which the
feature exists
\citep{kofman/etal:1991,wang/kamionkowski:2000,komatsu/etal:2003}. The
structure of non-Gaussianity from features is much more complex and
model-dependent than $\fnlKS$
\citep{chen/easther/lim:2007,chen/easther/lim:2008}.  

There is also a possibility that non-Gaussianity can be used
to test alternatives to inflation. In a collapsing universe followed by
a bounce (e.g., New Ekpyrotic scenario), 
 $\fnlKS$ is given by the {\it inverse} (as well as inverse-squared) of slow-roll parameters; thus, 
$\fnlKS$ as large as of order 10 to $10^2$ is a fairly generic
prediction of this class of models
\citep{koyama/etal:2007,buchbinder/khoury/ovrut:2008,lehners/steinhardt:2008,lehners/steinhardt:prep}.  

Using the angular bispectrum,\footnote{For a pedagogical
introduction to the bispectrum (3-point function) and trispectrum
(4-point function) and various topics on non-Gaussianity, see
\citet{komatsu:prep,bartolo/etal:2004}.} the harmonic transform of the angular
3-point correlation function, 
\citet{komatsu/etal:2002} have obtained the first
observational limit on $\fnlKS$ from the \cobe\ 4-year data
\citep{bennett/etal:1996}, finding $-3500<\fnlKS<2000$ (95\% CL). The
uncertainty was large due to a relatively large beam size of {\sl COBE},
which allowed us to go only to the maximum multipole of $l_{\rm
max}=20$. Since the signal-to-noise ratio of  $\fnlKS$ is proportional
to $l_{\rm max}$, it was expected that the \map\ data would yield a
factor of $\sim 50$ improvement over the \cobe\ data
\citep{komatsu/spergel:2001}. 

The full bispectrum analysis was not feasible with the \WMAP\ data, as
the computational cost scales as $N_{\rm pix}^{5/2}$, where $N_{\rm pix}$ is
the number of pixels, which is on the order of millions for the \WMAP\
data. The ``KSW'' estimator \citep{komatsu/spergel/wandelt:2005} has solved
this problem by inventing a cubic statistic that combines the triangle
configurations of the bispectrum optimally so that it is maximally sensitive to
$\fnlKS$.\footnote{Since the angular bispectrum is the
harmonic transform of the angular three-point function, it forms a
triangle in the harmonic space. While there are many possible triangles, 
the ``squeezed triangles,'' in which the two wave vectors are long and
one is short, are most sensitive to $\fnlKS$
\citep{Babich/creminelli/zaldarriaga:2004}.} The computational cost of
the KSW estimator scales as $N_{\rm pix}^{3/2}$. We give a detailed
description of the method that we use in this paper in
Appendix~\ref{sec:estimators}. 

We have applied this technique to
the \map\ 1-year and 3-year data, and found $-58<\fnlKS<134$
\citep[$l_{\rm max}=265$;][]{komatsu/etal:2003} and 
$-54<\fnlKS<114$ \citep[$l_{\rm
max}=350$;][]{spergel/etal:2007}, respectively, at 95\% CL. Creminelli
et al. performed an independent analysis of the \WMAP\ data and found
similar limits: $-27<\fnlKS<121$ \citep[$l_{\rm
max}=335$;][]{creminelli/etal:2006} and $-36<\fnlKS<100$
\citep[$l_{\rm max}=370$;][]{creminelli/etal:2007} for the 1-year and
3-year data, respectively. These constraints are slightly better than
the \map\ team's, as their estimator for $\fnlKS$
was improved from the original KSW estimator.

While these constraints are obtained from the KSW-like fast bispectrum
statistics, many groups have used the \map\ data to measure $\fnlKS$
using various other statistics, such as the Minkowski functionals
\citep{komatsu/etal:2003,spergel/etal:2007,gott/etal:2007,hikage/etal:prep},
real-space 
3-point function \citep{gaztanaga/wagg:2003,chen/szapudi:2005},
integrated bispectrum \citep{cabella/etal:2006}, 2-1 cumulant correlator
power spectrum \citep{chen/szapudi:2006}, local curvature
\citep{cabella/etal:2004}, and spherical Mexican hat wavelet
\citep{mukherjee/wang:2004}. The sub-orbital CMB experiments have also
yielded constraints on $\fnlKS$: MAXIMA \citep{santos/etal:2003}, VSA
\citep{smith/etal:2004}, Archeops \citep{curto/etal:2007}, and BOOMERanG
\citep{detroia/etal:2007}. 

We stress that it is important to use different statistical tools to
measure $\fnlKS$ if any signal is found, as different tools are
sensitive to different systematics. The analytical predictions for the
Minkowski functionals \citep{hikage/komatsu/matsubara:2006}
and the angular trispectrum \citep[the harmonic transform of the angular 4-point
correlation function;][]{okamoto/hu:2002,kogo/komatsu:2006} as a
function of $\fnlKS$ are available now.  Studies on the forms of the
trispectrum from inflation models have just begun, and some important
insights have been obtained
\citep{boubekeur/lyth:2006,huang/shiu:2006,byrnes/sasaki/wands:2006,seery/lidsey:2007,seery/lidsey/sloth:2007,arroja/koyama:2008}. It
is now understood that the trispectrum is at least as important as the
bispectrum in discriminating inflation models: some models do not
produce any bispectra but produce significant trispectra, and other
models produce similar amplitudes of the bispectra but produce very
different trispectra
\citep{huang/shiu:2006,buchbinder/khoury/ovrut:2008}.  

In this paper we shall use the  estimator that further improves upon
\citet{creminelli/etal:2006} by correcting an inadvertent numerical
error of a factor of 2 in their derivation \citep{yadav/etal:2008}.
 \citet{yadav/wandelt:2008} used
this estimator to measure $\fnlKS$ from the \WMAP\ 3-year data. 
We shall also use the Minkowski functionals to find a limit on $\fnlKS$.

In addition to $\fnlKS$, we shall also estimate the ``{\it equilateral}
non-linear coupling parameter,'' $\fnleq$, which
characterizes the amplitude of the three-point function (i.e., the
bispectrum) of the equilateral configurations, in which the lengths of
all the three wave vectors forming a triangle in Fourier space are
equal. This parameter is useful and highly complementary to the
local one: while $\fnlKS$ characterizes mainly the amplitude of the
bispectrum of the squeezed configurations, in which two wave vectors are
large and nearly equal and the other wave vector is small, and thus it
is fairly insensitive to the equilateral configurations, $\fnleq$ is
mainly sensitive to the equilateral configurations with little
sensitivity to the squeezed configurations. In other words, it is
possible that one may detect $\fnlKS$ without any
detection of $\fnleq$, and {\it vice versa}.  

These two parameters cover a fairly large class of models. For example,
$\fnleq$ can be generated from inflation models in which the scalar
field takes on the non-standard (non-canonical) kinetic form, such as
${\cal L}=P(X,\phi)$, where $X=(\partial\phi)^2$. In this class of
models, the effective sound speed of $\phi$ can be smaller than the
speed of light, $c_s^2=[1+2X(\partial^2P/\partial X^2)/(\partial
P/\partial X)]^{-1}<1$. 
While the sign of $\fnleq$ is negative for the DBI inflation,
$\fnleq\sim -1/c_s^2<0$ in the limit of $c_s\ll 1$, 
it can be positive or negative for more general models
\citep{seery/lidsey:2005,chen/etal:2007,cheung/etal:2008,li/wang/wang:2008}. Such models
can be realized in the context of String Theory via the non-canonical
kinetic action called the Dirac-Born-Infeld (DBI) form
\citep{alishahiha/silverstein/tong:2004}, and in the context of an
infrared modification of gravity called the ghost condensation
\citep{arkani-hamed/etal:2004}.  

The observational limits on $\fnleq$ have been obtained from the \map\
1-year and 3-year data as $-366<\fnleq<238$ \citep[$l_{\rm
max}=405$;][]{creminelli/etal:2006} and $-256<\fnleq<332$ \citep[$l_{\rm
max}=475$;][]{creminelli/etal:2007}, respectively. 

There are other forms, too. Warm inflation might produce a different
form of $f_{\rm NL}$\citep{moss/chun:2007,moss/graham:2007}. Also, the
presence of particles at the beginning of inflation, i.e., a departure
of the initial state of quantum fluctuations from the Bunch-Davies
vacuum, can result in an enhanced non-Gaussianity in the  ``flattened''
triangle configurations \citep{chen/etal:2007,holman/tolley:2008}. 
We do not
consider these forms of non-Gaussianity in this paper. 

In this paper we do not discuss the non-Gaussian signatures that
cannot be characterized by $\fnlKS$, $\fnleq$, or $\bsrc$ (the point
source bispectrum amplitude).
There have been many studies on non-Gaussian signatures in the \WMAP\
data in various forms \citep{chiang/etal:2003,chiang/naselsky/coles:2007,naselsky/etal:2007,park:2004,deoliveira-costa/etal:2004a,tegmark/deoliveira-costa/hamilton:2003,larson/wandelt:2004,eriksen/etal:2004,eriksen/etal:2004b,eriksen/etal:2004,eriksen/etal:2007c,copi/huterer/starkman:2004,schwarz/etal:2004,copi/etal:2006,copi/etal:2007,gordon/etal:2005,bielewicz/etal:2005,jaffe/etal:2005,jaffe/etal:2006,vielva/etal:2004,cruz/etal:2005,cruz/etal:2006,cruz/etal:2007,cruz/etal:2007b,cayon/jin/treaster:2005,bridges/etal:2007,wiaux/etal:2008,rath/schuecker/banday:2007,land/magueijo:2005,land/magueijo:2005b,land/magueijo:2007,rakic/schwarz:2007,park/park/gott:2007,bernui/etal:2007,hajian/souradeep:2003,hajian/souradeep/cornish:2005,hajian/souradeep:2006,prunet/etal:2005,hansen/banday/gorski:2004,hansen/etal:2004}, many of which are related to the
large-scale features at $l\lesssim 20$. We expect these features to be present
in the \WMAP\ 5-year temperature map, as the structure of CMB anisotropy
in the \WMAP\ data on such large angular scales has not changed very
much since the 3-year data.

\subsubsection{Analysis}
\label{sec:NG_analysis}
The largest concern in measuring primordial non-Gaussianity
from the CMB data is the potential contamination from the Galactic
diffuse foreground emission. To test how much the results would be
affected by this, we measure $f_{NL}$ parameters from the raw
temperature maps as well as from the foreground-reduced maps. 

We shall mainly use the {\it KQ75} mask, the new mask that is
recommended for tests of Gaussianity \citep{gold/etal:prep}. The
important difference between the new mask and the previous {\it Kp0}
mask \citep{bennett/etal:2003c} is that the new mask is defined by the
{\it  difference} between the K band map and the Internal Linear
Combination (ILC) map, and that between the Q band and ILC. Therefore,
the CMB signal was absent when the mask was defined, which removes any
concerns regarding a potential bias in the distribution of CMB on the
masked sky.\footnote{Previously, the {\it Kp0} mask was defined by the K
band map, which contains CMB as well as the foreground emission. By
cutting bright pixels in the K band map, it could be possible to cut
also the bright CMB pixels, introducing the negative skewness in the
distribution of CMB. Since we did not include isolated ``islands'' on
the high Galactic latitudes, some of which could be bright CMB spots, in
the final mask when we defined the {\it Kp0} mask, the skewness bias
mentioned above should not be as large as one would expect, if
any. Nevertheless, with the new definition of mask, the masked maps are
free from this type of bias. For more details on the definition of the
mask, see \citet{gold/etal:prep}.} 

To carry out tests of Gaussianity, one should use the {\it
KQ75} mask, which is slightly more conservative than {\it Kp0}, as the
{\it KQ75} mask cuts slightly more sky: we retain 71.8\% of the sky with
{\it KQ75}, while 76.5\% with {\it Kp0}. To see how sensitive we are to
the details of the mask, we also tried {\it Kp0} as well as the new mask
that is recommended for the power spectrum analysis, {\it KQ85},
which retains 81.7\% of the sky. The previous mask that corresponds to
{\it KQ85} is the {\it Kp2} mask, which retains 84.6\% of the sky.

In addition, we use the {\it KQ75p1} mask, which replaces the point
source mask of {\it KQ75} with the one that does not mask the sources
identified in the \WMAP\ K-band data. 
Our point source selection at K band removes more sources and sky in
regions with higher CMB flux.
We estimate the amplitude of this bias by using the {\it KQ75p1} mask
which does not use any \WMAP\ data for the point source identification.
The small change in $\fnlKS$ shows that this is a small bias.

The unresolved extra-galactic point sources also contribute to the
bispectrum
\citep{refregier/spergel/herbig:2000,komatsu/spergel:2001,argueso/gonzalez-nuevo/toffolatti:2003,serra/cooray:2008},
and they can bias our estimates of primordial non-Gaussianity parameters
such as $\fnlKS$ and $\fnleq$. We estimate the bias by measuring 
$\fnlKS$ and $\fnleq$ from Monte Carlo
simulations of point sources, and list them as $\Delta\fnlKS$ and
$\Delta\fnleq$ in Table~\ref{tab:local_clean} and \ref{tab:eq_clean}.
As the errors in these estimates of the bias are limited by the number of
Monte Carlo realizations (which is 300), one may obtain a better
estimate of the bias using more realizations.

We give a detailed description of our estimators for
$\fnlKS$, $\fnleq$, and $\bsrc$, the amplitude of the point source
bispectrum, as well as of Monte Carlo simulations in
Appendix~\ref{sec:estimators}. 

\subsubsection{Results: Bispectrum}
\label{sec:NG_results}

\begin{deluxetable}{cccccc}
\tablecolumns{6}
\small
\tablewidth{0pt}
\tablecaption{%
 Clean-map estimates and the corresponding 68\% intervals of the local
 form of primordial non-Gaussianity, 
$\fnlKS$, the point source bispectrum amplitude, $\bsrc$ (in units of
 $10^{-5}~\mu{\rm K}^3~{\rm sr}^2$), and
Monte-Carlo estimates of bias due to point sources, $\Delta \fnlKS$
}
\tablehead{\colhead{Band} & \colhead{Mask} &
 \colhead{$l_{\rm max}$} &  
\colhead{$\fnlKS$} & \colhead{$\Delta \fnlKS$}
& \colhead{$\bsrc$}}
\startdata
V+W & {\it KQ85}   & 400   & $50\pm 29$ & $1\pm 2$    & $0.26\pm 1.5$ \nl
V+W & {\it KQ85}   & 500   & $61\pm 26$ & $2.5\pm 1.5$& $0.05\pm 0.50$ \nl
V+W & {\it KQ85}   & 600   & $68\pm 31$ & $3\pm 2$    & $0.53\pm 0.28$ \nl
V+W & {\it KQ85}   & 700   & $67\pm 31$ & $3.5\pm 2$  & $0.34\pm 0.20$
 \nl
V+W & {\it Kp0}    & 500   & $61\pm 26$ & $2.5\pm 1.5$&   \nl
V+W & {\it KQ75p1}\footnote{This mask replaces the point-source mask in
 {\it KQ75} with the one that does not mask the sources identified in the \WMAP\ K-band data} & 500   & $53\pm 28$ & $4\pm 2$    &   \nl
V+W & {\it KQ75}   & 400   & $47\pm 32$ & $3\pm 2$    & $-0.50\pm 1.7$ \nl
V+W & {\it KQ75}   & 500   & $55\pm 30$ & $4\pm 2$    & $0.15\pm 0.51$ \nl 
V+W & {\it KQ75}   & 600   & $61\pm 36$ & $4\pm 2$    & $0.53\pm 0.30$ \nl
V+W & {\it KQ75}   & 700   & $58\pm 36$ & $5\pm 2$    & $0.38\pm 0.21$
\enddata
\label{tab:local_clean}
\end{deluxetable}

\begin{deluxetable}{cccc}
\tablecolumns{4}
\small
\tablewidth{0pt}
\tablecaption{%
 Null tests, frequency dependence, and raw-map estimates of the local
 form of primordial non-Gaussianity, $\fnlKS$, for $l_{\rm max}=500$
}
\tablehead{\colhead{Band} & \colhead{Foreground} & \colhead{Mask} &
\colhead{$\fnlKS$}}
\startdata
Q$-$W & Raw   & {\it KQ75} & $-0.53\pm 0.22$ \nl
V$-$W & Raw   & {\it KQ75} & $-0.31\pm 0.23$ \nl
Q$-$W & Clean & {\it KQ75} & $0.10\pm 0.22$ \nl
V$-$W & Clean & {\it KQ75} & $0.06\pm 0.23$ \nl
\hline
Q & Raw   &{\it KQ75p1}\footnote{This mask replaces the point-source mask in
 {\it KQ75} with the one that does not mask the sources identified in the \WMAP\ K-band data}& $-42\pm
 45$ \nl 
V & Raw   &{\it KQ75p1} & $38\pm 34$ \nl
W & Raw   &{\it KQ75p1} & $43\pm 33$ \nl
Q & Raw & {\it KQ75}  & $-42\pm 48$   \nl
V & Raw & {\it KQ75}  & $41\pm 35$ \nl
W & Raw & {\it KQ75}  & $46\pm 35$  \nl
Q & Clean &{\it KQ75p1} & $9\pm 45$ \nl
V & Clean &{\it KQ75p1} & $47\pm 34$ \nl
W & Clean &{\it KQ75p1} & $60\pm 33$\nl
Q & Clean & {\it KQ75}& $10\pm 48$ \nl
V & Clean & {\it KQ75}& $50\pm 35$ \nl
W & Clean & {\it KQ75}& $62\pm 35$ \nl
\hline
V+W & Raw & {\it KQ85}     & $ 9\pm 26$ \nl
V+W & Raw & {\it Kp0}     & $48\pm 26$ \nl
V+W & Raw & {\it KQ75p1}     & $41\pm 28$  \nl
V+W & Raw & {\it KQ75}     & $43\pm 30$
\enddata
\label{tab:local_null}
\end{deluxetable}

\begin{deluxetable}{ccccc}
\tablecolumns{5}
\small
\tablewidth{0pt}
\tablecaption{%
 Clean-map estimates and the corresponding 68\% intervals of the equilateral
 form of primordial non-Gaussianity, 
$\fnleq$, and
Monte-Carlo estimates of bias due to point sources, $\Delta \fnleq$
}
\tablehead{\colhead{Band}  & \colhead{Mask} &
 \colhead{$l_{\rm max}$} & \colhead{$\fnleq$} & \colhead{$\Delta \fnleq$}}
\startdata
V+W &  {\it KQ75} & 400   &$77\pm 146$ & $9\pm 7$ \nl
V+W &  {\it KQ75} & 500   &$78\pm 125$ & $14\pm 6$ \nl 
V+W & {\it KQ75} & 600   &$71\pm 108$ & $27\pm 5$ \nl
V+W & {\it KQ75} & 700   &$73\pm 101$ & $22\pm 4$
\enddata
\label{tab:eq_clean}
\end{deluxetable}

\begin{deluxetable}{cccc}
\tablecolumns{4}
\small
\tablewidth{0pt}
\tablecaption{%
 Point source bispectrum amplitude, $\bsrc$, for $l_{\rm max}=900$
}
\tablehead{\colhead{Band} & \colhead{Foreground} & \colhead{Mask} &
\colhead{$\bsrc$ [$10^{-5}~{\mu\rm K^3~sr^2}$]}}
\startdata
Q   & Raw & {\it KQ75p1}\footnote{This mask replaces the point-source mask in
 {\it KQ75} with the one that does not mask the sources identified in the \WMAP\ K-band data}   & $11.1\pm 1.3$\nl
V   & Raw & {\it KQ75p1}   & $0.83\pm 0.31$\nl
W   & Raw & {\it KQ75p1}   & $0.16\pm 0.24$\nl
V+W & Raw & {\it KQ75p1}   & $0.28\pm 0.16$ \nl
Q   & Raw & {\it KQ75}     & $6.0\pm 1.3$ \nl
V   & Raw & {\it KQ75}     & $0.43\pm 0.31$ \nl
W   & Raw & {\it KQ75}     & $0.12\pm 0.24$ \nl
V+W & Raw & {\it KQ75}     & $0.14\pm 0.16$ \nl
V+W & Raw & {\it KQ85}     & $0.20\pm 0.15$ \nl
\hline
Q   & Clean & {\it KQ75p1} & $8.7\pm 1.3$\nl
V   & Clean & {\it KQ75p1} & $0.75\pm 0.31$\nl
W   & Clean & {\it KQ75p1} & $0.16\pm 0.24$\nl
V+W & Clean & {\it KQ75p1} & $0.28\pm 0.16$ \nl
Q   & Clean & {\it KQ75}   & $4.3\pm 1.3$ \nl
V   & Clean & {\it KQ75}   & $0.36\pm 0.31$ \nl
W   & Clean & {\it KQ75}   & $0.13\pm 0.24$ \nl
V+W & Clean & {\it KQ75}   & $0.14\pm 0.16$ \nl
V+W & Clean & {\it KQ85}   & $0.13\pm 0.15$ 
\enddata
\label{tab:bsrc}
\end{deluxetable}

\begin{figure}[ht]
\centering \noindent
\includegraphics[width=8.5cm]{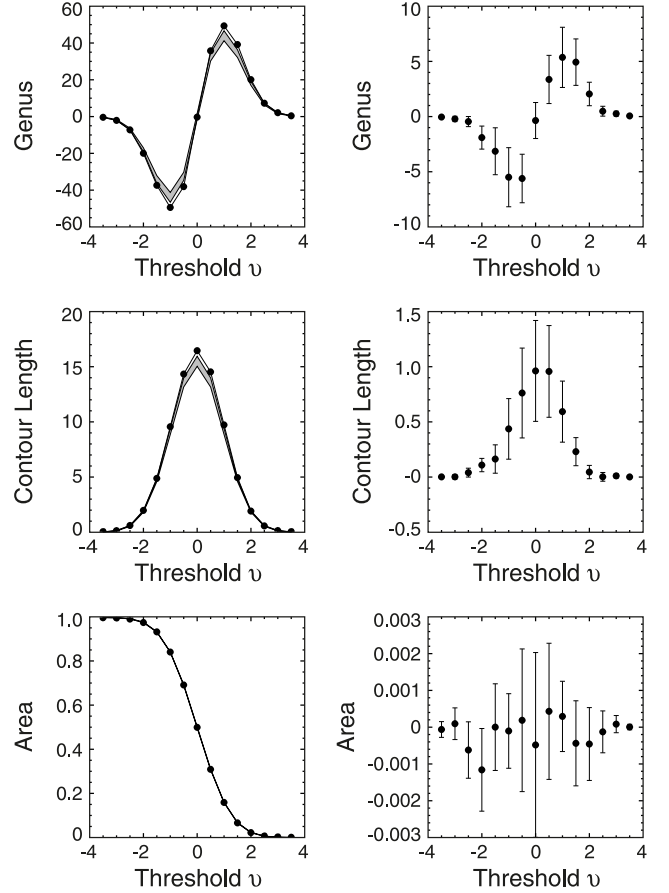}
\caption{%
Minkowski functionals from the \WMAP\ 5-year data, measured from the
 template-cleaned V$+$W map at $N_{\rm
 side}=128$ ($28'$ pixels) outside of the {\it KQ75} mask. 
From the top to bottom panels, we show the Euler characteristics (also
 known as the genus), the contour length, and the cumulative surface
 area, as a function of the threshold (the number of $\sigma$'s of hot and
 cold spots), $\nu\equiv \Delta  T/\sigma_0$. 
({\it Left}) The data (symbols) are fully consistent with
 the mean and dispersion of Gaussian realizations that include CMB and
 noise. The gray bands 
 show the  68\% intervals of Gaussian realizations.
({\it Right}) The residuals between the \WMAP\ data and the mean of the Gaussian
 realizations. Note that the residuals are highly correlated from bin to
 bin. From this result we find 
$\fnlKS=-57\pm 60$ (68\% CL). From $N_{\rm side}=64$ we find
$\fnlKS=-68\pm 69$ (68\% CL).
}
\label{fig:MF}
\end{figure}

\begin{deluxetable}{lcc}
\tablecolumns{4}
\small
\tablewidth{0pt}
\tablecaption{%
$\chi^2$ analysis of the Minkowski functionals for the template-cleaned
 V$+$W map. The results from the area, contour length, and Euler
 characteristics are combined
}
\tablehead{\colhead{$N_{\rm side}$} & \colhead{$\chi^2_{WMAP}/{\rm dof}$}
& $F(>\chi^2_{WMAP})$}
\startdata
256 & $51.5/45$ & 0.241\nl
128 & $40.0/45$ & 0.660\nl
64  & $54.2/45$ & 0.167\nl
32  & $46.8/45$ & 0.361\nl
16  & $44.7/45$ & 0.396\nl
8   & $61.3/45$ & 0.104
\enddata
\label{tab:MF}
\end{deluxetable}

In Table~\ref{tab:local_clean} we show our measurement of 
$\fnlKS$ from the template-cleaned V+W map \citep{gold/etal:prep} with 4
different masks, 
{\it KQ85}, {\it Kp0}, {\it KQ75p1}, and {\it KQ75}, in the increasing
order of the size of the mask.
For {\it KQ85} and  {\it KQ75} we show the results from different
maximum multipoles used in the analysis, 
$l_{\rm max}=400$, 500, 600, and 700. The \WMAP\ 5-year temperature data are
limited by cosmic variance to $l\sim 500$.

We find that both {\it KQ85} and {\it Kp0} for $l_{\rm max}=500$ 
show evidence for $\fnlKS>0$ at more than 95\% CL, 
$9<\fnlKS<113$ (95\% CL), before the point source bias correction, and
$6.5<\fnlKS<110.5$ (95\% CL) after the correction.
For a higher $l_{\rm max}$, $l_{\rm max}=700$, we still find evidence for 
$\fnlKS>0$, $1.5<\fnlKS<125.5$ (95\% CL), after the
correction.\footnote{The uncertainty for $l_{\rm max}>500$ is slightly
larger than that for $l_{\rm max}=500$ due to a small sub-optimality of
the estimator of $\fnlKS$ \citep{yadav/etal:2008}.}

This evidence is, however, reduced when we use larger masks, {\it
KQ75p1} and {\it KQ75}. For the latter we find 
$-5<\fnlKS<115$ (95\% CL) before the source bias correction, and 
$-9<\fnlKS<111$ (95\% CL) after the correction, which we take as our
best estimate. This estimate improves upon our previous estimate from
the 3-year data, $-54<\fnlKS<114$ \citep[95\% CL;][for $l_{\rm
max}=350$]{spergel/etal:2007}, by cutting much of the allowed region for
$\fnlKS<0$. To test whether the evidence for $\fnlKS>0$ can also be seen
with the {\it KQ75} mask, we need more years of \WMAP\ observations.

Let us study the effect of mask further. 
We find that the central value of $\fnlKS$ (without the source
correction) changes from 61 for {\it KQ85} to 55 for {\it KQ75} at $l_{max}=500$. Is this
change expected? To study this we have computed $\fnlKS$ from each of
the Monte Carlo realizations using {\it KQ85}  and {\it KQ75}.
We find the r.m.s. scatter of $\langle
(\fnlKS{}^{KQ85}-\fnlKS{}^{KQ75})^2\rangle_{\rm MC}^{1/2}=13$, 12, 15, and 15,
for $l_{max}=400$, 500, 600, and 700. Therefore, the  change in
 $\fnlKS$ measured from the \WMAP\ data is consistent with a statistical fluctuation.
For the other masks at $l_{max}=500$ we find
 $\langle
(\fnlKS{}^{Kp0}-\fnlKS{}^{KQ75})^2\rangle_{\rm MC}^{1/2}=9.7$ and
 $\langle
(\fnlKS{}^{KQ75p1}-\fnlKS{}^{KQ75})^2\rangle_{\rm MC}^{1/2}=4.0$.

In Table~\ref{tab:local_null} we summarize the results from various
tests. As a null test, we have measured $\fnlKS$ from the difference maps
such as Q$-$W and V$-$W, which are sensitive to non-Gaussianity in noise
and the residual foreground emission. Since the difference maps do not
contain the CMB signal, which is a source of a large cosmic variance in
the estimation of $\fnlKS$, the errors in the estimated $\fnlKS$ are
much smaller.  
Before the foreground cleaning (``Raw'' in the second column) we see
negative values of $\fnlKS$, which is consistent with the foreground
emission having positively skewed temperature distribution and
$\fnlKS>0$ mainly generating negative skewness.
We do not find any significant signal of $\fnlKS$ at more than 99\% CL
for raw maps, or at more than 68\% CL for cleaned maps, 
which indicates that the temperature maps are quite clean outside of the
{\it KQ75} mask.

From the results presented in Table~\ref{tab:local_null}, we find 
that the raw-map results yield more
scatter in  $\fnlKS$ estimated from various data combinations
than the clean-map results. From these studies we conclude that the
clean-map results are robust against the data combinations, as long as
we use only the V and W band data. 

In Table~\ref{tab:eq_clean} we show the equilateral bispectrum,
$\fnleq$, from the template-cleaned V+W map with the {\it KQ75} mask.
We find that the point source 
bias is much more significant for $\fnleq$: we detect the bias in
$\fnleq$ at more 
than the $5\sigma$ level for $l_{\rm max}=600$ and 700. 
After correcting for the bias, we find $-151<\fnleq<253$ (95\% CL;
$l_{\rm max}=700$) as our best estimate. Our estimate improves upon the
previous one, $-256<\fnleq<332$ \citep[95\% CL;][for $l_{\rm
max}=475$]{creminelli/etal:2006}, by reducing the allowed region from both
above and below by a similar amount.

Finally, the bispectrum from very high multipoles, e.g., $l_{\rm
max}=900$, can be used to estimate the amplitude of residual point
source contamination. One can use this information to check for a
consistency between the estimate of the residual point sources from the
power spectrum and that from the bispectrum. In Table~\ref{tab:bsrc} we
list our estimates of $\bsrc$. The raw maps and cleaned maps yield
somewhat different values, indicating a possible leakage from the
diffuse foreground to an estimate of $\bsrc$. Our best estimate in the Q
band is $\bsrc=4.3\pm 1.3~{\mu\rm K}^3~{\rm sr}^2$ (68\% CL). 
See \citet{nolta/etal:prep} for the comparison between $\bsrc$,
$C_{ps}$, and the point-source counts.

Incidentally, we also list $\bsrc$ from the {\it KQ75p1} mask, whose source
mask is exactly the same as we used for the first-year analysis.
We find $\bsrc=8.7\pm 1.3~{\mu\rm K}^3~{\rm sr}^2$ in the Q band, which is in an
excellent agreement with the first-year result, $\bsrc=9.5\pm
4.4~{\mu\rm K}^3~{\rm sr}^2$ \citep{komatsu/etal:2003}. 

\subsubsection{Results: Minkowski Functionals}
\label{sec:MF_results}

For the analysis of the Minkowski functionals, we follow the method
described in \citet{komatsu/etal:2003} and \citet{spergel/etal:2007}.
In Fig.~\ref{fig:MF} we show all of the three Minkowski functionals 
\citep{gott/etal:1990,mecke/buchert/wagner:1994,schmalzing/buchert:1997,schmalzing/gorski:1998,winitzki/kosowsky:1998} that
one can define on a two-dimensional sphere: the cumulative surface area
(bottom), the contour length (middle), and the Euler characteristics
(which is also known as the genus; top), as a function the
``threshold,'' $\nu$, which is the number of $\sigma$'s of hot and cold spots,
defined by 
\begin{equation}
 \nu\equiv \frac{\Delta T}{\sigma_0},
\end{equation}
where $\sigma_0$ is the standard deviation of the temperature data (which
includes both signal and noise) at a given resolution of the map that
one works with.  We compare the Minkowski functionals measured from the
\WMAP\ data with the mean and dispersion 
of Gaussian realizations that include CMB signal and noise. We use
the {\it KQ75} mask and the V$+$W-band map.

While Fig.~\ref{fig:MF} shows the results at resolution 7
($N_{\rm side}=128$), we have carried out Gaussianity tests using the
Minkowski functionals at six different resolutions from resolution 3
($N_{\rm side}=8$) to resolution 8 ($N_{\rm side}=256$). We find no
evidence for departures from Gaussianity at any resolutions, as
summarized in Table~\ref{tab:MF}: in this table we list the values of
$\chi^2$ of the Minkowski functionals relative to the Gaussian
predictions:
\begin{eqnarray}
\nonumber
 \chi^2_{WMAP} &=& \sum_{ij}\sum_{\nu_1\nu_2}
\left[F^i_{WMAP}-\langle F_{\rm
 sim}^i\rangle\right]_{\nu_1}\\
& &\times (\Sigma^{-1})_{\nu_1\nu_2}^{ij}
\left[F^j_{WMAP}-\langle F_{\rm
 sim}^j\rangle\right]_{\nu_2},
\end{eqnarray}
where $F^i_{WMAP}$ and $F^i_{\rm sim}$ are the $i$th Minkowski
functionals measured from the \WMAP\ data and Gaussian simulations,
respectively, the angular bracket denotes the average over
realizations, and $\Sigma_{\nu_1\nu_2}^{ij}$ is the covariance matrix
estimated from the simulations.
We use 15 different thresholds from $\nu=-3.5$ to $\nu=+3.5$,
as indicated by the symbols in Fig.~\ref{fig:MF}, and thus the number of
degrees of freedom in the fit is $15\times 3=45$. We show the values of
$\chi^2_{WMAP}$ and the degrees of freedom in the second column, and the
probability of having $\chi^2$ that is larger than the measured value,
$F(>\chi^2_{WMAP})$, in the third column. The smallest probability is
0.1 (at $N_{\rm side}=8$), and thus we conclude that the Minkowski
functionals measured from the \WMAP\ 5-year data are fully consistent
with Gaussianity. 

What do these results imply for $\fnlKS$? We find that the absence of 
non-Gaussianity at $N_{\rm side}=128$ and 64 gives the 68\% limits on 
$\fnlKS$ as 
$\fnlKS=-57\pm 60$ and $-68\pm 69$, respectively.
The 95\% limit from $N_{\rm side}=128$ is $-178 < \fnlKS < 64$.
The errors are larger than those from the bispectrum analysis given
in \S~\ref{sec:NG_results} by a factor of two, which is partly
because we have not used  
the Minkowski functional at all six resolutions to constrain $\fnlKS$.
For a combined analysis of the \WMAP\ 3-year data, see 
\citet{hikage/etal:prep}.

It is intriguing that the Minkowski functionals prefer a {\it negative}
value of $\fnlKS$, $\fnlKS\sim -60$, whereas the bispectrum prefers a
{\it positive} value, $\fnlKS\sim 60$. In the limit that non-Gaussianity
is weak, the Minkowski functionals are sensitive to three ``skewness
parameters:'' (1)  $\langle(\Delta T)^3\rangle$,
(2)  $\langle(\Delta T)^2[\partial^2(\Delta T)]\rangle$, and 
(3) $\langle[\partial (\Delta T)]^2[\partial^2(\Delta T)]\rangle$,
all of which can be written in terms of the weighted sum of the
bispectrum; thus, the Minkowski functionals are sensitive to some selected
configurations of the bispectrum
\citep{hikage/komatsu/matsubara:2006}. It would be important to study
where  an apparent ``tension'' between the Minkowski functionals and the KSW
estimator comes from. This example shows how important it is to use
different statistical tools to identify the origin of non-Gaussian
signals on the sky.

\subsection{Adiabaticity of primordial fluctuations}
 \label{sec:AD}
\subsubsection{Motivation}
\label{sec:AD_motivation}
``Adiabaticity'' of primordial fluctuations offers important tests of
inflation as well as clues to the origin of matter in the universe. 
The negative correlation between the temperature and E-mode polarization
(TE) at $l\sim 100$ is a generic signature of adiabatic super-horizon
fluctuations 
\citep{zaldarriaga/spergel:1997,peiris/etal:2003}. 
The improved measurement of the TE power spectrum as well as the
temperature power spectrum from the \WMAP\ 5-year data, combined with
the distance information from BAO and SN,
now provide tight limits on deviations of primordial fluctuations from
adiabaticity.  

Adiabaticity may be defined loosely as the following relation between
fluctuations in radiation density and those in matter density: 
\begin{equation}
 \frac{3\delta\rho_r}{4\rho_r} = \frac{\delta\rho_m}{\rho_m}.
\label{eq:adi}
\end{equation}
This version\footnote{A more general relation is
$\delta\rho_x/\dot{\rho}_x=\delta\rho_y/\dot{\rho}_y$, where $x$ and $y$
refer to some energy components. Using the energy conservation equation,
$\dot{\rho}_x=-3H(1+w_x)\rho_x$ (where $w_x$ is the equation of state
for the component $x$), one can recover Eq.~(\ref{eq:adi}), as
$w_r=1/3$ and $w_m=0$. For a recent discussion on this topic, see, e.g.,
\citet{weinberg:2003}.} of the condition
guarantees that the entropy density (dominated by radiation, $s_r\propto
\rho_r^{3/4}$) per matter particle is unperturbed, i.e.,
$\delta(s_r/n_m)=0$.  

There are two situations in which the adiabatic condition may be
satisfied: (1) there is only one degree of freedom in the system, e.g.,
both radiation and matter were created from decay products of a single
scalar field that was solely responsible for generating fluctuations,
and (2) matter and radiation were in thermal equilibrium before any
non-zero conserving quantum number (such as baryon number minus
lepton number, $B-L$) was created 
\citep[e.g.,][]{weinberg:2004}.  

Therefore, detection of any non-adiabatic fluctuations, i.e., any
deviation from the adiabatic condition (Eq.~[\ref{eq:adi}]), would imply
that there were multiple scalar fields during inflation, {\it and}
either matter (baryon or dark matter) was never in thermal equilibrium
with radiation, or a non-zero conserving quantum number associated with
matter was 
created well before the era of thermal equilibrium. In any case, the
detection of non-adiabatic fluctuations between matter and radiation
has a profound implication for the physics of inflation and, perhaps
more importantly, the origin of matter. 

For example, axions, a good candidate for dark matter, 
generate non-adiabatic fluctuations between dark matter and photons, as
axion density fluctuations could be produced during inflation
independent of curvature perturbations (which were generated from
inflaton fields, and responsible for CMB anisotropies that we observe
today), {\it and}  were not in thermal equilibrium with radiation in the
early universe \citep[see][for reviews]{kolb/turner:TEU,sikivie:2008}. We
can therefore place stringent limits on the properties of axions by
looking at a signature of deviation from the adiabatic relation in  the
CMB temperature and polarization anisotropies. 

In this paper we focus on the non-adiabatic perturbations between cold
dark matter (CDM) and CMB photons. Non-adiabatic
perturbations between baryons and photons are exactly the same as those
between CDM and photons, up to an overall constant; thus, we shall not
consider them separately in this paper. 
For  neutrinos and photons we consider only adiabatic perturbations.
In other words, we consider only three standard neutrino species (i.e., no
sterile neutrinos), and assume that the neutrinos were in thermal
equilibrium before the lepton number was generated. 

The basic idea behind this study is not new, and adiabaticity has been
constrained extensively using the \map\ data since the first year
release, including general (phenomenological) studies without 
references to specific models
\citep{peiris/etal:2003,crotty/etal:2003,bucher/etal:2004,moodley/etal:2004,lazarides/deaustri/trotta:2004,kurki-suonio/muhonen/valiviita:2005,beltran/etal:2005,dunkley/etal:2005b,bean/dunkley/pierpaoli:2006,trotta:2007,keskitalo/etal:2007},
as well as constraints on specific models such as  double inflation
\citep{silk/turner:1987,polarski/starobinsky:1992,polarski/starobinsky:1994},
axion
\citep{weinberg:1978,wilczek:1978,seckel/turner:1985,linde:1985,linde:1991,turner/wilczek:1991},
and curvaton
\citep{lyth/wands:2003,moroi/takahashi:2001,moroi/takahashi:2002,bartolo/liddle:2002},
all of which can be constrained from the limits on non-adiabatic
fluctuations
\citep{gordon/lewis:2002,gordon/malik:2004,beltran/etal:2004,lazarides:2005,parkinson/etal:2005,beltran/etal:2005b,beltran/garcia-bellido/lesgourgues:2007,kawasaki/sekiguchi:prep}. 

We shall use the \map\ 5-year data, combined with the distance
information from BAO and SN, to place more stringent limits on two types
of non-adiabatic CDM fluctuations: (i) axion-type, and (ii)
curvaton-type. Our study given below is similar to the one by
\citet{kawasaki/sekiguchi:prep} for the \map\ 3-year data. 

\subsubsection{Analysis}
\label{sec:AD_analysis}

\begin{figure*}[ht]
\centering \noindent
\includegraphics[width=18cm]{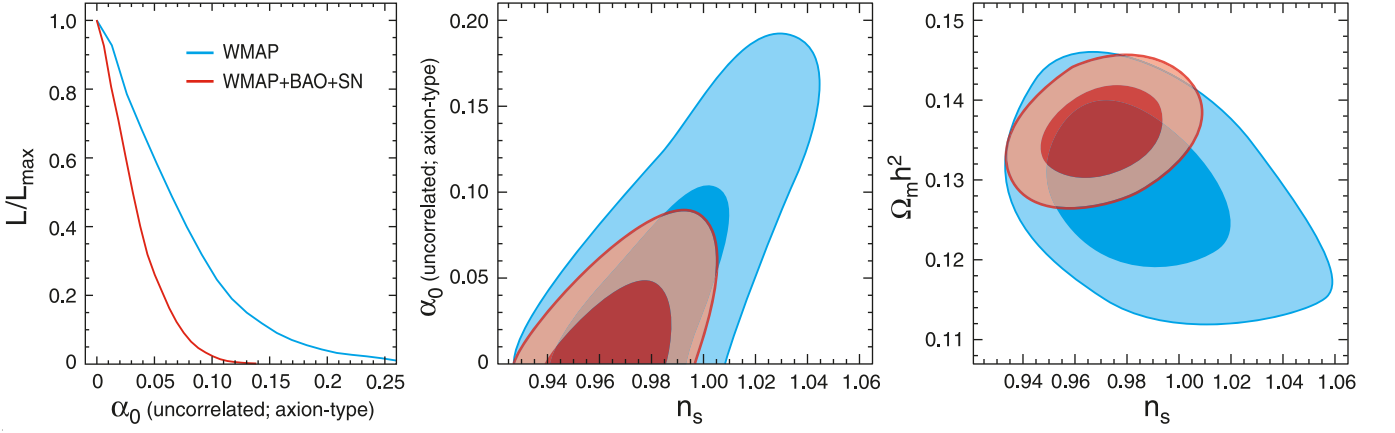}
\caption{%
Constraint on the axion entropy perturbation fraction, $\alpha_0$
(\S~\ref{sec:AD_results_axion}). 
 In all panels we show the WMAP-only results in blue and WMAP+BAO+SN in red.
({\it Left}) One-dimensional marginalized constraint on $\alpha_{0}$,
 showing WMAP-only and WMAP+BAO+SN. 
({\it Middle}) Joint two-dimensional  marginalized constraint (68\% and
 95\% CL),
 showing the correlation between $\alpha_0$ and $n_s$ for WMAP-only
 and WMAP+BAO+SN.
({\it Right}) Correlation between $n_s$ and $\Omega_mh^2$. The BAO
 and SN data help to reduce this correlation which, in turn, reduces
 correlation between $\alpha_0$ and $n_s$, resulting in a factor of 2.2
 better limit on  $\alpha_0$.
}
\label{fig:axion}
\end{figure*}

\begin{figure*}[ht]
\centering \noindent
\includegraphics[width=18cm]{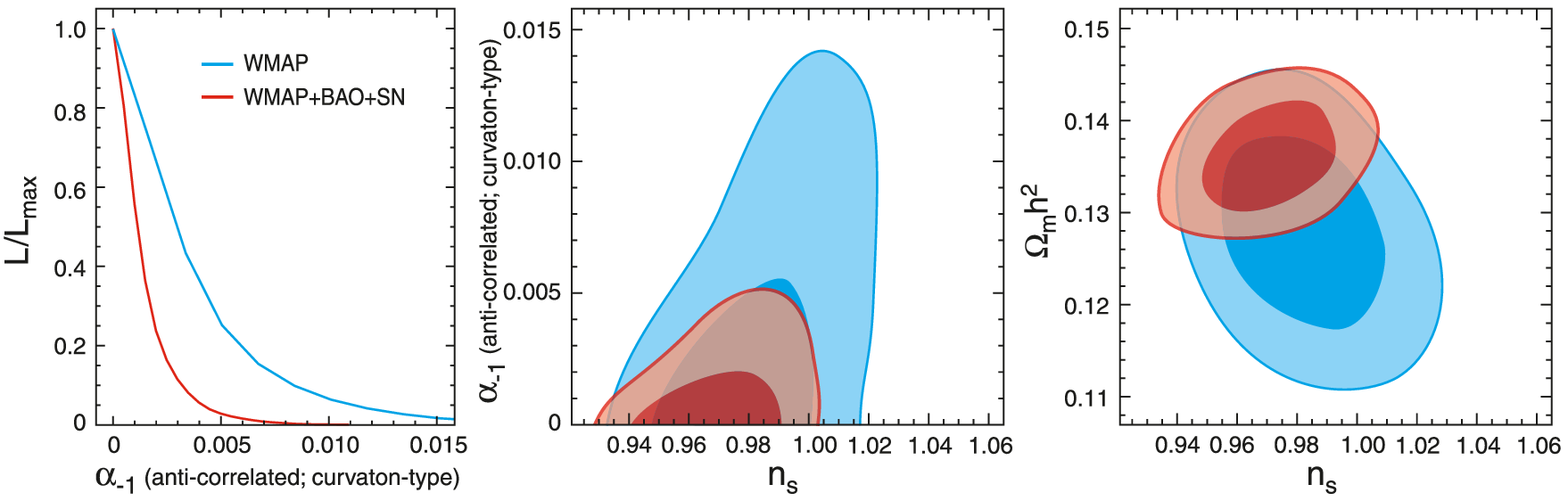}
\caption{%
Constraint on the curvaton entropy perturbation fraction, $\alpha_{-1}$
(\S~\ref{sec:AD_results_curvaton}).
 In all panels we show
 the WMAP-only results in blue and WMAP+BAO+SN in red.
({\it Left}) One-dimensional marginalized constraint on $\alpha_{-1}$,
 showing WMAP-only and WMAP+BAO+SN. 
({\it Middle}) Joint two-dimensional  marginalized constraint  (68\% and
 95\% CL),
 showing the correlation between $\alpha_{-1}$ and $n_s$ for  WMAP-only
 and WMAP+BAO+SN.
({\it Right}) Correlation between $n_s$ and $\Omega_mh^2$. The BAO
 and SN data help to reduce this correlation which, in turn, reduces
 correlation between $\alpha_{-1}$ and $n_s$, resulting in a factor of
 2.7 better limit on  $\alpha_{-1}$. These properties are similar to
 those of the axion dark matter presented in Fig.~\ref{fig:axion}.
}
\label{fig:curvaton}
\end{figure*}

 We define the non-adiabatic, or entropic, perturbation between the CDM
 and  photons,  ${\cal S}_{c,\gamma}$, as 
\begin{equation}
 {\cal S}_{c,\gamma} \equiv
\frac{\delta\rho_c}{\rho_c}-\frac{3\delta\rho_\gamma}{4\rho_\gamma},
\end{equation}
and report on the limits on the ratio of the power spectrum of ${\cal
S}_{c,\gamma}$, $P_{\cal S}(k)$, to the curvature perturbation, $P_{\cal
R}(k)$, at a given pivot wavenumber, $k_0$, given by
\citep[e.g.,][]{bean/dunkley/pierpaoli:2006} 
\begin{equation}
 \frac{\alpha(k_0)}{1-\alpha(k_0)}\equiv \frac{P_{\cal S}(k_0)}{P_{\cal
  R}(k_0)}. 
\label{eq:alphadef}
\end{equation}
We shall take $k_0$ to be $0.002~{\rm Mpc}^{-1}$.

While $\alpha$ parametrizes the ratio of the entropy power spectrum to
the curvature power spectrum, it may be more informative to quantify
``how much the adiabatic relation (Eq.~[\ref{eq:adi}]) can be
violated.'' To quantify this, we introduce the adiabaticity deviation
parameter, $\delta_{adi}$, given by
\begin{equation}
 \delta^{(c,\gamma)}_{adi}\equiv
  \frac{\delta\rho_c/\rho_c-3\delta\rho_\gamma/(4\rho_\gamma)}{\frac12[\delta\rho_c/\rho_c+3\delta\rho_\gamma/(4\rho_\gamma)]},
\label{eq:deltaadi}
\end{equation}
which can be used to say, ``the deviation from the adiabatic relation
between dark matter and photons must be less than
$100\delta^{(c,\gamma)}_{adi}$\%.'' The numerator is just
the definition of the entropy perturbation, ${\cal S}_{c,\gamma}$,
whereas the denominator is given by 
\begin{equation}
 \frac12\left(\frac{\delta\rho_c}{\rho_c}+\frac{3\delta\rho_\gamma}{4\rho_\gamma}\right)
= 3{\cal R} + O({\cal S}).
\end{equation}
Therefore, we find, up to the first order in ${\cal S}/{\cal R}$, 
\begin{equation}
 \delta^{(c,\gamma)}_{adi}\approx \frac{\cal S}{3\cal R}\approx
\frac{\sqrt{\alpha}}3,
\end{equation}
for $\alpha\ll 1$.

There could be a significant correlation between ${\cal S}_{c,\gamma}$
and ${\cal R}$
\citep{langlois:1999,langlois/riazuelo:2000,gordon/etal:2001}. We take
this into account by introducing the cross-correlation coefficient, as  
\begin{equation}
 -\beta(k_0)\equiv \frac{P_{S,\cal R}(k_0)}{\sqrt{P_{\cal S}(k_0)P_{\cal
  R}(k_0)}}, 
\label{eq:betacorr}
\end{equation}
where $P_{{\cal S},\cal R}(k)$ is the cross-correlation power spectrum.

Here, we have a negative sign on the left hand side because of the
following reason. In our notation that we used for Gaussianity analysis,
the sign convention of the curvature perturbation is such that it gives
temperature anisotropy on large scales (the Sachs--Wolfe limit) as
$\Delta T/T=-(1/5){\cal R}$. On the other hand, those who investigate
correlations between ${\cal S}$ and ${\cal R}$ usually use an opposite
sign convention for the curvature perturbation, such that the
temperature anisotropy is given by 
\begin{equation}
 \frac{\Delta T}{T} = \frac15{\tilde{\cal R}} - \frac25{\cal S}
\label{eq:entSW}
\end{equation}
on large angular scales \citep{langlois:1999}, where $\tilde{\cal
R}\equiv -{\cal R}$, and define the correlation coefficient by 
\begin{equation}
 \beta(k_0)= \frac{P_{S,\tilde{\cal R}}(k_0)}{\sqrt{P_{\cal S}(k_0)P_{\tilde{\cal
  R}}(k_0)}}.
\end{equation}
Therefore, in order to use the same sign convention for $\beta$ as most
of the previous work, we shall use Eq.~(\ref{eq:betacorr}), and
call $\beta=+1$ ``totally correlated,'' and $\beta=-1$ ``totally
anti-correlated,'' entropy perturbations. 

It is also useful to understand how the correlation or anti-correlation
affects the CMB power spectrum at low multipoles. By squaring
equation~(\ref{eq:entSW}) and taking the average, we obtain 
\begin{equation}
 \frac{\langle(\Delta T)^2\rangle}{T^2} =
\frac1{25}\left(P_{\tilde{\cal R}}+4P_{\cal
   S}-4\beta\sqrt{P_{\tilde{\cal R}}P_{\cal S}(k)}\right).
\end{equation}
Therefore, the ``correlation,'' $\beta>0$, reduces the temperature power
spectrum on low multipoles, whereas the ``anti-correlation,'' $\beta<0$,
increases the power. This point will become important when we interpret
our results: namely, models with $\beta<0$ will result in a positive
correlation between $\alpha$ and $n_s$ \citep{gordon/lewis:2002}. 
Note that this property is similar to that of the tensor mode: as
the tensor mode adds a significant power only to 
$l\lesssim 50$, the tensor-to-scalar ratio, $r$, is degenerate with
$n_s$ (see Fig.~\ref{fig:tens}).

Finally, we specify the power spectrum of $S$ as a pure power-law,
\begin{equation}
 P_{\cal S}(k)\propto k^{m-4},\qquad
 P_{{\cal S},{\cal R}}(k)\propto k^{(m+n_s)/2-4},
\end{equation}
in analogy to the curvature power spectrum, $P_{\cal R}(k)\propto
k^{n_s-4}$. Note that $\beta$ does not depend on $k$ for this choice of
$P_{{\cal S},{\cal R}}(k)$.  With this parametrization, it is
straightforward to compute the angular power spectra of the temperature and
polarization of CMB. 

In this paper we shall pay attention to two limiting cases that are
physically motivated: totally uncorrelated ($\beta=0$) entropy
perturbations -- axion-type
\citep{seckel/turner:1985,linde:1985,linde:1991,turner/wilczek:1991},
and totally anti-correlated ($\beta=-1$) entropy perturbations --
curvaton-type
\citep{linde/mukhanov:1997,lyth/wands:2003,moroi/takahashi:2001,moroi/takahashi:2002,bartolo/liddle:2002}. Then,
we shall use $\alpha_0$ to denote $\alpha$ for $\beta=0$ and
$\alpha_{-1}$ for $\beta=-1$. 

\subsubsection{Results: Implications for Axion}
\label{sec:AD_results_axion}
First, let us consider the axion case in which ${\cal S}$ and ${\cal R}$
are totally uncorrelated, i.e., $\beta=0$. This case represents the
entropy perturbation between photons and axions, with axions accounting
for some fraction of dark matter in the universe. For simplicity, we
take the axion perturbations to be scale invariant, i.e., $m=1$. In
Appendix~\ref{sec:axion} we show that this choice corresponds to taking
one of the slow-roll parameters, $\epsilon$, to be less than $10^{-2}$,
or adding a tiny amount of gravitational waves, $r\ll 0.1$, which justifies our
ignoring gravitational waves in the analysis. 

The left panel of Fig.~\ref{fig:axion} shows that we do not find any
evidence for the axion entropy perturbations. The limits are
\ensuremath{\alpha_{0} < 0.16\ \mbox{(95\% CL)}} and
\ensuremath{\alpha_{0} < 0.072\ \mbox{(95\% CL)}} for the
WMAP-only analysis and WMAP+BAO+SN, respectively. The latter limit is the most
stringent to date,  from which we find the adiabaticity deviation
parameter of  $\delta_{adi}^{c,\gamma}<0.089$ (Eq.~[\ref{eq:deltaadi}]);
thus, we conclude that the axion dark matter and photons should obey the
adiabatic relation (Eq.~[\ref{eq:adi}]) to 8.9\%, at the 95\% CL. 

We find that $n_s$ and $\alpha_0$ are strongly degenerate 
(see the middle panel of Fig.~\ref{fig:axion}).
It is easy to understand the direction of correlation.
As the entropy perturbation with a scale invariant spectrum adds power
to the temperature anisotropy on large angular scales only, the
curvature perturbation tries to compensate it by reducing power on large
scales with a larger tilt, $n_s$. However, since a larger $n_s$ produces
too much power on small angular scales, the fitting tries to increase
$\Omega_bh^2$ to suppress the second peak, and reduce $\Omega_ch^2$ to
suppress the third peak. Overall, $\Omega_mh^2$ needs to be reduced to
compensate an increase in $n_s$, as shown in the right panel of
Fig.~\ref{fig:axion}. 

Adding the distance information from the BAO and SN helps to break 
the correlation between $\Omega_mh^2$ and $n_s$ by constraining
$\Omega_mh^2$, independent of $n_s$. Therefore, with WMAP+BAO+SN
we find an impressive, factor of 2.2 improvement in the constraint on
$\alpha_{0}$. 

What does this imply for the axions? It has been shown that the limit on
the axion entropy perturbation can be used to place a constraint on the
energy scale of inflation which, in turn, leads to a stringent
constraint on the tensor-to-scalar ratio, $r$
\citep{kain:2006,beltran/garcia-bellido/lesgourgues:2007,sikivie:2008,kawasaki/sekiguchi:prep}. 

In Appendix~\ref{sec:axion} we study a particular axion cosmology called
the ``misalignment angle scenario,'' in which the Pecci-Quinn symmetry
breaking occurred during inflation, and {\it was never restored after
inflation}. In other words, we assume that the Pecci-Quinn symmetry
breaking scale set by the axion decay constant, $f_a$, which has been
constrained to be greater than $10^{10}$~GeV from the supernova 1987A
\citep{yao/etal:2006}, is at least greater than the reheating
temperature of the universe after inflation. This is a rather reasonable
assumption, as the reheating temperature is usually taken to be as low
as $10^8$~GeV in order to avoid overproduction of unwanted relics
\citep{pagels/primack:1982,coughlan/etal:1983,ellis/nanopoulos/quiros:1986}.
Such a low reheating temperature is natural also because a coupling
between inflaton and matter had to be weak; otherwise, it would
terminate inflation prematurely.  

There is another constraint. The Hubble parameter during inflation needs
to be smaller than $f_a$, i.e., $H_{inf}\lesssim f_a$; otherwise, the
Pecci-Quinn symmetry would be restored by quantum fluctuations
\citep{lyth/stewart:1992b}.

In this scenario axions acquired quantum fluctuations during inflation,
in the same way that inflaton fields would acquire fluctuations. These
fluctuations were then converted to mass density fluctuations when
axions acquired mass at the QCD phase transition at $\sim 200$~MeV. We
observe a signature of the axion mass density fluctuations via
CDM-photon entropy perturbations imprinted in the CMB temperature and
polarization anisotropies. 

We find that the tensor-to-scalar ratio $r$, the axion density,
$\Omega_a$, CDM density, $\Omega_c$, the phase of the Pecci-Quinn field
within our observable universe, $\theta_a$, and $\alpha_0$, are related
as (for an 
alternative expression that has $f_a$ left instead of $\theta_a$, see
Eq.~[\ref{eq:faleft}]) 
\begin{eqnarray}
 r &=& \frac{4.7\times 10^{-12}}{\theta_a^{10/7}}\left(\frac{\Omega_ch^2}{\gamma}\right)^{12/7}
\left(\frac{\Omega_c}{\Omega_a}\right)^{2/7}
\frac{\alpha_{0}}{1-\alpha_{0}},\\
&<&\frac{(0.99\times 10^{-13})\alpha_{0}}{\theta_a^{10/7}\gamma^{12/7}}
\left(\frac{\Omega_c}{\Omega_a}\right)^{2/7}
\label{eq:rlimitfromaxion}
\end{eqnarray}
where $\gamma\le 1$ is a ``dilution factor'' representing the amount by
which the axion density parameter, $\Omega_ah^2$, would have been
diluted due to a potential late-time entropy production by, e.g., decay
of some (unspecified) heavy particles, between 200~MeV and the epoch of
nucleosynthesis, 1~MeV. Here, we have used the limit on the CDM density
parameter, $\Omega_ch^2$, from the axion entropy perturbation model that
we consider here,
\ensuremath{\Omega_ch^2 = 0.1052^{+ 0.0068}_{- 0.0070}},  as well as 
 the observational fact that $\alpha_0\ll 1$.  

With our limit,
 \ensuremath{\alpha_{0} < 0.072\ \mbox{(95\% CL)}}, we
 find a limit on $r$ within this scenario as 
\begin{equation}  
r<\frac{6.6\times
 10^{-15}}{\theta_a^{10/7}\gamma^{12/7}}\left(\frac{\Omega_c}{\Omega_a}\right)^{2/7}.
\label{eq:rlimit}
\end{equation}
Therefore, in order for the axion dark matter scenario that we have
considered here to be compatible with $\Omega_c\sim \Omega_a$ and the
limits on the 
non-adiabaticity and $\Omega_ch^2$, the energy scale of inflation should
be low, and hence the gravitational waves are predicted to be
negligible, {\it unless} the axion density was diluted severely by
a late-time entropy production, $\gamma\sim 0.8\times 10^{-7}$ (for
$\theta_a\sim 1$), the axion phase (or the misalignment angle) within
our observable universe was close to zero, $\theta_a\sim 3\times
10^{-9}$ (for $\gamma\sim 1$), or both $\gamma$ and $\theta_a$ were
close to zero with lesser degree. All of these possibilities would
give $r\sim 0.01$, a value that could be barely detectable in the
foreseeable future.  
One can also reverse Eq.~(\ref{eq:rlimit}) to obtain
\begin{equation}  
\frac{\Omega_a}{\Omega_c}<\frac{3.0\times
 10^{-39}}{\theta_a^5\gamma^6}\left(\frac{0.01}{r}\right)^{7/2}.
\end{equation}
Therefore, the axion density would be negligible for the detectable $r$,
unless $\theta_a$ or $\gamma$ or both are tuned to be small. 

Whether such an extreme production of entropy is highly unlikely, or
such a tiny angle is an undesirable fine-tuning, can be debated. In any
case, it is clear that the cosmological observations, such as the CDM
density, entropy perturbations, and gravitational waves, can be used to place
a rather stringent limit on the axion cosmology based upon the
misalignment scenario, one of the most popular scenarios for axions to
become a dominant dark matter component in the universe.  

\subsubsection{Results: Implications for Curvaton}
\label{sec:AD_results_curvaton}
Next, let us consider one of the curvaton models in which ${\cal S}$ and
$\tilde{\cal R}$ are totally anti-correlated, i.e., $\beta=-1$
\citep{lyth/wands:2003,moroi/takahashi:2001,moroi/takahashi:2002,bartolo/liddle:2002}. One
can also write ${\cal S}$ as ${\cal S}=B{\cal R}=-B\tilde{\cal R}$ where
$B>0$; thus,
$B^2=\alpha_{-1}/(1-\alpha_{-1})$.\footnote{This variable, $B$, is the
same as $B$ used in 
\cite{gordon/lewis:2002}, including the sign convention.}
 We take the spectral index of the
curvaton entropy perturbation, $m$, to be the same as that of the
adiabatic perturbation, $n_s$, i.e., $n_s=m$. 

The left panel of Figure~\ref{fig:curvaton} shows that we do not
find any evidence for the curvaton entropy perturbations, either. The
limits, \ensuremath{\alpha_{-1} < 0.011\ \mbox{(95\% CL)}} and
\ensuremath{\alpha_{-1} < 0.0041\ \mbox{(95\% CL)}} for the
WMAP-only analysis and WMAP+BAO+SN, respectively, are more than a factor of 10
better than those for the axion perturbations. 
The WMAP-only limit is better than the previous limit by a factor of 4
\citep{bean/dunkley/pierpaoli:2006}. From the WMAP+BAO+SN limit, we find
the adiabaticity deviation parameter of
$\delta_{adi}^{(c,\gamma)}<0.021$  (Eq.~[\ref{eq:deltaadi}]); thus, we
conclude that the  curvaton dark matter and photons should obey the
adiabatic relation (Eq.~[\ref{eq:adi}]) to 2.1\% at the 95\% CL. 

Once again, adding the distance information from the BAO and SN helps to 
reduce the correlation between $n_s$ and $\Omega_mh^2$ (see the
right panel of Fig.~\ref{fig:curvaton}), and reduces the correlation
between $n_s$ and $\alpha_{-1}$. The directions in which these
parameters are degenerate are similar to those for the  axion case (see
Fig.~\ref{fig:axion}), as the entropy perturbation with $\beta=-1$ also
increases the CMB temperature power spectrum on large angular scales, as
we described in \S~\ref{sec:AD_analysis}. 

What is the implication for this type of curvaton scenario, in which
$\beta=-1$? This scenario would arise when CDM was created from the
decay products of the curvaton field. One then finds a prediction
\citep{lyth/ungarelli/wands:2003} 
\begin{equation}
 \frac{\alpha_{-1}}{1-\alpha_{-1}}\approx 
9\left(\frac{1-\rho_{\rm curvaton}/{\rho_{\rm total}}}
{\rho_{\rm curvaton}/\rho_{\rm total}}\right)^2,
\label{eq:lythrelation}
\end{equation}
where $\rho_{\rm curvaton}$ and  $\rho_{\rm total}$ are the curvaton
density and total density  at the curvaton decay,
respectively. Note that there would be no entropy perturbation if
curvaton dominated the energy density of the universe completely at the
decay. The 
reason is simple: in such a case {\it all} of the curvaton perturbation
would become the adiabatic perturbation, so would the CDM perturbation. 
Our limit,
\ensuremath{\alpha_{-1} < 0.0041\ \mbox{(95\% CL)}},  
indicates that $\rho_{\rm curvaton}/{\rho_{\rm total}}$ is close to
unity, which simplifies the relation (Eq.~[\ref{eq:lythrelation}]) to give
\begin{equation}
 \frac{\rho_{\rm curvaton}}{\rho_{\rm total}} \approx 1 -
  \frac{\sqrt{\alpha_{-1}}}3 = 1-\delta_{adi}^{(c,\gamma)}.
\end{equation}  
Note that it is the adiabaticity deviation parameter given by
Eq.~(\ref{eq:deltaadi}) that gives the deviation of 
$\rho_{\rm curvaton}/{\rho_{\rm total}}$ from unity. From this result
we find 
\begin{equation}
 1 \ge \frac{\rho_{\rm curvaton}}{\rho_{\rm total}}\gtrsim
  0.98\qquad\mbox{(95\% CL)}. 
\end{equation}

As we mentioned in \S~\ref{sec:NG_motivation}, the curvaton scenario is
capable of producting the local form of non-Gaussianity, and 
$\fnlKS$ is given by \citep[][and references therein\footnote{Note that
the sign convention of $\fnlKS$ in \citet{lyth/rodriguez:2005} is such
that $\fnlKS{}_{,WMAP}=-\fnlKS{}_{,\rm
theirs}$}]{lyth/rodriguez:2005} 
\begin{equation} 
\fnlKS = \frac{5\rho_{\rm total}}{4\rho_{\rm curvaton}}-\frac53
-\frac{5\rho_{\rm curvaton}}{6\rho_{\rm total}},
\end{equation}
which gives $-1.25\le\fnlKS({\rm curvaton})\lesssim -1.21$, for
\ensuremath{\alpha_{-1} < 0.0041\ \mbox{(95\% CL)}}.
While we need to add additional contributions from post-inflationary,
non-linear gravitational perturbations of order unity to this value in
order to compare with what we measure from CMB, the limit from the
curvaton entropy perturbation is  consistent with the
limit from the measured $\fnlKS$ (see \S~\ref{sec:NG_results}). 

However, should the future data reveal $\fnlKS\gg 1$, then either
this scenario would be ruled out \citep{beltran:prep,li/etal:prep}, or
the curvaton dark matter must have been in thermal equilibrium with
photons. 

For the other possibilities, including
possible baryon entropy perturbations, see \citet{gordon/lewis:2002}. 

\section{Probing parity violation of the universe: TB and EB
 correlation}
\label{sec:TB}
\subsection{Motivation}
\label{sec:TB_motivation} 
\begin{figure*}[ht]
\centering \noindent
\includegraphics[width=18cm]{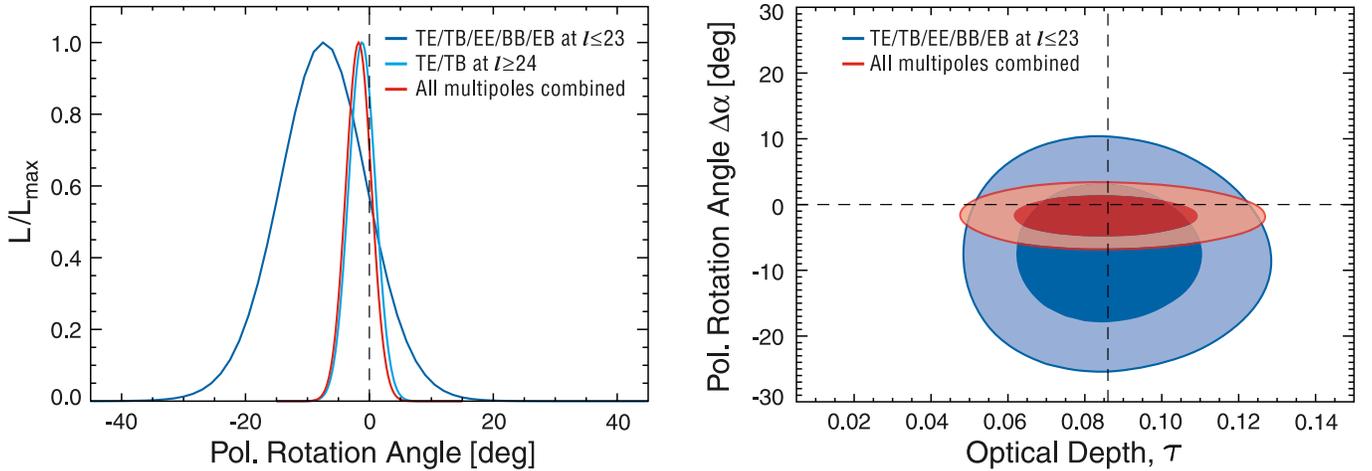}
\caption{%
 Constraint on the polarization rotation angle, $\Delta\alpha$, due to a
 parity-violating interaction that rotates the polarization angle of CMB
 (\S~\ref{sec:TB_results}). We 
 have used the polarization spectra (TE/TB/EE/BB/EB at $l\le 23$, and TE/TB
 at $l\geq 24$), and did not use the TT power spectrum. ({\it Left})
 One-dimensional marginalized constraint on $\Delta\alpha$ in units of
 degrees. The dark blue, light blue, and red curves show the
 limits from the 
 low-$l$ ($2\le l\le 23$), high-$l$ ($24\le l\le 450$), and combined
 ($2\le l\le 450$) analysis of the polarization data,
 respectively. ({\it Right}) 
 Joint two-dimensional marginalized constraint on $\tau$ and
 $\Delta\alpha$ (68\% and 95\% CL). The bigger contours are from the
 low-$l$ analysis, 
 while the smaller ones are from the combined analysis. The vertical
 dotted line shows the best-fitting optical depth in the absence of
 parity violation ($\tau=0.086$), whereas the horizontal dotted line
 shows $\Delta\alpha=0$ to guide eyes. 
}
\label{fig:parity}
\end{figure*}

Since the temperature and E-mode polarization are parity-even and the
B-mode polarization is parity-odd, the TB and EB correlations should
vanish in a universe that conserves parity
\citep{kamionkowski/kosowsky/stebbins:1997b,kamionkowski/kosowsky/stebbins:1997,seljak/zaldarriaga:1997,zaldarriaga/seljak:1997}. For
this reason the TB and EB correlations are usually used to check for
systematics, and not widely used as a cosmological probe. 

However,  parity is violated in the weak
interactions \citep{lee/yang:1956,wu/etal:1957}. Why can't parity be
violated at cosmological scales? 

Polarization of photons offers a powerful way of probing the
cosmological parity violation, or the ``cosmological birefringence''
\citep{lue/wang/kamionkowski:1999,carroll:1998}. Let us consider a parity-violating interaction
term in the Lagrangian such as the Chern-Simons term, ${\cal L}_{\rm
CS}=-(1/2)p_\alpha A_\beta\tilde{F}^{\alpha\beta}$, where
$F^{\alpha\beta}$ and $A_\beta$ are the usual electromagnetic tensor and
vector potential, respectively,
$\tilde{F}^{\alpha\beta}=(1/2)\epsilon^{\alpha\beta\mu\nu}F_{\mu\nu}$ is
a dual tensor, and $p_\alpha$ is an arbitrary timelike
four-vector.\footnote{See \citet{lepora:prep,klinkhamer:2000,adam/klinkhamer:2001}
for studies on a spacelike $p_\alpha$, including its signatures in CMB.}
\citet{carroll/field/jackiw:1990} have shown that the
Chern-Simons term makes two polarization states of photons propagate
with different group velocities, {\it causing the polarization plane to
rotate} by an angle $\Delta\alpha$.  

What would $p_\alpha$ be? We may take this to be a derivative of a light
scalar field, $p_\alpha=2(\partial_\alpha\phi)/M$, where $M$ is some
unspecified energy scale. In this case, the rotation angle is given by
$\Delta\alpha=\int \frac{dt}{a} \dot{\phi}/M=(\Delta\phi)/M$
\citep{carroll/field/jackiw:1990,carroll:1998,liu/lee/ng:2006,xia/etal:2008}. Such 
a field might have something to do with dark energy, for example. We are
therefore looking at a potential parity-violating interaction between
the visible section (i.e., photons) and dark sector (i.e., dark
energy). 

Such an unusual rotation of polarization vectors has been constrained by
observations of radio galaxies and quasars \citep{carroll:1998}: one of
the best data sets available today at a single redshift is 3C9 at $z=2.012$,
which gives a limit on the rotation angle, $\Delta\alpha=2^\circ\pm
3^\circ$ (68\% CL). There are about 10 measurements between $z=0.425$
and $z=2.012$, whose average is $\Delta\alpha=-0.6^\circ\pm 1.5^\circ$
(68\% CL). 

The rotation of polarization plane converts the E-mode polarization to
the B-mode. As a result, B modes can be produced from E modes even
if inflation did 
not produce much B modes. This is similar to the gravitational lensing effect,
which also produces B modes from E modes
\citep{zaldarriaga/seljak:1998}, but there 
is an important difference: the lensing does not violate parity, but
this interaction does. As a result, the lensing does not yield non-zero
TB or EB, but this interaction yields both TB and EB. 

We shall constrain $\Delta\alpha$ between the reionization epoch $z\sim
10$ and the present epoch, as well as $\Delta\alpha$ between the
decoupling epoch, $z\simeq 1090$, and the present epoch, using the TB
and EB spectra that we measure from the \map\ 5-year data. 

\subsection{Analysis}
\label{sec:TB_analysis}
Before we proceed, we should remember that the magnitude of polarization
rotation angle, $\Delta\alpha$, depends on the path length over which
photons experienced a parity-violating interaction. As pointed out
by \citet{liu/lee/ng:2006}, this leads to the polarization
angle that depends on $l$. We can divide this $l$-dependence in two regimes:
\begin{itemize}
 \item $l\lesssim 20$: the polarization signal was generated 
during reionization \citep{zaldarriaga:1997}. We are sensitive only to
       the polarization 
       rotation between the reionization epoch and present epoch.
 \item $l\gtrsim 20$: the polarization signal was generated at the
       decoupling epoch.  We are sensitive to the polarization rotation
       between the decoupling epoch and present epoch; thus, we have the
       largest path length in this case.
\end{itemize}
Below, we shall explore two cases separately. 
Note that we shall use only the polarization spectra: TE, TB, EE, BB,
and EB, and do not use the temperature spectrum, as the temperature
spectrum is not affected by the parity-violating interaction.

Moreover, for the analysis at $l\le 23$ we only vary the polarization angle
$\Delta\alpha$, and the optical depth, $\tau$, and fix the other
parameters at $\Omega_k=0$, $\Omega_bh^2=0.02265$, $\Omega_ch^2=0.1143$,
$H_0=70.1~{\rm km~s^{-1}~Mpc^{-1}}$, and $n_s=0.960$. 
At each value of $\tau$, we re-adjust the overall normalization of power
spectra such that the first peak of the temperature spectrum is held
fixed. For the analysis at $l\ge 24$ we fix $\tau$ at 0.085, and vary only
$\Delta\alpha$, as there is no correlation between $\Delta\alpha$ and
$\tau$ at high multipoles. We ignore EE, BB and EB at $l\ge 24$, as they
are much noisier than TE and TB and thus do not add much information.

When the polarization plane is rotated by $\Delta\alpha$, the intrinsic
(primordial) TE, EE, and BB spectra are converted into TE, TB, EE, BB,
and EB spectra as \citep{lue/wang/kamionkowski:1999,feng/etal:2005}
\begin{eqnarray}
 C_l^{TE,obs} &=& C_l^{TE}\cos(2\Delta\alpha),\\
 C_l^{TB,obs} &=& C_l^{TE}\sin(2\Delta\alpha),\\
 C_l^{EE,obs} &=& C_l^{EE}\cos^2(2\Delta\alpha) +
  C_l^{BB}\sin^2(2\Delta\alpha),\\
 C_l^{BB,obs} &=& C_l^{EE}\sin^2(2\Delta\alpha) +
  C_l^{BB}\cos^2(2\Delta\alpha),\\
 C_l^{EB,obs} &=& \frac12\left(C_l^{EE}-C_l^{BB}\right)
\sin(4\Delta\alpha),
\end{eqnarray}
where $C_l$'s are the primordial power spectra in the absence
of parity violation, while $C_l^{obs}$'s are what we would observe
in the presence of parity violation. To simplify the problem and
maximize our sensitivity to a potential signal of $\Delta\alpha$, we
ignore the primordial BB, and use only a reduced set:
\begin{eqnarray}
 C_l^{TE,obs} &=& C_l^{TE}\cos(2\Delta\alpha),\\
 C_l^{TB,obs} &=& C_l^{TE}\sin(2\Delta\alpha),\\
 C_l^{EE,obs} &=& C_l^{EE}\cos^2(2\Delta\alpha),\\
 C_l^{BB,obs} &=& C_l^{EE}\sin^2(2\Delta\alpha),\\
 C_l^{EB,obs} &=& \frac12C_l^{EE}\sin(4\Delta\alpha).
\end{eqnarray}
Therefore, TB and EB will be produced via the ``leakage'' from TE and
EE. Note that E and B are totally correlated in this case:
$(C_l^{EB,obs})^2=C_l^{EE,obs}C_l^{BB,obs}$.

Several groups have constrained $\Delta\alpha$ from the \map\ 3-year
data as well as from the BOOMERanG data
\citep{feng/etal:2006,liu/lee/ng:2006,kostelecky/mewes:2007,cabella/natoli/silk:2007,xia/etal:2008}. 
All but \citet{liu/lee/ng:2006} assume that $\Delta\alpha$ is constant at
all multipoles, which is acceptable when they consider the TB and EB data at
$l\gtrsim 20$, i.e., the BOOMERanG data and high-$l$ \map\
data. However, it requires care when one considers the low-$l$ \map\
data. Moreover, all of the authors used a Gaussian form of the
likelihood function for $C_l$, which is again acceptable at high multipoles, but
it is inaccurate at low multipoles. 

For the 5-year data release we have added capabilities of computing the
likelihood of TB and EB spectra at low multipoles, $2\le l\le 23$, exactly,
as well as that of TB spectrum at high multipoles, $24\le l\le 450$,
using the MASTER (pseudo-$C_l$) algorithm.
We shall use this code to obtain the limit on $\Delta\alpha$ from the
5-year \map\ polarization data. For the low-$l$ polarization we use the Ka,
Q, and V band data, whereas for the high-$l$ polarization we use the Q and V
band data.

\subsection{Results}
\label{sec:TB_results}
Fig.~\ref{fig:parity} shows our limit on $\Delta\alpha$ between 
(i) the reionization epoch and present epoch from the low-$l$
polarization data (dark blue), (ii) between
 the decoupling epoch and present epoch from the high-$l$
polarization data (light blue), and 
(iii) combined constraints from the low-$l$ and high-$l$ data assuming
a constant $\Delta\alpha$ across the entire multipole range (red). We
find no evidence for parity-violating interactions: the 
95\% CL (68\% CL) limits are $-22.2^\circ<\Delta\alpha<7.2^\circ$
($\Delta\alpha=-7.5^\circ\pm 7.3^\circ$) for (i),
$-5.5^\circ<\Delta\alpha<3.1^\circ$ ($\Delta\alpha=-1.2^\circ\pm
2.2^\circ$) for (ii), and $-5.9^\circ<\Delta\alpha<2.4^\circ$
($\Delta\alpha=-1.7^\circ\pm 2.1^\circ$) for (iii).  

The previous 95\% CL (68\% CL) limits on $\Delta\alpha$ are largely
based upon the high-$l$ TB and EB data from the \map\ 3-year data and/or
BOOMERanG: $-13.7^\circ<\Delta\alpha<1.9^\circ$
($\Delta\alpha=-6.0^\circ\pm 4.0^\circ$) \citep{feng/etal:2006},
$-25^\circ<\Delta\alpha<2^\circ$ 
($\Delta\alpha=-12^\circ\pm 7^\circ$) \citep{kostelecky/mewes:2007},
$-8.5^\circ<\Delta\alpha<3.5^\circ$ ($\Delta\alpha=-2.5^\circ\pm
3.0^\circ$) \citep{cabella/natoli/silk:2007}, and
$-13.8^\circ<\Delta\alpha<1.4^\circ$ ($\Delta\alpha=-6.2^\circ\pm
3.8^\circ$) \citep{xia/etal:2008}. Our limits from the \map\ 5-year data
are tighter than the previous ones by a factor of 1.5 to 2, and already
comparable to those from the polarization data of radio galaxies and
quasars (see \S~\ref{sec:TB_motivation}). Note that the radio galaxies
and quasars measure the rotation of polarization between up to $z=2$ and
the present epoch, whereas our limits measure the rotation between the
decoupling epoch, $z\simeq 1090$, and the present epoch.  

These results show that the TB and EB polarization data can provide
interesting limits on parity-violating interaction terms. The future
data will be able to place more stringent limits \citep{xia/etal:2008}.
In particular, adding the Ka and W band data to the high-$l$
polarization should improve our limit significantly.

\section{Dark energy}
\label{sec:DE}
\subsection{Motivation}
\label{sec:DE_motivation}
Dark energy is one of the most mysterious observations in physics today. The
issue is the following: when the luminosity distances out to Type Ia
supernovae \citep{riess/etal:1998,perlmutter/etal:1999} and the angular
diameter distances measured from the BAO \citep{eisenstein/etal:2005} 
as well as CMB \citep{bennett/etal:2003b} are put together in
the context of homogeneous and isotropic cosmological models, one cannot
fit these distances without having an accelerated expansion of the
universe today.  A straightforward interpretation of this result is that
we need an additional energy component in the universe that has 
a large negative pressure, which causes the expansion to accelerate.

However, we do not know much about dark energy. 
A study of review articles written over
the past twenty years reveals a growing
circle of ignorance
\citep{weinberg:1989,carroll/press/turner:1992,sahni/starobinsky:2000,padmanabhan:2003,peebles/ratra:2003,padmanabhan:2005,copeland/sami/tsujikawa:2006}:  
physicists first struggled to
understand why the cosmological constant or vacuum energy term was so
close to zero, then to understand why it was non-zero.   Cosmologists then
explored the possibility that dark energy was dynamical,
e.g., in a form of some light scalar field
\citep{ford:1987,wetterich:1988,ratra/peebles:1988,peebles/ratra:1988,fujii/nishioka:1990,chiba/sugiyama/nakamura:1997,caldwell/dave/steinhardt:1998,copeland/liddle/wands:1998,ferreira/joyce:1998,zlatev/wang/steinhardt:1999}. 
Recently,
there has been significant interest in modifications to General
Relativity, in the context of explaining the acceleration of the universe
\citep{dvali/gabadadze/porrati:2000,deffayet/dvali/gabadadze:2002}. 

Currently, the properties of dark energy  are mainly constrained by the
distance information. There are other promising ways of finding dark
energy independent of distances: the expansion rate of the universe at
higher ($z\gtrsim 0.5$) redshifts, the Integrated Sachs-Wolfe
(ISW) effect, and a slow-down of the growth of large-scale structure in
the universe due to dark energy. While these tools are powerful in
principle, the current data are not accurate enough to 
distinguish between the effects of dark energy and spatial curvature of
the universe, owing to the degeneracy between them
\citep[e.g.,][]{nesseris/perivolaropoulos:2008,ho/etal:prep,giannantonio/etal:prep}.

Indeed, the properties of dark energy, such as the
present-day density and its evolution, e.g., the equation of
state parameter $w$, are degenerate with the spatial curvature of the universe,
$\Omega_k$. In this section we shall explore both flat and curved
universes when we report on our limits on the dark energy properties.

In \S~\ref{sec:w0flat} and \S~\ref{sec:w0curv} we explore constraints on
time-independent (i.e., constant) equation of state, $w$, assuming 
flat ($\Omega_k=0$) and curved ($\Omega_k\neq 0$) geometries,
respectively. In \S~\ref{sec:wmapprior} we introduce a set of ``\WMAP\
distance priors,'' and use them to explore a wider range of model space
that has time-dependent equation of state, $w=w(z)$. 
Throughout \S~\ref{sec:wmapprior} we use the distance
information only to 
constrain the properties of dark energy.
We thus assume the standard homogeneous and isotropic
Friedmann-Lemaitre-Robertson-Walker universe, and do not
consider modifications of gravity or local
inhomogeneity, as the distance information alone cannot discriminate
between these models and the accelerated expansion due to dark energy.
Finally, in \S~\ref{sec:wmapnorm} we introduce a ``\WMAP\ normalization
prior.''

\subsection{Constant equation of state: Flat universe}
\label{sec:w0flat}
\begin{figure*}[ht]
\centering \noindent
\includegraphics[width=18cm]{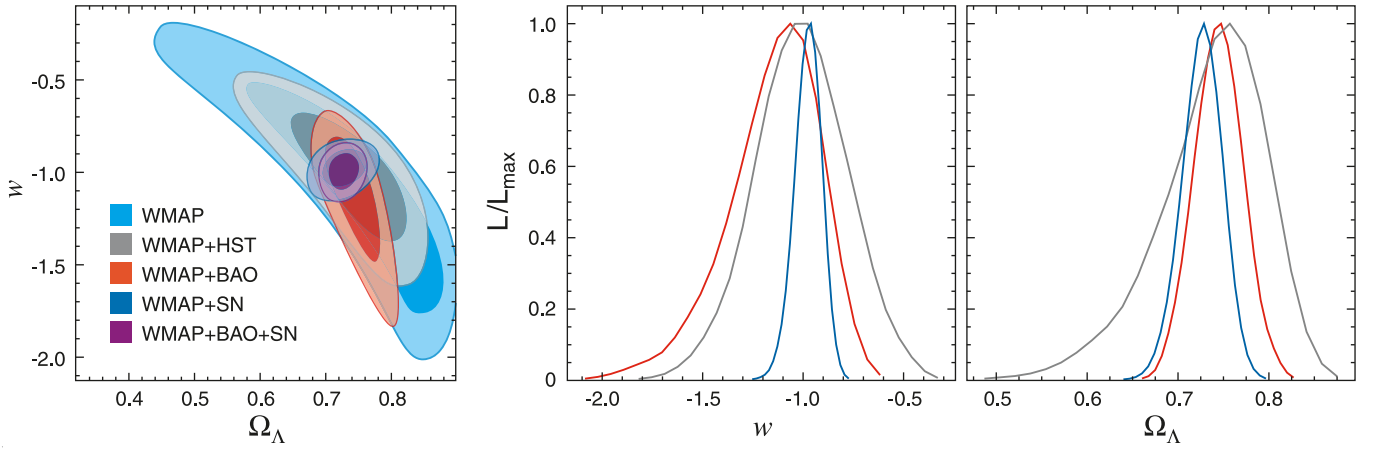}
\caption{%
 Constraint on the time-independent
 (constant) dark energy equation of state, $w$, and the present-day dark
 energy density, $\Omega_\Lambda$, assuming a flat universe,
 $\Omega_k=0$ (\S~\ref{sec:w0flat}). 
 Note that we have imposed a prior on $w$, $w>-2.5$.
 ({\it Left}) 
 Joint two-dimensional marginalized distribution of $w$ and $\Omega_k$.
 The contours show the 68\% and 95\% CL.
 The WMAP-only  constraint (light blue) is compared with WMAP+HST (gray),
 WMAP+BAO (red),  WMAP+SN (dark blue), and WMAP+BAO+SN (purple). This figure
 shows how powerful  a combination of the \map\  data and 
 the current SN data is  for constraining $w$.  
({\it Middle}) 
 One-dimensional marginalized constraint on $w$ for a flat universe
 from WMAP+HST (gray), WMAP+BAO (red), and WMAP+SN (dark blue). The
 WMAP+BAO+SN result (not shown) is essentially the same as WMAP+SN.
({\it Right}) 
 One-dimensional marginalized constraints on $\Omega_\Lambda$ for a flat
 universe from WMAP+HST (gray), WMAP+BAO (red), and WMAP+SN (dark
 blue). The WMAP+BAO+SN result (not shown) is 
 essentially the same as WMAP+SN.
 See Fig.~\ref{fig:okw} for the constraints on $w$ for non-flat
 universes.
 Note that neither BAO nor SN alone is able to constrain $w$:
 they need the \map\ data for lifting the degeneracy. Note also that
 BAO+SN is unable to lift the degeneracy either, as BAO needs the
 sound horizon size measured by the \map\ data.
}
\label{fig:okwflat}
\end{figure*}

What are we doing by assuming a flat universe, when we constrain the
dark energy equation of state, $w$? 
Most inflation models in which the inflationary periods last for much longer
 than 60 $e$-folds predict
$\Omega_k\sim 10^{-5}$, which is three orders of magnitude below the
current constraint (see \S~\ref{sec:OK}). 
In this subsection, we use a
``strong inflation prior,'' imposing a flatness prior, and explore dark
energy models in the context of such inflation models. We shall explore
curved universes in \S~\ref{sec:w0curv}.

Figure~\ref{fig:okwflat} shows the constraints on $w$ and the
present-day dark energy density, $\Omega_\Lambda$. 
The \WMAP\ data alone cannot constrain this parameter space very well, as
certain combinations of $w$ and $\Omega_\Lambda$ can produce very
similar angular diameter distances out to the decoupling epoch.

The HST prior helps a little bit
\ensuremath{-0.47<1+w<0.42\ \mbox{(95\% CL)}} by constraining
$\Omega_\Lambda$: the \map\ data measure $\Omega_mh^2$, and a
flatness prior imposes a constraint, $\Omega_\Lambda=1-(\Omega_mh^2)/h^2$;
thus, an additional constraint on $h$ from the HST Key Project helps
determine $\Omega_\Lambda$ better. 

The current angular diameter distance measurements from the BAO do not
quite break the degeneracy between $w$ and $\Omega_\Lambda$, as they
constrain the distances at relatively low redshifts, $z=0.2$ and 0.35,
whereas the transition from matter to dark energy domination, which is
sensitive to $w$, happens at earlier times. (Therefore, the future 
BAO surveys at higher redshifts should be more sensitive to $w$.)
The WMAP+BAO yields
\ensuremath{-0.68<1+w<0.21\ \mbox{(95\% CL)}}.\footnote{The
68\% limit is \ensuremath{w = -1.15^{+ 0.21}_{- 0.22}}
(WMAP+BAO; $\Omega_k=0$).}

Finally, the Type Ia supernova data break the degeneracy nicely, as
their constraint on this parameter space is nearly orthogonal to what is
determined by the CMB data: WMAP+SN yields
\ensuremath{-0.12<1+w<0.14\ \mbox{(95\% CL)}}.\footnote{The
68\% limit is \ensuremath{w = -0.977^{+ 0.065}_{- 0.064}} (WMAP+SN; $\Omega_k=0$).}

With a flatness prior, the constraint on $w$ from SN is so powerful that
WMAP+SN is 
similar to WMAP+BAO+SN. We conclude that, when the equation of
state does not depend on redshifts, dark energy is consistent with
vacuum energy, with
\ensuremath{-0.12<1+w<0.13\ \mbox{(95\% CL)}}\footnote{The
68\% limit is \ensuremath{w = -0.992^{+ 0.061}_{- 0.062}}
(WMAP+BAO+SN; $\Omega_k=0$).} (from WMAP+BAO+SN), in the context of a
flat universe at 
the level of curvature that  is predicted by long-lasting inflation models.

\subsection{Constant equation of state: Curved universe}
\label{sec:w0curv}
\begin{figure*}[ht]
\centering \noindent
\includegraphics[width=18cm]{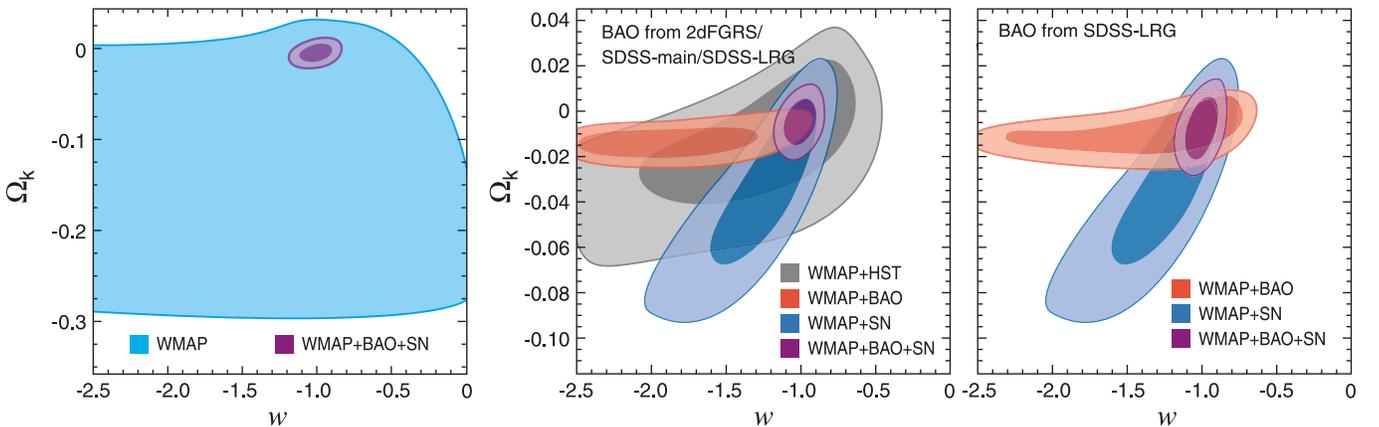}
\caption{%
 Joint two-dimensional marginalized constraint on the time-independent
 (constant) dark energy equation of state, $w$, and the curvature parameter,
 $\Omega_{k}$  (\S~\ref{sec:w0curv}). Note that we have imposed a prior
 on $w$, $w>-2.5$.  The contours show the 68\% and 95\% CL.
 ({\it Left})  The WMAP-only constraint (light blue; 95\% CL) compared
 with WMAP+BAO+SN (purple; 68\% and 95\% CL). This figure shows how
 powerful the extra distance information 
 from BAO and SN is for constraining $\Omega_k$ and $w$ simultaneously.
 ({\it Middle}) A blow-up of the left panel, showing WMAP+HST (gray),
 WMAP+BAO (red), WMAP+SN (dark blue), and WMAP+BAO+SN (purple). This figure
 shows that we need both BAO and SN to constrain $\Omega_k$ and $w$
 simultaneously: WMAP+BAO fixes $\Omega_k$, and WMAP+SN fixes $w$.
 ({\it Right}) The same as the middle panel, but with the BAO prior 
 re-weighted by a weaker BAO
 prior from the SDSS LRG sample \citep{eisenstein/etal:2005}. The BAO
 data used in the other panels combine the SDSS main and LRG, as
 well as the 2dFGRS data \citep{percival/etal:2007c}. The constraints
 from these are similar, and thus our results are not sensitive
 to the exact form of the BAO data sets. 
 Note that neither BAO nor SN alone is able to constrain $w$ or $\Omega_k$:
 they need the \map\ data for lifting the degeneracy. Note also that
 BAO+SN is unable to lift the degeneracy either, as BAO needs the
 sound horizon size measured by the \map\ data.
}
\label{fig:okw}
\end{figure*}

In this subsection we do not assume a flat universe, but do assume a
constant equation of state. (For time-dependent equation of state, see
\S~\ref{sec:wz}.)
As we discussed in
\S~\ref{sec:OK}, the \map\ data alone are unable to place meaningful
constraints on the spatial curvature of the universe; however, two or
more distance or expansion rate measurements break the degeneracy between
 $\Omega_k$ and
$\Omega_m$. 
As Fig.~\ref{fig:ok} shows, the combination of
the \WMAP\ measurement of the distance to
the decoupling epoch at $z\simeq 1090$, and
the distance measurements out to $z=0.2$ and 0.35 from BAO, strongly
constrains the curvature, at the level of 1--2\%.   

However, when dark energy is dynamical, we need 
three distance indicators that cover a wide range of
redshift.  As the
current SN data cover a wider range in redshifts, $0.02\lesssim z\le
1.7$, than the BAO data, the SN data help to constrain the evolution
of dark energy, i.e., $w$. 

Figure~\ref{fig:okw} shows the constraints on $w$ and $\Omega_k$ from
the \map\ 5-year data alone, WMAP+HST, WMAP+BAO, WMAP+SN, as well as
WMAP+BAO+SN. The middle panel is particularly illuminating. The WMAP+BAO
combination fixes $\Omega_k$, nearly independent of $w$. \footnote{For
the WMAP+BAO limit, there is a long degenerate valley with a significant 
volume at $w < -1$.  Models anywhere in this valley are good fits to
both data sets.  It is dangerous to marginalize over these degenerate
parameters as conclusions are very sensitive to the choice and the form
of priors.} 
The WMAP+SN combination yields a degeneracy line that is 
tilted with respect to the WMAP+BAO line. 
The WMAP+BAO and WMAP+SN lines
intersect at $\Omega_k\sim 0$ and $w\sim -1$, and the combined constraints are
\ensuremath{-0.0179<\Omega_k<0.0081\ \mbox{(95\% CL)}}
and
\ensuremath{-0.14<1+w<0.12\ \mbox{(95\% CL)}}.\footnote{The
68\% limits are
\ensuremath{\Omega_k = -0.0049^{+ 0.0066}_{- 0.0064}} 
and
\ensuremath{w = -1.006^{+ 0.067}_{- 0.068}} 
(WMAP+BAO+SN).}
It is remarkable that the limit on $\Omega_k$ is as good as that for 
a vacuum energy model,
\ensuremath{-0.0178<\Omega_k<0.0066\ \mbox{(95\% CL)}}. This
is because the BAO and SN yield constraints on $\Omega_k$ and $w$ that
are  complementary to each other, breaking the degeneracy effectively.  

These limits give the lower bounds to the curvature radii of the
observable universe as $R_{\rm curv}>33~h^{-1}$Gpc and $R_{\rm
curv}>22~h^{-1}$Gpc for negatively and positively curved universes,
respectively. 

Is the apparent ``tension'' between the WMAP+BAO limit and the WMAP+SN
limit in Fig.~\ref{fig:okw} the signature of new physics?  We have
checked this by the BAO distance scale out to $z=0.35$ from
the SDSS LRG sample, obtained by \citet{eisenstein/etal:2005}, 
instead of the $z=0.2$ and $z=0.35$ constraints based on the combination
of SDSS LRGs 
with the SDSS main sample and 2dFGRS \citep{percival/etal:2007c}.  While is
it not an independent check, it does provide some measurement of the
sensitivity of the constraints to the details of the BAO data set. 

The right panel of Fig.~\ref{fig:okw} shows that the results are not
sensitive to the exact form of the BAO data sets.\footnote{To obtain
the WMAP+BAO contours in the right panel of Fig.~\ref{fig:okw}, we have
re-weighted the WMAP+BAO data in the middle 
panel of Fig.~\ref{fig:okw} by the likelihood ratio of $L(\mbox{Eisenstein's BAO})/L(\mbox{Percival's BAO})$. As a result the contours do
not extend to $w\sim 0$; however, the contours would extend more to $w\sim 0$
if we ran a Markov Chain Monte Carlo from the beginning with Eisenstein
et al.'s BAO.}  
Eisenstein et al.'s
BAO prior is a bit weaker than Percival et al.'s, and thus the WMAP+BAO
contours extend more to $w\gtrsim -1$. The important point is that the
direction of degeneracy does not change.  Therefore, the combined limits
from WMAP, SN and Eisenstein et al.'s BAO, 
\ensuremath{-0.15<1+w<0.13\ \mbox{(95\% CL)}},
and
\ensuremath{-0.0241<\Omega_k<0.0094\ \mbox{(95\% CL)}}
 are similar to those with
Percival et al.'s BAO,
\ensuremath{-0.14<1+w<0.12\ \mbox{(95\% CL)}}, and
\ensuremath{-0.0179<\Omega_k<0.0081\ \mbox{(95\% CL)}}. 
As expected, a weaker BAO prior resulted in a weaker limit on
$\Omega_k$.

While the above argument suggests that there is no serious tension
between WMAP+BAO and WMAP+SN constraints, would it be possible that the
tension, if any, could be caused by the \map\ data? 
As the BAO data use the sound horizon size measured by the \map\ data,
$r_s(z_d)$, 
some systematic errors causing the mis-calculation of 
$r_s(z_d)$ could lead to a mis-interpretation of the BAO data.
The current measurement errors in $r_s(z_d)/D_V(z)$ from the BAO data
are 2.9\% at $z=0.2$ and 3.0\% at $z=0.35$. On the other hand, \map\
measures $r_s(z_d)$ with 1.3\% accuracy (see Table~\ref{tab:ruler}).
We are confident that the systematic error in $r_s(z_d)$, if any, is
smaller than the statistical error; thus, it is unlikely that \map\
causes a mis-interpretation of the BAO data.

From these studies, we are able to place rather stringent, simultaneous
limits on $\Omega_k$ (to 1--2\% level, depending upon the sign), and $w$
(to 14\% level). The spatial curvature is consistent with zero, and the
dark energy is consistent with vacuum energy. How does this conclusion
change when we allow $w$ to vary?

\subsection{\WMAP\ distance priors for testing dark  energy models}
\label{sec:wmapprior}
\subsubsection{Motivation}
\label{sec:wmapprior_motivation}
\begin{deluxetable}{lrrr}
\tablecolumns{4}
\small
\tablewidth{0pt}
\tablecaption{%
\WMAP\ distance priors obtained from the \map\ 5-year fit to models with spatial
 curvature and dark energy. The correlation coefficients are:
$r_{l_A,R}=0.1109$, 
$r_{l_A,z_*}=0.4215$, and
$r_{R,z_*}=0.6928$.
}
\tablehead{ & \colhead{5-year ML\footnote{Maximum likelihood values
 (recommended)}} & \colhead{5-year Mean\footnote{Mean of the likelihood}} &
 \colhead{Error, $\sigma$} 
}
\startdata
$l_A(z_*)$ & 302.10 & 302.45 & 0.86 \nl
$R(z_*)$   & 1.710 & 1.721  & 0.019 \nl
$z_*$\footnote{Equation~(\ref{eq:zstar})} & 1090.04 & 1091.13 & 0.93
\enddata
\label{tab:wmap_prior}
\end{deluxetable}
\begin{deluxetable}{lrrr}
\tablecolumns{4}
\small
\tablewidth{0pt}
\tablecaption{Inverse covariance matrix for the \WMAP\ distance priors}
\tablehead{ & \colhead{$l_A(z_*)$} & \colhead{$R(z_*)$} & \colhead{$z_*$}}
\startdata
$l_A(z_*)$ & $1.800$ &   $27.968$ &  $-1.103$ \nl
$R(z_*)$   &         & $5667.577$ & $-92.263$ \nl
$z_*$      &         &            & $2.923$
\enddata
\label{tab:wmap_prior_cov}
\end{deluxetable}
\begin{figure*}[ht]
\centering \noindent 
\includegraphics[width=16cm]{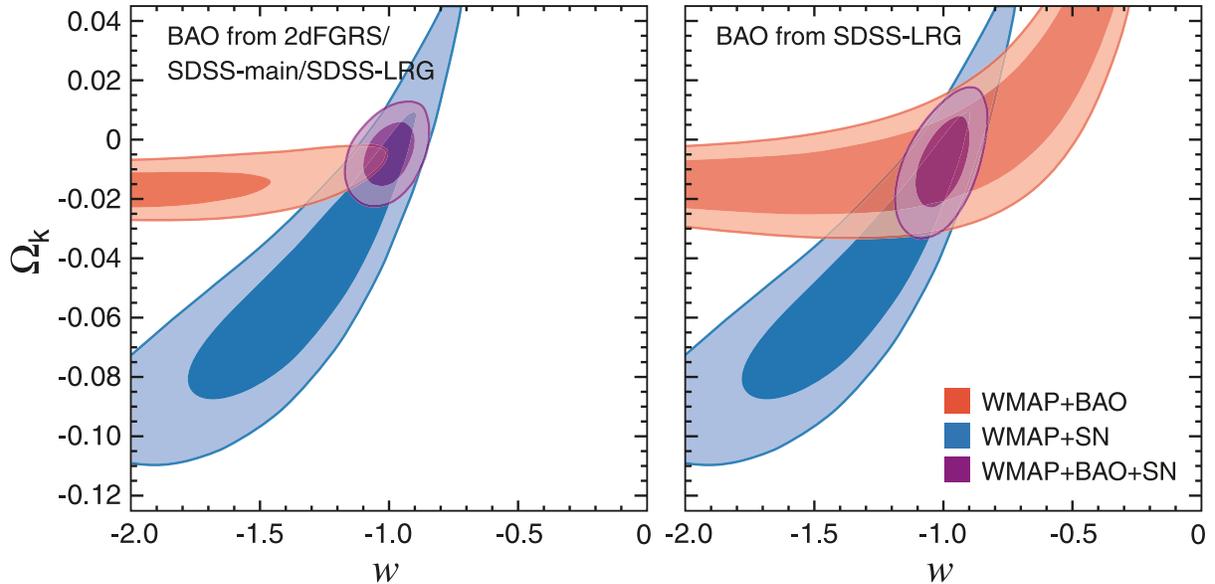}
\caption{%
 Joint two-dimensional marginalized constraint on the time-independent
 (constant) dark energy equation of state, $w$, and the curvature parameter,
 $\Omega_{k}$, derived solely from the \WMAP\ distance priors ($l_A$,
 $R$, $z_*$) (see \S~\ref{sec:wmapprior_motivation}), combined with
 either BAO (red) or SN (dark blue) or both (purple). 
 The contours show the $\Delta\chi^2_{\rm total}=2.30$ (68.3\% CL) and 
 $\Delta\chi^2_{\rm total}=6.17$  (95.4\% CL).
 The left (BAO data from \citet{percival/etal:2007c}) and right (BAO data
 from \citet{eisenstein/etal:2005}) panels should be compared with the
 middle and right panels of Fig.~\ref{fig:okw}, respectively, which have been
 derived from the full \WMAP\ data combined with the same BAO and SN
 data. While the \WMAP\ distance priors capture most of the information
 in this parameter space, the constraint is somewhat weaker than 
 that  from the full analysis.
}
\label{fig:wmapprior}
\end{figure*}
\begin{deluxetable}{lrrrr}
\tablecolumns{5}
\small
\tablewidth{0pt}
\tablecaption{Inverse covariance matrix for the extended \WMAP\ distance
 priors. The maximum likelihood value of $\Omega_bh^2$ is
 $100\Omega_bh^2=2.2765$.} 
\tablehead{ & \colhead{$l_A(z_*)$} & \colhead{$R(z_*)$} &
 \colhead{$z_*$} & \colhead{$100\Omega_bh^2$}}
\startdata
$l_A(z_*)$       & $31.001$ &   $-5015.642$ &    $183.903$ &    $2337.977$ \nl
$R(z_*)$         &          &  $876807.166$ & $-32046.750$ & $-403818.837$ \nl
$z_*$            &          &               &   $1175.054$ &   $14812.579$ \nl
$100\Omega_bh^2$ &          &               &              &  $187191.186$  
\enddata
\label{tab:wmap_prior_cov2}
\end{deluxetable}
%    31.00057807955674        -5015.641541635505         183.9028944744792          2337.977426799118     
%   -5015.641541635503         876807.1655143860        -32046.75007651475         -403818.8368599184     
%    183.9028944744792        -32046.75007651475         1175.053555246274          14812.57860037952     
%    2337.977426799117        -403818.8368599185         14812.57860037952          187191.1863414606     

Dark energy influences the distance scales as well as the growth of
structure. The CMB power spectra are sensitive to both, although
sensitivity to the growth of structure is fairly limited, as it
influences the CMB power spectrum via the ISW effect at low multipoles
($l\lesssim 10$), whose precise measurement is hampered by a large
cosmic variance.  

On the other hand, CMB is sensitive to the distance to the
decoupling epoch via the locations of peaks and troughs of the acoustic
oscillations, which can be measured precisely. More specifically, CMB
measures two distance ratios: (i) the angular diameter distance to the
decoupling epoch divided by the sound horizon size at the decoupling
epoch, $D_A(z_*)/r_s(z_*)$, and (ii) the angular diameter distance to
the decoupling epoch divided by the Hubble horizon size at the
decoupling epoch, $D_A(z_*)H(z_*)/c$. This consideration motivates our
using these two distance ratios to constrain various dark energy models,
in the presence of the spatial curvature, 
on the basis of distance information
\citep{wang/mukherjee:2007,wright:2007b}.

We shall quantify the first distance ratio, $D_A(z_*)/r_s(z_*)$, by the
``acoustic scale,'' $l_{\rm A}$, defined by
\begin{equation}
 l_{\rm A}\equiv (1+z_*)\frac{\pi D_A(z_*)}{r_s(z_*)},
\label{eq:la}
\end{equation}
where a factor of $(1+z_*)$ arises because $D_A(z_*)$ is the proper
(physical) angular diameter distance (Eq.~[\ref{eq:da}]), whereas
$r_s(z_*)$ is the comoving sound horizon at $z_*$ (Eq.~[\ref{eq:rs}]).
Here, we shall use the fitting function of $z_*$ proposed by
\citet{hu/sugiyama:1996}:
\begin{equation}
 z_*=1048\left[1+0.00124(\Omega_{b}h^2)^{-0.738}\right]
\left[1+g_1\left(\Omega_{m}h^2\right)^{g_2}\right],
\label{eq:zstar}
\end{equation}
where
\begin{eqnarray}
 g_1&=&\frac{0.0783(\Omega_bh^2)^{-0.238}}{1+39.5(\Omega_bh^2)^{0.763}},\\
 g_2&=&\frac{0.560}{1+21.1(\Omega_bh^2)^{1.81}}.
 \end{eqnarray}
Note that one could also use the peak of the probability of last
scattering of photons, i.e., the peak of the visibility function, to
define the decoupling epoch, which we denote as $z_{dec}$.
Both quantities yield similar values. We shall adopt $z_*$ here
because it is easier to compute, and therefore it allows one to
implement the \WMAP\ distance priors in a straightforward manner.

The second distance ratio, $D_A(z_*)H(z_*)/c$, is often called the 
``shift parameter,'' $R$, given by \citep{bond/efstathiou/tegmark:1997}
\begin{equation}
 R(z_*)\equiv \frac{\sqrt{\Omega_{m}H_0^2}}{c}(1+z_*)D_A(z_*).
\label{eq:shift}
\end{equation}
This quantity is different from $D_A(z_*)H(z_*)/c$ by a factor of 
$\sqrt{1+z_*}$, and also ignores the contributions from radiation,
curvature, or dark energy to $H(z_*)$. Nevertheless, 
we shall use $R$ to follow the convention in the 
literature.

We give the 5-year \WMAP\ constraints on $l_A$, $R$, and $z_*$ that we 
recommend as the \WMAP\ distance priors for constraining dark energy models.
However, we note an important caveat. 
As pointed out by
\citet{elgaroy/multamaki:2007} and \citet{corasaniti/melchiorri:2008}, the
derivation of the \WMAP\ distance priors requires us to assume the underlying
cosmology first, as all of these quantities are {\it derived parameters}
from fitting the CMB power spectra. Therefore, one must be careful about
which model one is testing. Here, we give the \WMAP\ distance priors,
assuming the following model:
\begin{itemize}
 \item The standard Friedmann-Lemaitre-Robertson-Walker universe with
       matter, radiation, dark energy, as well as spatial curvature.
 \item Neutrinos with the effective
       number of neutrinos equal to 3.04, and the minimal mass
       ($m_\nu\sim 0.05$~eV).
 \item Nearly power-law primordial power spectrum of curvature
       perturbations, $\left|dn_s/d\ln k\right|\ll 0.01$.
 \item Negligible primordial gravitational waves relative to the curvature
       perturbations, $r\ll 0.1$. 
 \item Negligible entropy fluctuations relative to the curvature
       perturbations, $\alpha\ll 0.1$. 
\end{itemize}

In Fig.~\ref{fig:wmapprior} we show the constraints on $w$ and
$\Omega_k$ from  the \WMAP\ distance priors (combined with BAO and SN).
We find a good agreement with the full MCMC results. (Compare the
middle and right panels of Fig.~\ref{fig:okw} with the left and right
panels of Fig.~\ref{fig:wmapprior}, respectively.) 
The constraints from the \WMAP\ distance priors are slightly weaker than
the full MCMC, as the distance priors use only a part of the information
contained in the \WMAP\ data.

Of course, the agreement between Fig.~\ref{fig:okw} and \ref{fig:wmapprior}
 does not guarantee that these priors yield good results for the other,
 more complex dark 
energy models with a time-dependent $w$; however, the previous studies
indicate that, under the assumptions given above, these priors can be
used to constrain a wide variety of dark energy models
\citep{wang/mukherjee:2007,elgaroy/multamaki:2007,corasaniti/melchiorri:2008}.
See also \citet{li/etal:prepb} for the latest comparison between the
 \WMAP\ distance priors and the full analysis.

Here is the prescription for using the \WMAP\ distance priors.
\begin{itemize}
 \item[(1)] For a given $\Omega_bh^2$ and $\Omega_mh^2$, compute $z_*$
	    from Eq.~(\ref{eq:zstar}). 
 \item[(2)] For a given $H_0$, $\Omega_mh^2$, $\Omega_rh^2$ (which
	    includes $N_{\rm eff}=3.04$), $\Omega_\Lambda$, and $w(z)$,
	    compute the expansion rate, $H(z)$, from
	    Eq.~(\ref{eq:hubble}), as well as the comoving sound horizon size at
	    $z_*$, $r_s(z_*)$, from Eq.~(\ref{eq:rs}).  
 \item[(3)] For a given $\Omega_k$ and $H(z)$ from the previous step,
	    compute the proper angular diameter distance, $D_A(z)$, from 
	    Eq.~(\ref{eq:da}). 
 \item[(4)] Use Eq.~(\ref{eq:la}) and (\ref{eq:shift}) to compute
	    $l_A(z_*)$ and $R(z_*)$, respectively.
 \item[(5)] Form a vector containing $x_i=(l_A,R,z_*)$ in this
	    order.
 \item[(6)] Use Table~\ref{tab:wmap_prior} for the data vector,
	    $d_i=(l_A^{WMAP}, R^{WMAP}, z_*^{WMAP})$. We recommend the
	    maximum likelihood (ML) values.
 \item[(7)] Use Table~\ref{tab:wmap_prior_cov} for the inverse covariance
	    matrix, $(C^{-1})_{ij}$. 
 \item[(8)] Compute the likelihood, $L$, as $\chi^2_{WMAP}\equiv -2\ln
	    L=(x_i-d_i)(C^{-1})_{ij}(x_j-d_j)$. 
 \item[(9)] Add this to the favorite combination of the cosmological
	    data sets. In this paper we add $\chi^2_{WMAP}$ to the BAO and SN
	    data, i.e., $\chi^2_{\rm
	    total}=\chi^2_{WMAP}+\chi^2_{BAO}+\chi^2_{SN}$. 
 \item[(10)] Marginalize the posterior distribution over $\Omega_bh^2$,
	     $\Omega_mh^2$, and $H_0$ with uniform priors. Since the
	     \WMAP\ distance priors combined with the BAO and SN data
	     provide tight constraints on these parameters, the
	     posterior distribution of these parameters is close to a
	     Gaussian distribution. Therefore, the marginalization is
	     equivalent to minimizing $\chi^2_{\rm total}$ with respect
	     to $\Omega_bh^2$, $\Omega_mh^2$, and $H_0$ \citep[see
	     also][]{cash:1976,wright:2007b}. We use a 
	     downhill simplex method for minimization ({\sf amoeba}
	     routine in Numerical Recipes
	     \citep{press/etal:NRIC:2e}). The marginalization over
	     $\Omega_bh^2$, 
	     $\Omega_mh^2$, and $H_0$ leaves us
	     with the the marginalized posterior distribution of the dark 
	     energy function, $w(a)$, and the curvature parameter,
	     $\Omega_k$.
\end{itemize}
Note that this prescription eliminates the need for running the Markov
Chain Monte Carlo entirely, and thus the computational cost for
evaluating the posterior distribution of $w(a)$ and $\Omega_k$ is not
demanding. 
In \S~\ref{sec:wz} we shall apply the \WMAP\ distance priors to
constrain the dark energy equation of state that depends on redshifts,
$w=w(z)$.

For those who wish to include an additional prior on $\Omega_bh^2$, we
give the inverse covariance matrix for the ``extended'' \WMAP\ distance priors:
$(l_A(z_*),R(z_*),z_*,100\Omega_bh^2)$, as well as the maximum likelihood
value of $100\Omega_bh^2$, in Table~\ref{tab:wmap_prior_cov2}.
We note, however, that it is sufficient to use the reduced set,
$(l_A(z_*),R(z_*),z_*)$, as the extended \WMAP\ distance priors give very
similar constraints on dark energy \citep[see][]{wang:prep}.

\subsubsection{Application of the \WMAP\ distance priors: 
Variable equation of state}
\label{sec:wz}
\begin{figure*}[ht]
\centering \noindent 
\includegraphics[width=18cm]{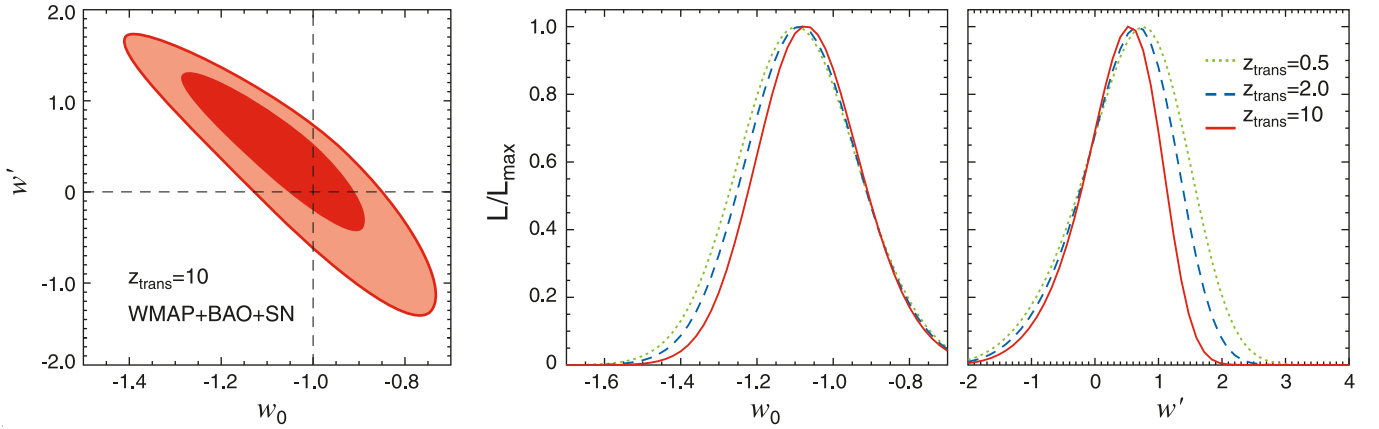}
\caption{%
Constraint on models of time-dependent dark energy equation of state,
 $w(z)$ (Eq.~[\ref{eq:wz}]), derived from the \WMAP\ distance priors
 ($l_A$, $R$, and $z_*$) combined with the BAO and SN distance data
 (\S~\ref{sec:wz}).
 There are three parameters: $w_0$
 is the value of $w$ at the present epoch, $w_0\equiv w(z=0)$, $w'$ is
 the first derivative of $w$ with respect to $z$ at $z=0$, 
$w'\equiv \left.dw/dz\right|_{z=0}$, and $z_{\rm trans}$ is the
 transition redshift above which $w(z)$ approaches to $-1$. Here, we assume
 flatness of the universe, $\Omega_k=0$.
 ({\it Left}) Joint two-dimensional marginalized distribution of $w_0$ and
 $w'$ for 
 $z_{\rm trans}=10$. The constraints are similar for the other values of
 $z_{\rm trans}$.
 The contours show the $\Delta\chi^2_{\rm total}=2.30$ (68.3\% CL) and 
 $\Delta\chi^2_{\rm total}=6.17$  (95.4\% CL).
 ({\it Middle}) One-dimensional marginalized distribution of $w_0$
 for $z_{\rm trans}=0.5$ (dotted), 2 (dashed), and 10 (solid).
 ({\it Middle}) One-dimensional marginalized distribution of $w'$
 for $z_{\rm trans}=0.5$ (dotted), 2 (dashed), and 10 (solid).
 The constraints are similar for all $z_{\rm trans}$.
 We do not find evidence for the evolution of dark energy.
 Note that neither BAO nor SN alone is able to constrain $w_0$ or $w'$:
 they need the \map\ data for lifting the degeneracy. Note also that
 BAO+SN is unable to lift the degeneracy either, as BAO needs the
 sound horizon size measured by the \map\ data.
}
\label{fig:wz}
\end{figure*}
\begin{figure*}[ht]
\centering \noindent 
\includegraphics[width=18cm]{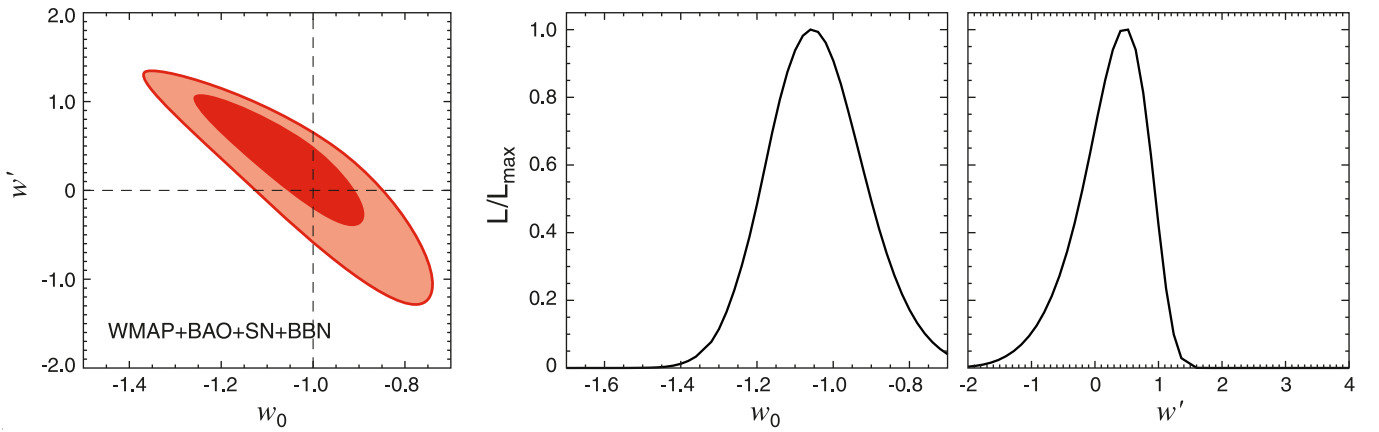}
\caption{%
Constraint on the linear evolution model of dark energy
 equation of state, 
 $w(z)=w_0+w'z/(1+z)$, derived from the \WMAP\ distance priors
 ($l_A$, $R$, and $z_*$) combined with the BAO and SN distance data as
 well as the Big Bang Nucleosynthesis (BBN) prior (Eq.~[\ref{eq:BBN}]).
 Here, we assume
 flatness of the universe, $\Omega_k=0$.
 ({\it Left}) Joint two-dimensional marginalized distribution of $w_0$ and
 $w'$. 
 The contours show the $\Delta\chi^2_{\rm total}=2.30$ (68.3\% CL) and 
 $\Delta\chi^2_{\rm total}=6.17$  (95.4\% CL).
 ({\it Middle}) One-dimensional marginalized distribution of $w_0$.
 ({\it Middle}) One-dimensional marginalized distribution of $w'$.
 We do not find evidence for the evolution of dark energy.
 Note that \citet{linder:2003}  defines $w'$ as
 the derivative of $w$ at $z=1$, whereas we define it as the derivative
 at $z=0$. They are related by $w'_{\rm linder}=0.5w'_{WMAP}$.
}
\label{fig:wright}
\end{figure*}

In this subsection we explore a time-dependent equation of state of dark
energy, $w(z)$. We use the following parametrized form:
\begin{equation}
 w(a) = \frac{a\tilde{w}(a)}{a+a_{\rm trans}} - \frac{a_{\rm
  trans}}{a+a_{\rm trans}},
\label{eq:wz}
\end{equation}
where $\tilde{w}(a)=\tilde{w}_0+(1-a)\tilde{w}_a$.
We give a motivation, derivation, and detailed discussion on this form
of $w(a)$ in Appendix~\ref{sec:eos}. This form has a number of desirable
properties: 
\begin{itemize}
 \item $w(a)$ approaches to $-1$ at early times, $a<a_{\rm trans}$, where
       $a_{\rm trans}=1/(1+z_{\rm trans})$ is the ``transition epoch,''
       and $z_{\rm trans}$ is the transition redshift. Therefore, the
       dark energy density tends to a constant value at $a<a_{\rm trans}$. 
 \item The dark energy density remains totally sub-dominant relative to the
       matter density at the decoupling epoch.
 \item We recover the widely used linear form, $w(a)=w_0+(1-a)w_a$
       \citep{chevallier/polarski:2001,linder:2003}, at late times,
       $a>a_{\rm trans}$. 
 \item The early-time behaviour is consistent with some of scalar field
       models classified as the ``thawing models,''
       \citep{caldwell/linder:2005} in which a scalar field was moving
       very slowly at early times, and then began to move faster at recent
       times. 
 \item Since the late-time form of $w(a)$ allows $w(a)$ to go below
       $-1$, our form is more general than 
       models based upon a single scalar field.
 \item The form is simple enough to give a closed, analytical form of
       the effective equation of state, $w_{\rm eff}(a)=(\ln
       a)^{-1}\int_0^{\ln a}d\ln a'~w(a')$ (Eq.~\ref{eq:weffnewform}), which determines the
       evolution of the dark energy density, $\rho_{\rm de}(a)=\rho_{\rm
       de}(0)a^{-3[1+w_{\rm eff}(a)]}$; thus, it allows one to compute
       the evolution of the expansion rate and cosmological distances
       easily. 
\end{itemize}

While this form contains three free parameters, $\tilde{w}_0$,
$\tilde{w}_a$ and $z_{\rm trans}$, we shall give constraints on the
present-day value of 
$w$, $w_0\equiv w(a=1)$, and the first derivative of $w$ at present,
$w'\equiv \left.dw/dz\right|_{z=0}$, instead of $\tilde{w}_0$ and
$\tilde{w}_a$, as they can be compared to the previous results in the
literature more directly. We find that the results are not sensitive to
the exact values of $z_{\rm trans}$.

In Fig.~\ref{fig:wz} we present the constraint on $w_0$ and $w'$ that we
have derived from the \WMAP\ distance priors ($l_A$, $R$, $z_*$),
combined with the BAO and SN data. 
Note that we have assumed a flat universe in this analysis,
although it is straightforward to include the spatial curvature. 
\citet{wang/mukherjee:2007} and \citet{wright:2007b} show that the
two-dimensional distribution 
extends more towards south-east, i.e., $w>-1$ and $w'<0$, when the
spatial curvature is allowed.

The 95\% limit on $w_0$ for $z_{\rm trans}=10$ is
$-0.33<1+w_0<0.21$\footnote{The 68\% intervals are
 $w_0=-1.06\pm 0.14$ and $w'=0.36\pm 0.62$ (WMAP+BAO+SN;
 $\Omega_k=0$).}.
Our results are consistent with the previous work using the \WMAP\
3-year data \citep[see][for recent work and references
therein]{zhao/etal:2007,wang/mukherjee:2007,wright:2007b,lazkoz/nesseris/perivolaropoulos:prep}. 
The \map\ 5-year data help tighten the upper limit on $w'$ and the lower limit
on $w_0$, whereas the lower limit on $w'$ and the upper limit on $w_0$ 
come mainly from the Type Ia supernova data. 
As a result, the lower limit on $w'$ and the upper limit on $w_0$ are
sensitive to whether we include the systematic errors in the supernova
data. For this investigation, see Appendix~\ref{sec:sn}.

Alternatively, one may take the linear form, $w(a)=w_0+(1-a)w_a$,
literally and extend it to an arbitrarily high redshift. 
This can result in an undesirable situation in 
which the dark energy is as important as the radiation density at the
epoch of the Big Bang Nucleosynthesis (BBN); however, one can constrain such a
scenario severely using the limit on the expansion rate from BBN
\citep{steigman:2007}. We follow \citet{wright:2007b} to adopt a
Gaussian prior on 
\begin{eqnarray}
\nonumber
& & \sqrt{1+\frac{\Omega_\Lambda (1+z_{\rm BBN})^{3[1+w_{\rm eff}(z_{\rm
  BBN})]}}{\Omega_m(1+z_{\rm BBN})^3+\Omega_r(1+z_{\rm
  BBN})^4+\Omega_k(1+z_{\rm BBN})^2}}\\
& &  = 0.942 \pm 0.030,	
\label{eq:BBN}
\end{eqnarray}
where we have kept $\Omega_m$ and $\Omega_k$ for definiteness, but they
are entirely negligible compared to the radiation density at the
redshift of BBN, $z_{\rm BBN}=10^9$.
Figure~\ref{fig:wright} shows the constraint on $w_0$ and $w'$ for the
linear evolution model derived from the \WMAP\ distance priors, the BAO
and SN data, and the BBN prior. The 95\% limit on $w_0$ is
$-0.29<1+w_0<0.21$\footnote{The 68\% intervals  
are $w_0=-1.04\pm 0.13$ and $w'=0.24\pm 0.55$ (WMAP+BAO+SN+BBN;
$\Omega_k=0$).}, which is similar to what we have obtained above.

\subsection{\WMAP\ normalization prior} 
\label{sec:wmapnorm}
\begin{deluxetable}{lc}
\tablecolumns{2}
\small
\tablewidth{0pt}
\tablecaption{%
Amplitude of curvature perturbations, ${\cal R}$, measured by \WMAP\ at
 $k_{WMAP}=0.02~{\rm Mpc}^{-1}$ 
}
\tablehead{ \colhead{Model} & \colhead{$10^9\times \Delta_{\cal
 R}^2(k_{WMAP})$}  
}
\startdata
\mbox{$\Omega_k=0$ and $w=-1$} & $2.211\pm 0.083$ \nl
\mbox{$\Omega_k\neq 0$ and $w=-1$} & $2.212\pm 0.084$ \nl
\mbox{$\Omega_k=0$ and $w\neq -1$} & $2.208\pm 0.087$ \nl
\mbox{$\Omega_k\neq 0$ and $w\neq -1$} & $2.210\pm 0.084$ \nl
\mbox{$\Omega_k=0$, $w=-1$ and $m_\nu>0$} & $2.212\pm 0.083$ \nl
\mbox{$\Omega_k=0$, $w\neq -1$ and $m_\nu>0$} & $2.218\pm 0.085$ \nl
\hline
\WMAP\ Normalization Prior & $2.21\pm 0.09$
\enddata
\label{tab:wmapnorm}
\end{deluxetable}
\begin{figure*}[ht]
\centering \noindent 
\includegraphics[width=18cm]{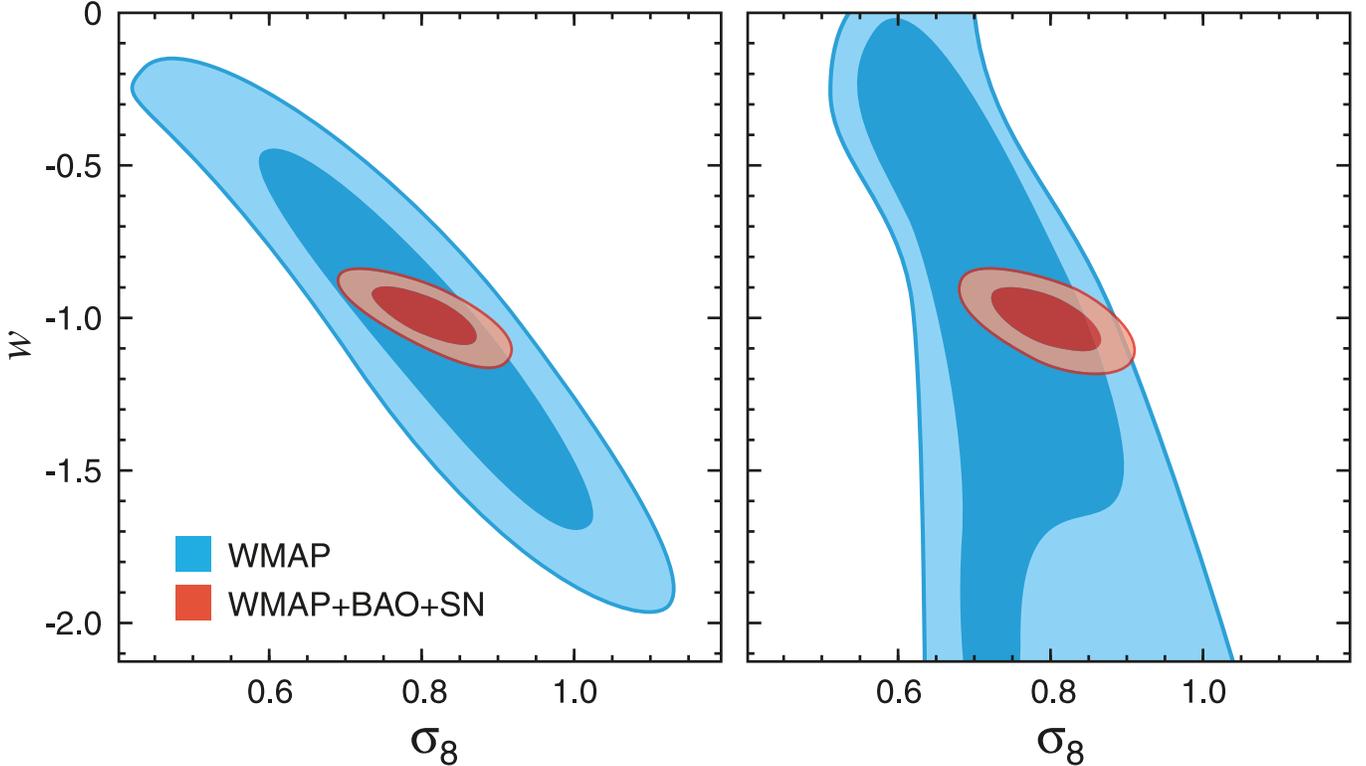}
\caption{%
 Predictions for the present-day amplitude of matter fluctuations,
 $\sigma_8$, as a function of the (constant) dark energy equation of
 state parameter, $w$, 
 derived from the full \WMAP\ data 
 (blue) as well as from WMAP+BAO+SN (red).
 The contours show the 68\% and 95\% CL.
 ({\it Left})  Flat universe, $\Omega_k=0$.
 ({\it Right}) Curved universe, $\Omega_k\neq 0$.
}
\label{fig:wmapnorm}
\end{figure*}

So far, we have been mainly using the distance information to constrain
the properties of 
dark energy; however, this is not the only information that
one can use to constrain the properties of dark energy.
The amplitude of fluctuations is  a
powerful tool for  distinguishing  between dark
energy and modifications to gravity
\citep{ishak/upadhye/spergel:2006,koyama/maartens:2006,amarzguioui/etal:2006,dore/etal:2007,linder/cahn:2007,upadhye:2007,zhang/etal:2007,yamamoto/etal:2007,chiba/takahashi:2007,bean/etal:2007b,hu/sawicki:2007,song/hu/sawicki:2007,daniel/etal:2008,jain/zhang:prep,bertschinger/zukin:prep,amin/wagoner/blandford:prep,hu:prep},
as well as for determining the mass of neutrinos 
\citep{hu/eisenstein/tegmark:1998,lesgourgues/pastor:2006}.

The microwave background observations measure the amplitude of fluctuations
at the decoupling epoch. By combining this measurement with the
amplitude measured from various low redshift tracers, one can learn more
about the dark energy properties and the mass of neutrinos.

The overall amplitude of CMB anisotropy is set by the amplitude of
primordial curvature perturbations, ${\cal R}$. For example, on  very
large angular scales where the Sachs--Wolfe limit can be used, the
temperature anisotropy is given by $\Delta T/T=-{\cal R}/5$ or, in
terms of the curvature perturbation during the matter era, $\Phi$, it is
given by $\Delta T/T=-\Phi/3$. On small angular scales where the
acoustic physics must be taken into account, we have the acoustic
oscillation whose
amplitude is also given by ${\cal R}$.

This motivates our reporting the ``\WMAP\ normalization,'' a measurement
of the overall normalization of the curvature perturbations expressed
in terms of $\Delta^2_{\cal R}(k_{WMAP})$, where $\Delta^2_{\cal
R}(k)\equiv k^3P_{\cal R}(k)/(2\pi^2)$ is a contribution to the total variance
of ${\cal R}$, $\langle {\cal R}^2\rangle$, per logarithmic interval of
$k$ (see also Eq.~[\ref{eq:pR}] and descriptions below it).

Here, $k_{WMAP}$ is different from $k_0=0.002~{\rm Mpc}^{-1}$ that we
used to define $n_s$, $dn_s/d\ln k$, $r$, and $\Delta^2_{\cal R}(k_0)$
reported in Table~\ref{tab:summary} or \ref{tab:ns}.
The goal in this subsection is to give the normalization that is {\it as
model independent as possible}. 

At $k_0=0.002~{\rm Mpc}^{-1}$, for example, 
we find $10^9\Delta_{\cal R}^2(k_0)=2.48$, 2.41, and 2.46 for
a flat $\Lambda$CDM model, a curved $\Lambda$CDM model, and a flat $\Lambda$CDM
model with massive neutrinos. The scatter between these values comes
solely from the fact that $k_0=0.002~{\rm Mpc}^{-1}$ is not a right
place to define the normalization. In other words, this is not the pivot
scale of the \WMAP\ data.

We find that $k_{WMAP}=0.02~{\rm Mpc}^{-1}$, i.e., a factor of 10 larger
than $k_0$, gives similar values of $\Delta_{\cal R}^2(k_{WMAP})$ for a
wide range of models, as summarized in Table~\ref{tab:wmapnorm}.
From these results, we give the \WMAP\ normalization prior:
\begin{equation}
 \Delta_{\cal R}^2(k_{WMAP}) = (2.21\pm 0.09)\times 10^{-9},
\label{eq:wmapnorm}
\end{equation}
which is valid for models with $\Omega_k\neq 0$, $w\neq -1$, or
$m_\nu> 0$. However, we find that these normalizations cannot be used 
for the models that have the tensor modes, $r>0$, or the running index,
$dn_s/d\ln k\neq 0$. We failed to find a universal normalization for
these models. Nevertheless, our \WMAP\ normalization given by
Eq.~(\ref{eq:wmapnorm}) is still valid for a wide range of cosmological models.

How can one use the \WMAP\ normalization? 
In order to predict the linear matter density power spectrum, $P_{\rm
lin}(k)$, one needs to relate the primordial curvature perturbations, 
${\cal R}_{\mathbf k}$, to the linear matter density fluctuations at
arbitrary redshifts,
$\delta_{m,{\mathbf k}}(z)$.
From Einstein's equations, we find
\citep[see e.g., Appendix C of][]{takada/komatsu/futamase:2006}
\begin{equation}
 \delta_{m,{\mathbf k}}(z) = \frac{2k^3}{5H_0^2\Omega_m}{\cal R}_{\mathbf
  k}T(k)D(k,z),
\end{equation}
where $D(k,z)$ and $T(k)$ is the linear growth rate and the matter
transfer function normalized such that 
$T(k)\rightarrow 1$ as $k\rightarrow 0$,  and
$(1+z)D(k,z)\rightarrow 1$ 
 as $k\rightarrow 0$ during the matter era (e.g., $z=30$, where the
 radiation density is less than 1\% of the matter density), respectively. 
Note that $D(k,z)$ does not depend on $k$ when neutrinos are massless;
however, it depends on $k$ when they are massive
\citep[e.g.,][]{hu/eisenstein:1998}. 
The linear matter density power spectrum is given by
\begin{eqnarray}
\nonumber
 \frac{k^3 P_{\rm lin}(k,z)}{2\pi^2} &=&  (2.21\pm 0.09)\times 10^{-9}
\left(\frac{2k^2}{5H_0^2\Omega_m}\right)^2\\
&\times& D^2(k,z) T^2(k)\left(\frac{k}{k_{WMAP}}\right)^{n_s-1}.
\end{eqnarray}

One application of the \WMAP\ normalization is the computation of the
present-day normalization of matter fluctuations, which is commonly
expressed in terms of $\sigma_8$, given by
\begin{eqnarray}
\nonumber
 \sigma_8^2 &=& (2.21\pm 0.09)\times 10^{-9}
\left(\frac{2}{5\Omega_m}\right)^2\\
\nonumber
&\times& \int \frac{dk}{k}
D^2(k,z=0)T^2(k)\frac{k^4}{H_0^4}\left(\frac{k}{k_{WMAP}}\right)^{n_s-1}\\
&\times& \left[\frac{3\sin(kR)}{(kR)^3}-\frac{3\cos(kR)}{(kR)^2}\right]^2,
\end{eqnarray}
where $R=8~h^{-1}{\rm Mpc}$.
Both the dark energy properties (or modified gravity) and the mass of
neutrinos change  
the value of $D(k,z=0)$. The transfer function, $T(k)$, is much less
affected, as long as neutrinos were still relativistic at the decoupling
epoch, and the dark energy or modified gravity effect was unimportant at
the decoupling epoch. 

Ignoring the mass of neutrinos and modifications to gravity, one can
obtain the growth 
rate by solving the following differential equation
\citep{wang/steinhardt:1998,linder/jenkins:2003}:
\begin{eqnarray}
\nonumber
& &\frac{d^2g}{d\ln a^2} 
+ \left[\frac52+\frac12\left(\Omega_k(a)-3w_{\rm
					      eff}(a)\Omega_{de}(a)\right)\right]\frac{dg}{d\ln
a} \\
&+&\left[2\Omega_k(a)+\frac32(1-w_{\rm eff}(a))\Omega_{de}(a)\right]g(a) =0,
\end{eqnarray}
where 
\begin{eqnarray}
 g(a)&\equiv& \frac{D(a)}{a} = (1+z)D(z),\\
 \Omega_k(a) &\equiv& \frac{\Omega_kH_0^2}{a^2H^2(a)},\\
 \Omega_{de}(a) &\equiv& \frac{\Omega_\Lambda H_0^2}{a^{3[1+w_{\rm
  eff}(a)]}H^2(a)},\\
 w_{\rm eff}(a) &\equiv& \frac1{\ln a}\int_0^{\ln a}d\ln a'~w(a').
\end{eqnarray}
During the matter era, $g(a)$ does not depend on $a$; thus, the natural
choice for the 
initial conditions are $g(a_{\rm initial})=1$ and $\left. dg/d\ln
a\right|_{a=a_{\rm initial}}=0$, where $a_{\rm initial}$ must be taken
during the matter era, e.g., $a_{\rm initial}=1/31$ (i.e., $z=30$).

In Fig.~\ref{fig:wmapnorm} we show the predicted values of $\sigma_8$ as
a function of $w$ in a flat universe (left panel) and curved universes
(right panel). (See the middle panel of Fig.~\ref{fig:mnu} for
$\sigma_8$ as a function of the mass of neutrinos.)
Here, we have used the full information in the \WMAP\ data. 
The normalization information alone is unable to give meaningful 
predictions for $\sigma_8$, which depends not only on $\Delta^2_{\cal
R}(k_{WMAP})$, but also on the other cosmological parameters via 
$T(k)$ and $D(k,z=0)$, especially $w$ and $\Omega_mh^2$.
While the predictions from the \WMAP\ data alone are still weak, 
adding the extra distance information from the BAO and SN data helps 
improve the predictions. We find
\ensuremath{\sigma_8 = 0.807^{+ 0.045}_{- 0.044}} for a flat universe, and 
\ensuremath{\sigma_8 = 0.795\pm 0.046} for
curved universes.

By combining these results with $\sigma_8$  measured from various low
redshift tracers, one can reduce the remaining correlation between $w$ and
$\sigma_8$ to obtain a better limit on $w$. The precision of
the current data from weak lensing surveys is comparable to these
predictions, e.g.,  
$\sigma_8(\Omega_m/0.25)^{0.64}=0.785\pm 0.043$
\citep[Canada-France-Hawaii Telescope Legacy Survey
 (CFHTLS);][]{fu/etal:2008}. The weak lensing surveys will soon become
powerful enough to yield smaller uncertainties in $\sigma_8$ than 
predicted from WMAP+BAO+SN.

\section{Neutrinos}
\label{sec:neutrino}
In this section, we shall use the \map\ data, combined with the distance
information from BAO and SN observations, to place limits on the total
mass of massive neutrino species (\S~\ref{sec:massnu}), as well as on
the effective number of neutrino-like species that were still
relativistic at the decoupling epoch (\S~\ref{sec:neff})

\subsection{Neutrino Mass}
\label{sec:massnu}
\subsubsection{Motivation}
\label{sec:massnu_motivation}
The existence of non-zero neutrino masses has been established firmly by
the experiments detecting atmospheric neutrinos
\citep{hirata/etal:1992,fukuda/etal:1994,fukuda/etal:1998,allison/etal:1999,ambrosio/etal:2001},
solar neutrinos
\citep{davis/harmer/hoffman:1968,cleveland/etal:1998,hampel/etal:1999,abdurashitov/etal:1999,fukuda/etal:2001b,fukuda:2001,ahmad/etal:2002,ahmed/etal:2004},
reactor neutrinos \citep{eguchi/etal:2003,araki/etal:2005}, and
accelerator beam neutrinos
\citep{ahn/etal:2003,michael/etal:2006}. These experiments have placed
stringent limits on the squared mass {\it differences} between the
neutrino mass eigenstates: $\Delta m^2_{21}\simeq 8\times 10^{-5}~{\rm
eV}^2$  from the solar and reactor experiments, and $\Delta
m^2_{32}\simeq 3\times 10^{-3}~{\rm eV}^2$ from the atmospheric and
accelerator beam experiments. 

One needs different experiments to measure the {\it absolute}
masses. The next-generation tritium $\beta$-decay experiment,
KATRIN\footnote{\sf http://www-ik.fzk.de/\~\ katrin}, is expected to
reach the electron neutrino mass of as small as $\sim 0.2$~eV. Cosmology
has also been providing useful limits on the mass of neutrinos
\citep[see][for
reviews]{dolgov/etal:2002,elgaroy/etal:2005,tegmark:2005,lesgourgues/pastor:2006,fukugita:2006,hannestad:2006b}. Since
the determination of the neutrino mass is of fundamental importance in
physics, there is enough motivation to pursue cosmological constraints
on the neutrino mass. 

How well can  CMB constrain the mass of neutrinos? We do not expect
massive neutrinos to affect the CMB power spectra very much (except
through the gravitational lensing effect), if they were still
relativistic at the decoupling epoch, $z\simeq 1090$. This means that, for
massive neutrinos to affect the CMB power spectra, {\it at least} one of
the neutrino masses must be greater than the mean energy of relativistic
neutrinos per particle at $z\simeq 1090$ when the photon temperature of
the universe was $T_\gamma\simeq 3000~{\rm K}\simeq 0.26$~eV. Since the
mean energy of relativistic neutrinos is given by $\langle
E\rangle=(7\pi^4 T_\nu)/(180\zeta(3))\simeq 3.15T_\nu =
3.15(4/11)^{1/3}T_\gamma$, we need {\it at least} one neutrino species
whose mass satisfies $m_\nu>3.15(4/11)^{1/3}T_\gamma \simeq 0.58$~eV;
thus, it would not be possible to constrain the neutrino mass using the CMB data alone, if the mass of the heaviest neutrino species is below this value. 

If the neutrino mass eigenstates are degenerate with the effective
number of species equal to 3.04, this argument suggests that $\sum
m_\nu\sim 1.8$~eV would be the limit to which the CMB data are
sensitive. \citet{ichikawa/fukugita/kawasaki:2005} argue that  $\sum
m_\nu\sim 1.5$~eV would be the limit for the CMB data alone, which is
fairly close to the value given above. 

In order to go beyond $\sim 1.5$~eV, therefore, one needs to combine the
CMB data with the other cosmological probes. We shall combine the \map\
data with the distance information from BAO and SN to place a limit on
the neutrino mass. We shall not use the galaxy power spectrum in this
paper, and therefore our limit on the neutrino mass is free from the
uncertainty in the galaxy bias. We discuss this in more detail in the
Analysis section below. 

\subsubsection{Analysis}
\label{sec:massnu_analysis}
\begin{figure*}[ht]
\centering \noindent 
\includegraphics[width=18cm]{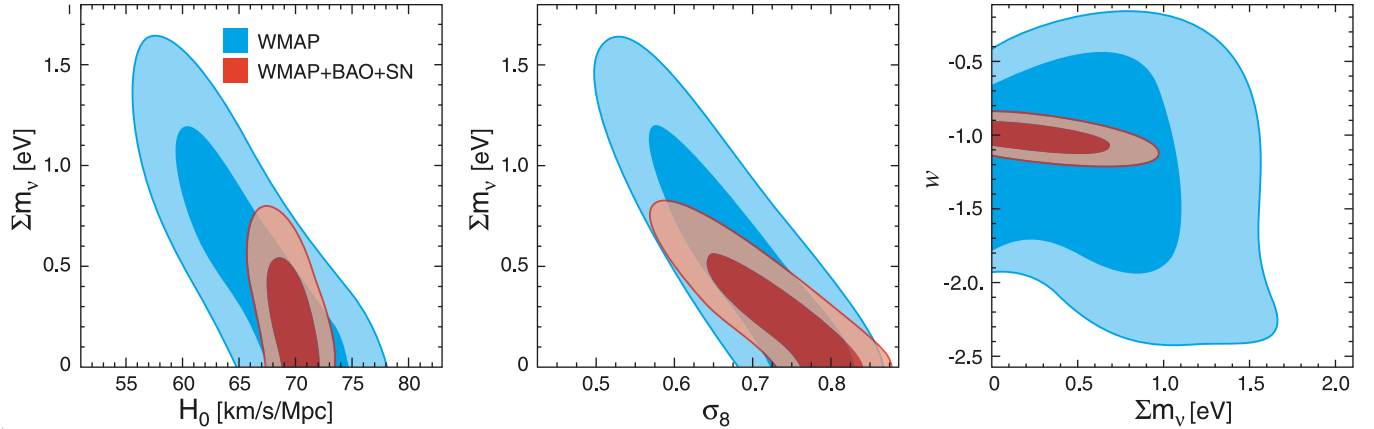}
\caption{%
Constraint on the total mass of neutrinos, $\sum m_\nu$
 (\S~\ref{sec:massnu_results}). 
 In all panels we show
 the WMAP-only results in blue and WMAP+BAO+SN in red.
 ({\it Left}) 
 Joint two-dimensional marginalized distribution of  $H_0$ and $\sum
 m_\nu$ (68\% and 95\% CL). The additional distance information from BAO
 helps reduce 
 the correlation between $H_0$ and $\sum m_\nu$.  ({\it Middle})
 The \map\ data, combined with the distances from BAO and SN, predict the
 present-day amplitude of matter fluctuations, $\sigma_8$, as a function
 of $\sum m_\nu$. An independent determination of $\sigma_8$ will help
 determine $\sum m_\nu$ tremendously. ({\it Right}) Joint
 two-dimensional marginalized distribution of $w$ and $\sum m_\nu$. No
 significant correlation is observed. Note that we have a prior on $w$,
 $w>-2.5$, and thus the WMAP-only lower limit on $w$ in this panel
 cannot be trusted. 
}
\label{fig:mnu}
\end{figure*}

We assume that, for definiteness, the neutrino mass eigenstates are {\it
degenerate}, which means that all of the three neutrino species have
equal masses.\footnote{While the current cosmological data are not yet
sensitive to the mass of {\it individual} neutrino species, i.e., the
mass hierarchy, this situation may change in the future, with high-$z$
galaxy  redshift surveys or  weak lensing
surveys \citep{takada/komatsu/futamase:2006,slosar:2006,hannestad/wang:2007,kitching/etal:2008,abdalla/rawlings:2007}.}
We measure the neutrino mass density parameter, $\Omega_\nu h^2$, and convert
it to the total mass, $\sum m_\nu$, via 
\begin{equation} 
\sum m_\nu \equiv N_\nu m_\nu = 94~{\rm eV}(\Omega_\nu h^2),
\end{equation}
where $N_\nu$ is the number of massive neutrino species. We take it to
be 3.04. Note that in this case the mass density parameter is the sum of
baryons, CDM, and neutrinos: $\Omega_m=\Omega_b+\Omega_c+\Omega_\nu$. 

Since the release of the 1-year \citep{spergel/etal:2003} and 3-year
\citep{spergel/etal:2007} results on the cosmological analysis of the
\map\ data, there have been a number of studies with regard to the
limits on the mass of neutrinos
\citep{hannestad:2003,elgaroy/lahav:2003,allen/schmidt/bridle:2003,tegmark/etal:2004,barger/marfatia/tregre:2004,hannestad/raffelt:2004,crotty/lesgourgues/pastor:2004,seljak/etal:2005b,seljak/etal:2005,ichikawa/fukugita/kawasaki:2005,hannestad:2005b,lattanzi/ruffini/vereshchagin:2005,hannestad/raffelt:2006,goobar/etal:2006,feng/etal:2006,lesgourgues/etal:2007}.
These analyses reached different limits depending upon (1) the choice of
data sets and (2) the parameters in the cosmological model. 

The strongest limits quoted on neutrino masses come from combining CMB
measurements with measurements of the amplitude of density fluctuations
in the recent universe.  Clustering of galaxies and Ly$\alpha$
forest observations have been used to obtain some of the strongest
limits on neutrino masses
\citep{seljak/etal:2005,seljak/slosar/mcdonald:2006,viel/haehnelt/lewis:2006}.
As the neutrino mass increases, the amplitude of mass fluctuations on
small scales decreases 
\citep[see also the middle panel of
Fig.~\ref{fig:mnu}]{bond/efstathiou/silk:1980,bond/szalay:1983,ma:1996},  
which can be used to weigh neutrinos in the universe
\citep{hu/eisenstein/tegmark:1998,lesgourgues/pastor:2006}.

These analyses are sensitive to correctly calculating the relationship
between the level of observed fluctuations in galaxies (or gas) and the
mass fluctuation with the strongest limits coming from the smallest
scales in the analyses. With the new \WMAP\ data, these limits are
potentially even stronger.  There are,
however, several potential concerns with this limit: there is already
``tension'' between the high level of fluctuations seen in the Lyman
alpha forest and the amplitude of mass fluctuations inferred from \WMAP\
\citep{lesgourgues/etal:2007}; the relationship between gas temperature
and density appears to be more complicated than assumed in the previous
Ly$\alpha$ forest analysis \citep{kim/etal:2007,bolton/etal:2008}; 
and additional astrophysics could potentially be the source of some of
the small scale fluctuations seen in the Ly$\alpha$ forest data.  Given
the power of the Ly$\alpha$ forest data, it is important to address
these issues; however, they are beyond the scope of this paper.  

In this paper, we take the more conservative approach and use only the
\map\ data and the distance measures to place limits on the neutrino masses.
Our approach is more conservative than \citet{goobar/etal:2006}, who
have found a limit of 0.62 eV on the sum of the neutrino mass from the
\WMAP\ 3-year data, the SDSS-LRG BAO measurement, the SNLS supernova
data, and the shape of the galaxy power spectra from the SDSS main
sample and 2dFGRS. While we use the \WMAP\ 5-year data, the SDSS+2dFGRS
BAO measurements, and the union supernova data, we do not use the
shape of the galaxy power spectra. See \S~\ref{sec:analysis_ex} in this
paper or \citet{dunkley/etal:prep} for more detail on this choice.

In summary, we do not use the amplitude or shape of the matter power
spectrum, but rely exclusively on the CMB data and the distance
measurements. As a result our limits are weaker than the strongest
limits in the literature. 

Next, let us discuss (2), the choice of parameters. A few correlations
between the neutrino mass and other cosmological parameters have been
identified. \citet{hannestad:2005b} has found that the limit on the
neutrino mass degrades significantly when the dark energy equation of
state, $w$, is allowed to vary \citep[see also Fig.~18 of
][]{spergel/etal:2007}.
 This correlation would arise only when the
amplitude of the galaxy or Ly$\alpha$ forest power spectrum was
included, as the dark energy equation of state influences the growth
rate of the structure formation. Since we do not include them, our limit
on the neutrino mass is not degenerate with $w$. We shall come back to
this point later in the Results section. Incidentally, our limit is not
degenerate with the running index,  $dn_s/d\ln k$, or the
tensor-to-scalar ratio, $r$.  

\subsubsection{Results}
\label{sec:massnu_results}
Figure~\ref{fig:mnu} summarizes our limits on the sum of neutrino
masses, $\sum m_\nu$. 

With the \map\ data alone we find
\ensuremath{\sum m_\nu < 1.3\ \mbox{eV}\ \mbox{(95\% CL)}} for the
$\Lambda$CDM model in which $w=-1$, and
\ensuremath{\sum m_\nu < 1.5\ \mbox{eV}\ \mbox{(95\% CL)}} for $w\ne -1$. (We
assume a flat universe in both cases.) These constraints are very
similar, which means that $w$ and $\sum m_\nu$ are not degenerate. We
show this more explicitly on the right panel of Fig.~\ref{fig:mnu}.   

When the BAO and SN data are added, our limits improve significantly, 
by a factor 2, to 
\ensuremath{\sum m_\nu < 0.67\ \mbox{eV}\ \mbox{(95\% CL)}} for
$w=-1$, and 
\ensuremath{\sum m_\nu < 0.80\ \mbox{eV}\ \mbox{(95\% CL)}} for
$w\ne -1$.  Again, we do not observe much correlation between $w$ and
$\sum m_\nu$. While the distances out to {\it either} BAO or SN 
cannot reduce correlation between $\Omega_m$ (or $H_0$) and $w$, 
a combination of the two can reduce this correlation effectively, leaving
little correlation left on the right panel of Fig.~\ref{fig:mnu}. 

What information do BAO and SN add to improve the limit on $\sum
m_\nu$? It's the Hubble constant, $H_0$, as shown in the left panel of
Fig.~\ref{fig:mnu}. This effect has been explained by
\citet{ichikawa/fukugita/kawasaki:2005} as follows.

The massive neutrinos modify the CMB power spectrum
by their changing the matter-to-radiation ratio at the decoupling epoch.
If the sum of degenerate neutrino masses is below 1.8~eV, the
neutrinos were still relativistic at the decoupling epoch.
However, {\it they are definitely non-relativistic at the present
epoch}, as the neutrino oscillation experiments have shown that
at least one neutrino species is heavier than 0.05~eV. This means that
the $\Omega_m$ that we measure must be the sum of $\Omega_b$, $\Omega_c$,
and $\Omega_\nu$; however, at the decoupling epoch, neutrinos were still
relativistic, and thus the matter density at the decoupling epoch was
actually smaller than a naive extrapolation from the present value.

As the matter-to-radiation ratio was smaller than one would
naively expect, it would accelerate the decay of gravitational potential
around the decoupling epoch. This leads to an enhancement in the
so-called  early integrated Sachs--Wolfe (ISW)
effect.
The larger $\sum m_\nu$ is, the larger early ISW becomes, as
long as the neutrinos were still relativistic at the decoupling epoch,
i.e., $\sum m_\nu\lesssim 1.8$~eV. 

The large ISW causes the first peak position to
shift to lower multipoles by adding  power at $l\sim 200$; however, this
shift can be absorbed by a reduction in the value of
$H_0$.\footnote{This is similar to what happens to the curvature
constraint from the 
CMB data alone. A positive curvature model, $\Omega_k<0$, shifts the
acoustic peaks to lower multipoles; however, this shift can be absorbed
by a reduction in the value of $H_0$. As a result, a closed universe
with $\Omega_k\sim -0.3$ and $\Omega_\Lambda\sim 0$ is still a good fit,
if Hubble's constant is as low as $H_0\sim 30~{\rm km/s/Mpc}$ \citep{spergel/etal:2007}.} 
This is why
$\sum m_\nu$ and $H_0$ are 
anti-correlated \citep[see][for a further discussion on this subject]{ichikawa/fukugita/kawasaki:2005}.

It is the BAO distance that provides a better limit on $H_0$, as BAO is
an absolute distance indicator. The SN is totally insensitive to $H_0$,
as their absolute magnitudes have been marginalized over (SN is a
relative distance indicator); however, 
the SN data do help reduce the correlation between $w$ and $H_0$ when $w$ is
allowed to vary. As a result, we have equally tight limits on $\sum
m_\nu$ regardless of $w$. 

Our limit,
\ensuremath{\sum m_\nu < 0.67\ \mbox{eV}\ \mbox{(95\% CL)}} (for
$w=-1$), is weaker than the best limit quoted in the literature, as
we have not used the information on the amplitude of fluctuations traced
by the large-scale structure. The middle panel of  Fig.~\ref{fig:mnu}
shows how the \WMAP\ data combined with BAO and SN predict the
present-day amplitude of matter fluctuations, $\sigma_8$, as a function
of $\sum m_\nu$. From this it is clear that an accurate, {\it
independent} measurement of $\sigma_8$ will reduce the correlation between
$\sigma_8$ and $\sum m_\nu$, and provide a significant improvement in
the limit on $\sum m_\nu$. 

Improving upon our understanding of non-linear astrophysical effects
such as those raised by \citet{bolton/etal:2008} for the Ly$\alpha$
forest data and  \citet{sanchez/cole:2008} for the SDSS and 2dFGRS data
is a promising way to improve upon the numerical value of the limit, as
well as the robustness of the limit, on the mass of neutrinos.

\subsection{Effective number of neutrino species}
\label{sec:neff}
\begin{figure*}[ht]
\centering \noindent 
\includegraphics[width=18cm]{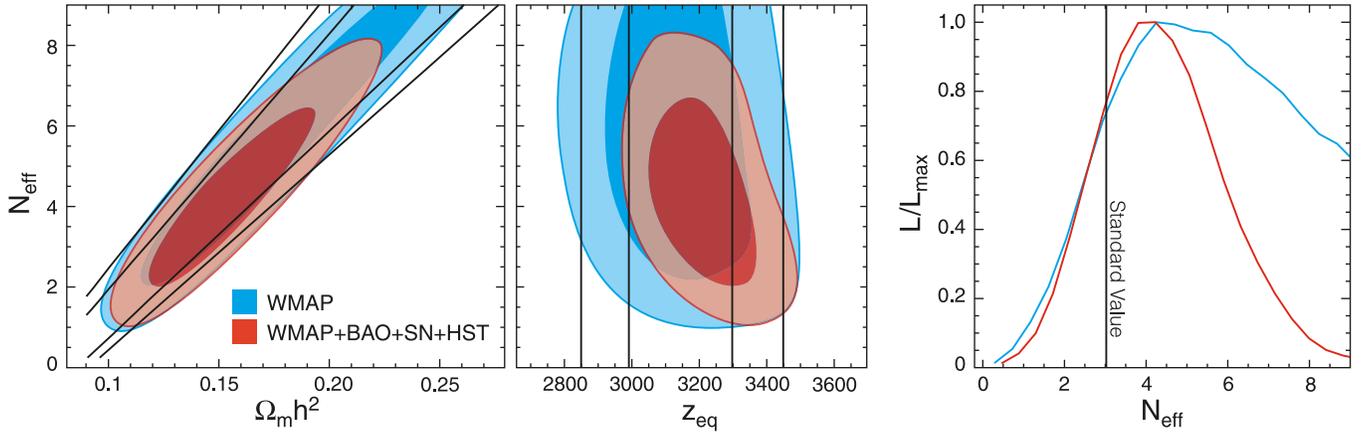}
\caption{%
 Constraint on the effective number of neutrino species, $N_{\rm
 eff}$ (\S~\ref{sec:neff_results}). Note that we have imposed a prior on
 $N_{\rm eff}$, $0<N_{\rm 
 eff}<10$. 
 In all panels we show
 the WMAP-only results in blue and WMAP+BAO+SN in red. ({\it Left}) 
 Joint two-dimensional marginalized distribution (68\% and 95\% CL),
 showing a strong 
 degeneracy between $\Omega_mh^2$ and  $N_{\rm eff}$. This degeneracy
 line is given by the equality redshift, $1+z_{\rm
 eq}=\Omega_m/\Omega_r=(4.050\times 10^4)\Omega_mh^2/(1+0.2271N_{\rm
 eff})$. The thick solid lines show the 68\% and 95\% limits calculated
 from the WMAP-only limit on $z_{\rm eq}$: 
 \ensuremath{z_{\rm eq} = 3141^{+ 154}_{- 157}} (68\% CL). The 95\%
 CL contours do not follow the 
 lines below $N_{\rm eff}\sim 1.5$ but close there, which shows 
a strong evidence for the cosmic neutrino background from its effects
 on the CMB power spectrum via the 
 neutrino anisotropic stress. The BAO and SN provide an independent
 constraint on $\Omega_mh^2$, which helps reduce the degeneracy between
 $N_{\rm eff}$ and $\Omega_mh^2$.
 ({\it Middle}) When we transform the horizontal axis of the left panel to
 $z_{\rm eq}$, we observe no degeneracy. The vertical solid lines show
 the one-dimensional marginalized 68\% and 95\% distribution calculated from 
 the WMAP-only limit on $z_{\rm eq}$:
 \ensuremath{z_{\rm eq} = 3141^{+ 154}_{- 157}} (68\% 
 CL). Therefore, the left panel is simply 
 a rotation of this panel using a relation between $z_{\rm eq}$,
 $\Omega_mh^2$, and $N_{\rm eff}$.
 ({\it Right}) One-dimensional marginalized distribution of $N_{\rm
 eff}$ from WMAP-only and WMAP+BAO+SN+HST. Note that a gradual decline of the
 likelihood toward $N_{\rm eff}\gtrsim 6$ for the WMAP-only constraint
 should not be trusted, as it is affected by the hard prior, $N_{\rm
 eff}<10$. The WMAP+BAO+SN+HST constraint is robust. This figure shows
 that the lower limit on $N_{\rm eff}$ is coming solely from the \WMAP\ data.
 The 68\% interval from WMAP+BAO+SN+HST,
 \ensuremath{N_{\rm eff} = 4.4\pm 1.5},
  is consistent with the standard value, $3.04$, which is shown by the
 vertical line.
}
\label{fig:neff}
\end{figure*}
\subsubsection{Motivation}
\label{sec:neff_motivation}
While the absolute mass of neutrinos is unknown, the number of neutrino
species is well known: it's three. 
The high precision measurement of the decay width of $Z$ into neutrinos 
(the total decay width minus the decay width to quarks and charged
leptons), carried out by LEP using the production of $Z$ in $e^+e^-$ collisions, 
has yielded $N_\nu = 2.984 \pm 0.008$ \citep{yao/etal:2006}.
However, are there any {\it other}
particles that we do not know yet,
and that were relativistic at the photon decoupling epoch? 

Such extra relativistic particle species can change the expansion rate
of the universe during the radiation era. As a result, they change the predictions
from the Big Bang Nucleosynthesis (BBN) for the abundance of light
elements such as helium and deuterium \citep{steigman/schramm/gunn:1977}. One can use this property to
place a tight bound on the relativistic degrees of freedom, expressed in
terms of the ``effective number of neutrino species,'' $N_{\rm eff}$
(see Eq.~[\ref{eq:neff}] below for the precise definition).  
As the BBN occurred at the energy of $\sim 0.1$~MeV, which is later than
the neutrino decoupling epoch immediately followed by $e^+e^-$ annihilation, 
the value of $N_{\rm eff}$ for 3 neutrino species is slightly larger
than 3. With other subtle corrections included, the current standard
value is $N_{\rm eff}^{\rm standard}=3.04$
\citep{dicus/etal:1982,gnedin/gnedin:1998,dolgov/hansen/semikoz:1999,mangano/etal:2002}. 
The 2$\sigma$ interval for $N_{\rm eff}$ from the observed helium abundance, $Y_P=0.240\pm
0.006$, is $1.61< N_{\rm eff}< 3.30$ \citep[see][for a recent summary]{steigman:2007}.

Many people have been trying to find evidence for the extra relativistic
degrees of freedom in the universe, using the cosmological probes
such as CMB and the large-scale structure
\citep{pierpaoli:2003a,hannestad:2003,crotty/lesgourgues/pastor:2003,crotty/lesgourgues/pastor:2004,barger/etal:2003,trotta/melchiorri:2005,lattanzi/ruffini/vereshchagin:2005,cirelli/strumia:2006,hannestad:2006,ichikawa/kawasaki/takahashi:2007,mangano/etal:2007,hamann/etal:2007,debernardis/etal:2008}.
There is a strong motivation to seek the answer for the following 
question, ``can we detect the cosmic neutrino background, and confirm
that the signal is consistent with the expected number of neutrino species that
we know?'' Although we cannot detect the cosmic neutrino background
directly yet, there is a possibility that we can detect it indirectly by
looking for the signatures of neutrinos in the CMB power spectrum.

In this section we shall revisit this classical problem by using the
\WMAP\ 5-year data as well as the distance information from BAO and SN,
and Hubble's constant measured by HST.

\subsubsection{Analysis}
\label{sec:neff_analysis}
It is common to write the energy density of neutrinos (including
anti-neutrinos), when they were still relativistic, in terms of the
effective number of neutrino species, $N_{\rm eff}$,  as 
\begin{equation}
 \rho_\nu = N_{\rm eff}\frac{7\pi^2}{120}T_\nu^4,
\end{equation}
where $T_\nu$ is the temperature of neutrinos.
How do we measure $N_{\rm eff}$ from CMB?

The way that we use CMB to determine $N_{\rm eff}$ is relatively
simple. The relativistic particles that stream freely influence CMB in two ways:
(1) their energy density changing the matter-radiation equality epoch,
and (2) their anisotropic stress acting as an additional source for the
gravitational potential via Einstein's equations.
Incidentally, the relativistic particles that {\it do not} stream
freely, but interact with matter frequently, do not have a significant
anisotropic stress because they isotropize themselves via interactions
with matter; thus, anisotropic stress of photons before the decoupling
epoch was very small. Neutrinos, on the other hand, decoupled from
matter much earlier ($\sim 2$~MeV), and thus their anisotropic stress
was significant at the decoupling epoch.

The amount of the early ISW effect changes as the equality redshift
changes. The earlier the equality epoch is, the more the ISW effect
CMB photons receive. This effect can be measured via the height of the
third acoustic peak relative to the first peak. Therefore, the equality
redshift, $z_{\rm eq}$, is one of the fundamental observables that one
can extract from the CMB power spectrum. 

One usually uses $z_{\rm eq}$ to determine $\Omega_mh^2$ from the CMB
power spectrum, without noticing that it is actually $z_{\rm eq}$ that
they are measuring. However, the conversion from $z_{\rm eq}$ to
$\Omega_mh^2$ is automatic only when one knows the radiation content of
the universe exactly -- in other words, when one knows $N_{\rm eff}$
exactly: 
\begin{equation}
 1+z_{\rm eq} = \frac{\Omega_m}{\Omega_{r}}
=\frac{\Omega_mh^2}{\Omega_\gamma h^2}\frac1{1+0.2271N_{\rm eff}},
\label{eq:zeq}
\end{equation}
where $\Omega_\gamma h^2=2.469\times 10^{-5}$ is the present-day photon
energy density parameter for $T_{\rm cmb}=2.725$~K. Here, we have used
the standard relation between the photon temperature and neutrino
temperature, $T_\nu=(4/11)^{1/3}T_\gamma$, derived from the entropy
conservation across the electron-positron annihilation \citep[see,
e.g.,][]{weinberg:GAC,kolb/turner:TEU}.

However, if we do not know $N_{\rm eff}$ precisely, it is not possible to
use $z_{\rm eq}$ to measure $\Omega_mh^2$. In fact, we lose our ability
to measure $\Omega_mh^2$ from CMB almost completely if we do not know
$N_{\rm eff}$. Likewise, if we do not know $\Omega_mh^2$ precisely, it is
not possible to use $z_{\rm eq}$ to measure $N_{\rm eff}$. As a result,
$N_{\rm eff}$ and $\Omega_mh^2$ are linearly correlated (degenerate),
with the width of the degeneracy line given by the uncertainty in our
determination of $z_{\rm eq}$.

The distance information from BAO and SN provides us with an independent
constraint on $\Omega_mh^2$, which helps to reduce the degeneracy
between $z_{\rm eq}$ and  $\Omega_mh^2$.

The anisotropic stress of neutrinos also leaves distinct signatures in
the CMB power spectrum, which is not degenerate with  $\Omega_mh^2$
\citep{hu/etal:1995,bashinsky/seljak:2004}. \citet{trotta/melchiorri:2005}
\citep[see also][]{melchiorri/serra:2006} have reported on evidence for the
neutrino anisotropic stress at slightly more than 95\% CL. They have 
parametrized the anisotropic stress by the viscosity parameter, $c_{\rm
vis}^2$ \citep{hu:1998}, and found $c_{\rm vis}^2>0.12$ (95\%
CL). However, they had to combine the \WMAP\ 1-year data with the SDSS
data to see the evidence for non-zero $c_{\rm vis}^2$.

In \citet{dunkley/etal:prep} we report on the lower limit to $N_{\rm eff}$
solely from the \WMAP\ 5-year data. In this paper we shall combine the
\WMAP\ data with the distance information from BAO and SN as well as
Hubble's constant from HST to find the best-fitting value of $N_{\rm eff}$.

\subsubsection{Results}
\label{sec:neff_results}
Figure~\ref{fig:neff} shows our constraint on $N_{\rm eff}$. The
contours in the left panel lie on the expected linear correlation
between $\Omega_mh^2$ and $N_{\rm eff}$ given by
\begin{equation}
 N_{\rm eff} = 3.04 + 
7.44 \left(\frac{\Omega_mh^2}{0.1308}\frac{3139}{1+z_{\rm eq}}-1\right),
\label{eq:neff}
\end{equation}
which follows from equation~(\ref{eq:zeq}). 
(Here, $\Omega_mh^2=0.1308$ and $z_{\rm eq}=3138$ are the maximum
likelihood values from the simplest $\Lambda$CDM model.)
The width of the degeneracy
line is given by the accuracy of our determination of $z_{\rm eq}$,
which is given by \ensuremath{z_{\rm eq} = 3141^{+ 154}_{- 157}}
(WMAP-only) for this model. Note that the mean value of $z_{\rm eq}$ for the
simplest $\Lambda$CDM model with $N_{\rm eff}=3.04$ is 
\ensuremath{z_{\rm eq} = 3176^{+ 151}_{- 150}}, which is close. 
This confirms that $z_{\rm eq}$ is one of the fundamental observables,
and $N_{\rm eff}$ is merely a secondary parameter that can be derived from
$z_{\rm eq}$. The middle panel of Fig.~\ref{fig:neff} shows this
clearly: $z_{\rm eq}$ is determined independently of $N_{\rm eff}$. For
each value of $N_{\rm eff}$ along a constant $z_{\rm eq}$ line, there is
a corresponding $\Omega_mh^2$ that gives the same value of $z_{\rm eq}$
along the line. 

However, the contours do not extend all the way down to $N_{\rm eff}=0$,
although equation~(\ref{eq:neff}) predicts that $N_{\rm eff}$ should
go to zero when $\Omega_mh^2$ is sufficiently small. This indicates that
we are seeing the effect of the neutrino anisotropic stress at a high
significance. While we need to repeat the analysis of
\citet{trotta/melchiorri:2005} in order to prove that our finding of
$N_{\rm eff}>0$ comes from the neutrino anisotropic stress, we believe
that there is a strong evidence that we see non-zero $N_{\rm eff}$
via the effect of neutrino anisotropic stress, rather than via $z_{\rm eq}$.

While the \WMAP\ data alone can give a lower limit on $N_{\rm eff}$ 
\citep{dunkley/etal:prep}, they cannot give an upper limit owing to the strong
degeneracy with $\Omega_mh^2$. Therefore, we use the BAO, SN, and HST
data to break the degeneracy. We find
\ensuremath{N_{\rm eff} = 4.4\pm 1.5}
(68\%) from WMAP+BAO+SN+HST, which is fully consistent with the standard
value, 3.04 (see the right panel of Fig.~\ref{fig:neff}).

\section{Conclusion}
\label{sec:conclusion}
\begin{figure*}[ht]
\centering \noindent 
\includegraphics[width=18cm]{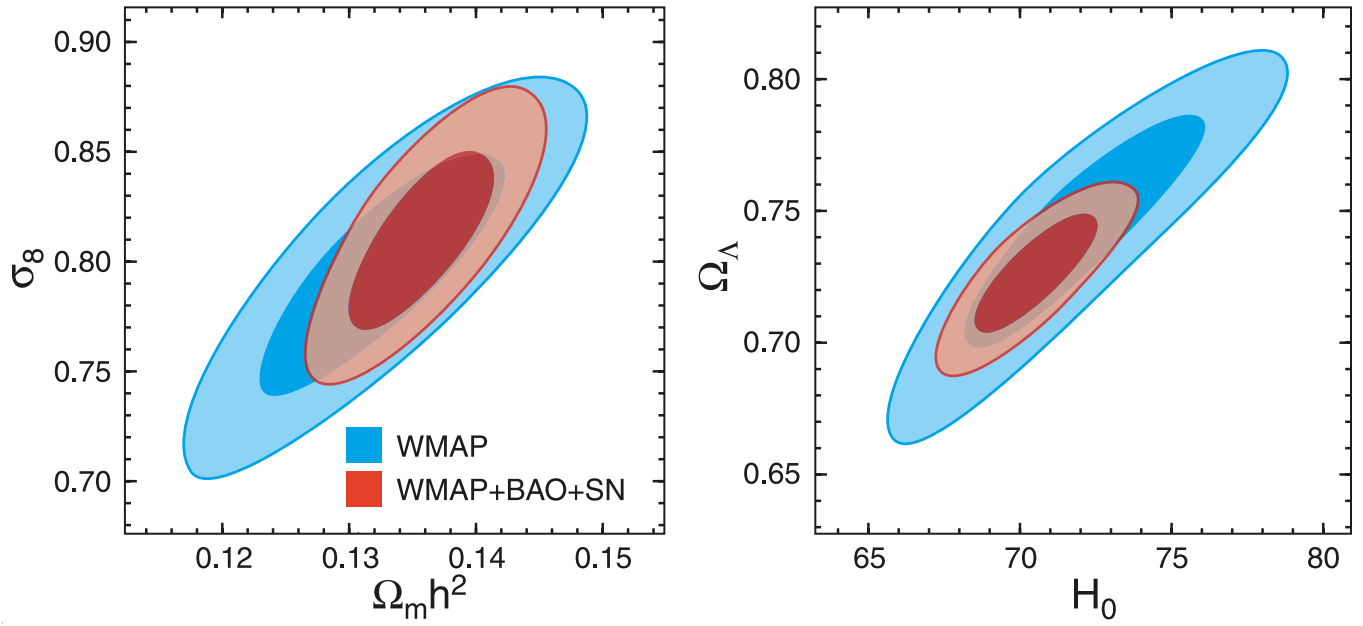}
\caption{%
 Four representative cosmological parameters that have improved
 significantly by adding the BAO and SN data. (See also
 Table~\ref{tab:summary}.) The contours show the 68\% and 95\% CL.
 The WMAP-only constraint is shown in blue, while WMAP+BAO+SN in red.
 ({\it Left}) The distance information from BAO and SN provides a better
 determination of $\Omega_mh^2$, which results in 40\% better
 determination of $\sigma_8$. ({\it Right}) The BAO data, being the
 absolute distance indicator, provides a better determination of $H_0$,
 which results in a factor of 2  better determination of
 $\Omega_\Lambda$, $\Omega_b$, and $\Omega_c$. 
}
\label{fig:lcdm}
\end{figure*}

With 5 years of integration, the \map\ temperature and polarization
data have improved significantly. An improved determination of the third
acoustic peak has enabled us to reduce the uncertainty in the amplitude
of matter fluctuation, parametrized by $\sigma_8$, by a factor of 1.4
from the \map\ 3-year result.
The E-mode polarization is now detected at 5 standard
deviations \citep[c.f., 3.0 standard deviations for the 3-year
data; ][]{page/etal:2007}, which rules out an instantaneous reionization
at $z_{\rm reion}=6$ at the 3.5$\sigma$ level.
Overall, the \map\ 5-year data continue to 
support the simplest, 6-parameter $\Lambda$CDM model \citep{dunkley/etal:prep}.

In this paper, we have explored our ability to limit deviations from the
simplest picture, namely, non-Gaussianity, non-adiabatic fluctuations, 
non-zero gravitational waves, non-power-law spectrum, non-zero curvature,
dynamical dark energy, parity-violating interactions, non-zero neutrino
mass, and non-standard number 
of neutrino species. Detection of any of these items will lead us
immediately to the new era in cosmology, and a better understanding of
the physics of our universe.

From these studies, we conclude that we have not detected any convincing
deviations from the simplest 6-parameter $\Lambda$CDM model at the level
greater than 99\% CL. By combining \WMAP\ data with the distance
information from BAO and SN, we have improved the 
accuracy of the derived cosmological parameters.
As the distance information provides strong constraints on the matter
density (both BAO and SN) and Hubble's constant (BAO), 
the uncertainties in $\Omega_mh^2$ and $H_0$ have been reduced by a factor
of 1.7 and 2, respectively, from the WMAP-only limits. 
The better determination of $H_0$ reduces
the uncertainty in $\Omega_\Lambda$ (as well as $\Omega_b$ and
$\Omega_c$) by a factor of 2, and the better determination of
$\Omega_mh^2$ reduces the uncertainty in $\sigma_8$ by a factor of
 1.4. These results are presented visually in Fig.~\ref{fig:lcdm}.
See also Table~\ref{tab:summary} for the summary of
the cosmological parameters of the $\Lambda$CDM model.
The addition of BAO and SN does not improve determinations of 
$\Omega_bh^2$ or $\tau$ as expected. Since $n_s$ is mainly degenerate
with $\Omega_bh^2$ and $\tau$, with the former being more degenerate, 
the addition of BAO and SN does not improve our determination of $n_s$,
when we consider the simplest 6-parameter $\Lambda$CDM model.

To find the limits on various deviations from the simplest model, we have
explored the parameter space by combining the \map\ 5-year data with the
distance measurements from the Baryon Acoustic Oscillations (BAO) and
Type Ia supernova (SN) observations. 
Here, we summarize significant findings from our analysis
(see also Table~\ref{tab:deviation}):
\begin{itemize}
 \item {\it Gravitational waves and primordial power spectrum}: Improved
       simultaneous constraint on the 
       amplitude of primordial gravitational waves and the shape of the
       primordial power spectrum (from WMAP+BAO+SN).
       In terms of the tensor-to-scalar
       ratio, $r$, we have found 
       \ensuremath{r < 0.22\ \mbox{(95\% CL)}},
       which is the tightest bound to date. A
       blue primordial spectrum index, $n_s>1$, now begins to be
       disfavoured even in the presence of gravitational waves.
       We find no evidence for the running index, $dn_s/d\ln
       k$. The parameter space allowed for inflation models has shrunk
       significantly since the 3-year data release
       (\S~\ref{sec:summary_inflation}), most notably for models that produce
       significant gravitational waves, such as chaotic or power-law
       inflation models, and models that produce $n_s>1$, such as hybrid
       inflation models.
 \item {\it Dark energy and curvature}: Improved simultaneous constraint on
       the dark energy equation of state, $w$, and the spatial curvature of
       the universe, $\Omega_k$ (from WMAP+BAO+SN). We find
       \ensuremath{-0.0179<\Omega_k<0.0081\ \mbox{(95\% CL)}}
       and
       \ensuremath{-0.14<1+w<0.12\ \mbox{(95\% CL)}}.
       The curvature radius of the universe should
       be greater than $R_{\rm curv}>22$ and $33~h^{-1}$Gpc for positive
       and negative curvature, respectively. The combination of \map,
       BAO, and SN is particularly powerful for constraining $w$ and
       $\Omega_k$ simultaneously. 
 \item {\it Time-dependent equation of state}: Using the \WMAP\ distance
       priors ($l_A$, $R$, $z_*$) combined with the BAO and SN distance
       data, we have constrained time-dependent $w$. The present-day
       value of $w$, $w_0$, is constrained as $-0.33<1+w_0<0.21$ (95\%
       CL), for a flat universe. 
 \item {\it Non-Gaussianity}: Improved constraints on primordial non-Gaussianity
       parameters, $-9<\fnlKS<111$ and $-151 < \fnleq < 253$ (95\%
       CL), from the \map\ temperature data. The Gaussianity tests show 
       that the primordial fluctuations are Gaussian to the 0.1\% level,
       which provides the strongest evidence for the quantum origin of the
       primordial fluctuations. 
 \item {\it Non-adiabaticity}: Improved constraints on non-adiabatic
       fluctuations. The photon and matter fluctuations are found to
       obey the adiabatic relation to 8.9\% and 2.1\% for axion- and
       curvaton-type non-adiabatic fluctuations, respectively.
 \item {\it Parity violation}: \map's limits on the TB and EB
       correlations indicate that parity-violating interactions that
       couple to photons could not have rotated the polarization
       angle by more than $-5.9^\circ<\Delta\alpha<2.4^\circ$ between
       the decoupling epoch and present epoch. 
 \item {\it Neutrino mass}: With the \map\ data combined with the
       distance information from BAO and SN, we find a limit on the
       neutrino mass,
       \ensuremath{\sum m_\nu < 0.67\ \mbox{eV}\ \mbox{(95\% CL)}}, 
       which is better than the WMAP-only limit by a factor of 2, owing
       to an additional constraint on $H_0$ provided by BAO. The limit does not
       get worse very much even when $w$ is allowed to vary, as the SN
       data reduce correlation between $H_0$ and $w$ effectively. Since we
       rely only on the CMB data and distance information, our limit is
       not sensitive to our understanding of non-linear
       astrophysical effects in the large-scale structure data.
 \item {\it Number of neutrino species}:
       With the \WMAP\ data alone we find evidence for non-zero $N_{\rm eff}$
       \citep{dunkley/etal:prep}, which is likely coming from our
       measurement of the effect of neutrino anisotropic stress on the CMB
       power spectrum. With the BAO, SN, and HST added, we break the
       degeneracy between $\Omega_mh^2$ and $N_{\rm eff}$, and find 
       \ensuremath{N_{\rm eff} = 4.4\pm 1.5}
       (68\% CL) and 
       \ensuremath{1.8<N_{\rm eff}<7.6\ \mbox{(95\% CL)}},
       which are consistent with the standard value, $N_{\rm
       eff}=3.04$; thus, we do not find any evidence for the extra
       relativistic species in the universe.
\end{itemize}

The limits that we have obtained from our analysis in this paper
are already quite stringent; however, we emphasize that 
they should still be taken as a proto-type of what we can 
achieve in the future, including more integration of the \map\
observations. 

A smaller noise level in the temperature data will reduce
the uncertainty in non-Gaussianity parameters. 
An improved determination of the TE spectrum increases our sensitivity to
non-adiabatic fluctuations as well as to the primordial gravitational
waves. 
The E-mode polarization will be more
dominated by the signal, to the point where we begin to constrain the
detailed history of the reionization of the universe beyond a simple
parametrization. Our limit on the B-mode polarization continues to
improve, which will provide us with a better understanding of the
polarized foreground. The improved TB and EB correlations will provide better
limits on the cosmological birefringence.

While we have
chosen to combine the \WMAP\ data mainly with the distance indicators, 
one should be able to put even more
stringent limits on important parameters such as $r$, $n_s$, $dn_s/d\ln
k$, $w(z)$, $m_\nu$, and $N_{\rm eff}$, by including the other
data sets that are sensitive to the {\it amplitude} of density fluctuations,
such as the amplitude of  
the galaxy power spectrum, Ly$\alpha$
forest, weak lensing, and cluster 
abundance. With the Ly$\alpha$ forest data from
\citet{seljak/slosar/mcdonald:2006}, for example, the limit  on 
the running index improves from  
\ensuremath{-0.068<dn_s/d\ln{k}<0.012\ \mbox{(95\% CL)}}
 to $-0.034 < dn_s/d\ln k < 0.011$ (95\% CL) for $r=0$, and 
\ensuremath{-0.11484<dn_s/d\ln{k}<-0.00079\ \mbox{(95\% CL)}}
 to $-0.0411 < dn_s/d\ln k < 0.0067$ (95\% CL) for $r\ne 0$.
Better understanding of systematic errors in these methods will be
crucial in improving our understanding of the physics of our universe.
Hints for new physics may well be hidden in the
deviations from the simplest 6-parameter $\Lambda$CDM model. 

\acknowledgements
The \map\ mission is made possible by the support of the Science Mission
Directorate Office at NASA Headquarters.  This research was additionally
supported by NASA grants NNG05GE76G, NNX07AL75G S01, LTSA03-000-0090,
ATPNNG04GK55G, and ADP03-0000-092.  
EK acknowledges support from an Alfred P. Sloan Research Fellowship.
We wish to
thank Mike Greason and Nils Odegard for their help on the analysis of
the \WMAP\ data, and Britt Griswold for the artwork. We thank
Uros Seljak, Anze Slosar, and Patrick McDonald for providing 
 the Ly$\alpha$ forest likelihood code, Will Percival for
 useful information on the implementation of the BAO data, Eric
 Hivon for providing a new HEALPix routine to remove the monopole and
 dipole from the  weighted temperature maps, and Kendrick Smith for his
 help on implementing the bispectrum optimization algorithm.
We thank Alex Conley for valuable comments on our original treatment of
the Type Ia supernova data, and Marek Kowalski, Julien Guy, and Anze Slosar
for making the union supernova data available to us. 
Computations for the analysis of non-Gaussianity in \S~\ref{sec:NG} were
carried out by the Terascale Infrastructure for Groundbreaking
Research in Engineering and Science (TIGRESS) at 
the Princeton Institute for Computational Science and Engineering
(PICSciE). 
This research has made use of NASA's Astrophysics Data System
Bibliographic Services.  We acknowledge use of the HEALPix
\citep{gorski/etal:2005}, CAMB \citep{lewis/challinor/lasenby:2000}, and
CMBFAST \citep{seljak/zaldarriaga:1996} packages. 

\appendix
\section{Fast cubic estimators}
\label{sec:estimators}
We use the following estimators for $\fnlKS$, $\fnleq$, and $b_{\rm
src}$ (the amplitude of the point-source bispectrum):
\begin{eqnarray}
\label{eq:offdiagonal1}
 \fnlKS = (F^{-1})_{11}S_1 + (F^{-1})_{12}S_2 + (F^{-1})_{13}S_3,\\
\label{eq:offdiagonal2}
 \fnleq = (F^{-1})_{21}S_1 + (F^{-1})_{22}S_2 + (F^{-1})_{23}S_3,\\
\label{eq:offdiagonal3}
 \bsrc = (F^{-1})_{31}S_1 + (F^{-1})_{32}S_2 + (F^{-1})_{33}S_3,
\end{eqnarray}
where $F_{ij}$ is the Fisher matrix given by
\begin{equation}
 F_{ij} \equiv \sum_{2\le l_1\le l_2\le l_3}
\frac{B^{(i)}_{l_1l_2l_3}B^{(j)}_{l_1l_2l_3}}{\tilde{C}_{l_1}\tilde{C}_{l_2}\tilde{C}_{l_3}}.
\end{equation}
Here, $B_{l_1l_2l_3}^{(i)}$ are theoretically calculated angular
bispectra (given below), where $i=1$ is used for $\fnlKS$, $i=2$ for
$\fnleq$, and $i=3$ for 
$\bsrc$. The denominator of $F_{ij}$ contains the total
power spectrum including the CMB signal ($C_l^{cmb}$) and noise ($N_l$),
$\tilde{C}_l\equiv C_l^{cmb}b_l^2+N_l$, and $b_l$ is the beam transfer
function given in \citet{hill/etal:prep}. 

While this formula allows one to estimate  $\fnlKS$, $\fnleq$, and $b_{\rm
src}$ simultaneously, we find that a simultaneous estimation does not 
change the results significantly. Therefore, we  use
\begin{equation}
 \fnlKS = S_1/F_{11},\qquad 
 \fnleq = S_2/F_{22},\qquad 
 \bsrc =  S_3/F_{33},
\end{equation}
to simplify the analysis, as well as to make the comparison with the
previous work easier. 
However, we do take into account a potential leakage of the point
sources into $\fnlKS$ and $\fnleq$ by using the Monte Carlo simulation
of point sources, as described later. These Monte-Carlo estimates of the
bias in $\fnlKS$ and $\fnleq$ due to the source contamination agree
roughly with contributions from the off-diagonal terms in
Eq.~(\ref{eq:offdiagonal1}) and (\ref{eq:offdiagonal2}).

Assuming white noise, which is a good approximation at high multipoles
in which noise becomes important, one can compute the noise power
spectrum analytically as
\begin{equation}
 N_l = \Omega_{\rm pix}\int \frac{d^2\hat{\mathbf n}}{4\pi f_{\rm
  sky}}\frac{\sigma_0^2M(\hat{\mathbf 
  n})}{N_{\rm obs}(\hat{\mathbf n})},
\end{equation}
where $\Omega_{\rm pix}\equiv 4\pi/N_{\rm pix}$ is the solid
angle per pixel, 
$M(\hat{\mathbf n})$ is the {\it KQ75} mask, $f_{\rm sky}=0.718$
is the fraction of sky retained by the {\it KQ75} mask, $\sigma_0$ is the rms
noise per observation, and $N_{\rm obs}(\hat{\mathbf n})$ is the number
of observations per pixel.

The angular bispectra are given by 
\begin{eqnarray}
 B^{(1)}_{l_1l_2l_3} &=& 2 I_{l_1l_2l_3}\int_0^\infty r^2 dr
  \left[\alpha_{l_1}(r)\beta_{l_2}(r)\beta_{l_3}(r) +
   (\mbox{cyc.})\right],\\
 B^{(2)}_{l_1l_2l_3} &=& -3B^{(1)}_{l_1l_2l_3} +
6 I_{l_1l_2l_3}\int_0^\infty r^2 dr\left\{
\left[\beta_{l_1}(r)\gamma_{l_2}(r)\delta_{l_3}(r)
+ (\mbox{cyc.})\right] 
-2\delta_{l_1}(r)\delta_{l_2}(r)\delta_{l_3}(r)
\right\},\\
B^{(3)}_{l_1l_2l_3} &=& I_{l_1l_2l_3},
\end{eqnarray}
where
\begin{equation}
 I_{l_1l_2l_3}\equiv \sqrt{\frac{(2l_1+1)(2l_2+1)(2l_3+1)}{4\pi}}
\left(\begin{array}{ccc}l_1&l_1&l_3\\0&0&0\end{array}\right).
\end{equation}

Various functions in $B^{(i)}_{l_1l_2l_3}$ are given by
\begin{eqnarray}
 \alpha_l(r)&\equiv&\frac{2}{\pi}\int k^2dk g_{Tl}(k)j_l(kr),\\
 \beta_l(r)&\equiv&\frac{2}{\pi}\int k^2dk P_\Phi(k)g_{Tl}(k)j_l(kr),\\
 \gamma_l(r)&\equiv&\frac{2}{\pi}\int k^2dk P^{1/3}_\Phi(k)g_{Tl}(k)j_l(kr),\\
 \delta_l(r)&\equiv&\frac{2}{\pi}\int k^2dk P^{2/3}_\Phi(k)g_{Tl}(k)j_l(kr).
\end{eqnarray}
Here, $P_\Phi(k)\propto k^{n_s-4}$ is the primordial 
power spectrum of Bardeen's curvature perturbations, and
$g_{Tl}(k)$ is the radiation transfer function that gives the angular
power spectrum as $C_l=(2/\pi)\int k^2 dk P_\Phi(k)g_{Tl}^2(k)$.

The skewness parameters, $S_i$, are given by
\citep{komatsu/spergel/wandelt:2005,babich:2005,creminelli/etal:2006,yadav/etal:2008}
\begin{eqnarray}
\nonumber
 S_1&\equiv& 4\pi\int r^2dr\int \frac{d^2\hat{\mathbf
  n}}{w_3}\left[A(\hat{\mathbf n},r)B^2(\hat{\mathbf n},r)\right.\\
\label{eq:s1}
& &\left.-2B(\hat{\mathbf
  n},r)\langle A_{\rm sim}(\hat{\mathbf n},r)B_{\rm sim}(\hat{\mathbf n},r)\rangle_{\rm MC}
-A(\hat{\mathbf n},r)\langle B_{\rm sim}^2(\hat{\mathbf n},r)\rangle_{\rm
MC}\right],\\
\nonumber
  S_2&\equiv& -3S_1 + 24\pi\int r^2dr\int 
\frac{d^2\hat{\mathbf
  n}}{w_3}\left\{
\left[B(\hat{\mathbf n},r)C(\hat{\mathbf n},r)D(\hat{\mathbf n},r)
-B(\hat{\mathbf n},r)\langle C_{\rm sim}(\hat{\mathbf n},r)D_{\rm sim}(\hat{\mathbf
n})\rangle_{\rm MC}\right.\right.\\
\nonumber
& &\left.-C(\hat{\mathbf n},r)\langle B_{\rm sim}(\hat{\mathbf n},r)D_{\rm
	  sim}(\hat{\mathbf 
n})\rangle_{\rm MC}
-D(\hat{\mathbf n},r)\langle B_{\rm sim}(\hat{\mathbf n},r)C_{\rm sim}(\hat{\mathbf
n},r)\rangle_{\rm MC}
\right] \\
\label{eq:s2}
& &\left.-\frac13\left[D^3(\hat{\mathbf n},r)-3D(\hat{\mathbf n},r)\langle
		  D^2_{\rm sim}(\hat{\mathbf n},r)\rangle_{\rm
		  MC}\right]
\right\}\\
S_3&\equiv& \frac{2\pi}{3}\int\frac{d^2\hat{\mathbf n}}{w_3}
\left[E^3(\hat{\mathbf n})-3E(\hat{\mathbf n})\langle
		  E^2_{\rm sim}(\hat{\mathbf n})\rangle_{\rm MC}\right],
\end{eqnarray}
where $w_3$ is the sum of the weighting function cubed:
\begin{equation}
 w_3\equiv \int d^2\hat{\mathbf n}~W^3(\hat{\mathbf n}).
\end{equation}
For a uniform weighting, the weighting function is simply given by 
the {\it KQ75} mask, i.e., $W(\hat{\mathbf n})=M(\hat{\mathbf n})$,
which gives $w_3=4\pi f_{\rm sky}$.
For our measurements of $\fnlKS$, $\fnleq$, and $\bsrc$, we shall use a
``combination signal-plus-noise weight,'' given by
\begin{equation}
 W({\mathbf n}) = \frac{M(\hat{\mathbf n})}{\sigma_{\rm
  cmb}^2+\sigma_0^2/N_{\rm obs}(\hat{\mathbf n})},
\label{eq:weightfunction}
\end{equation}
where $\sigma^2_{\rm cmb}\equiv (1/4\pi)\sum_l (2l+1)C_l^{\rm cmb}b_l^2$
is the CMB signal variance, $\sigma_0$ is the rms noise per observation,
and $N_{\rm obs}(\hat{\mathbf n})$ is the number of observations per pixel.
This combination weighting yields a nearly optimal performance for
$\bsrc$ and $\fnleq$, whereas it results in a minor improvement in 
$\fnlKS$ over the uniform weighting. 
The bracket, $\langle\rangle_{\rm MC}$, denotes the average over Monte Carlo
realizations, and ``sim'' denotes that these are the filtered maps 
of the Monte Carlo realizations.

The filtered temperature maps, 
$A$, $B$, $C$, $D$, and $E$, are given by
\begin{eqnarray}
 A(\hat{\mathbf n},r)&\equiv&
  \sum_{l=2}^{l_{\rm max}}\sum_{m=-l}^l\alpha_l(r)
\frac{b_l}{\tilde{C}_l}a_{lm}Y_{lm}(\hat{\mathbf n}),\\
 B(\hat{\mathbf n},r)&\equiv&
  \sum_{l=2}^{l_{\rm max}}\sum_{m=-l}^l\beta_l(r)
\frac{b_l}{\tilde{C}_l}a_{lm}Y_{lm}(\hat{\mathbf
  n}),\\
 C(\hat{\mathbf n},r)&\equiv&
  \sum_{l=2}^{l_{\rm max}}\sum_{m=-l}^l\gamma_l(r)
\frac{b_l}{\tilde{C}_l}a_{lm}Y_{lm}(\hat{\mathbf
  n}),\\
 D(\hat{\mathbf n},r)&\equiv&
  \sum_{l=2}^{l_{\rm max}}\sum_{m=-l}^l\delta_l(r)
\frac{b_l}{\tilde{C}_l}a_{lm}Y_{lm}(\hat{\mathbf
  n}),\\
 E(\hat{\mathbf n})&\equiv&
  \sum_{l=2}^{l_{\rm max}}\sum_{m=-l}^l\frac{b_l}{\tilde{C}_l}a_{lm}Y_{lm}(\hat{\mathbf
  n}),
\end{eqnarray}
respectively. Here, $l_{\rm max}$ is the maximum multipole that we use
in the analysis. We vary $l_{\rm max}$ to see how much the results depend on
$l_{\rm max}$. 

Eq.~(\ref{eq:s1}) and (\ref{eq:s2}) involve the integrals over the
conformal distances, $r$. We evaluate these integrals as
\begin{equation}
 \int r^2 dr [\dots](r) =
\sum_i (w_i)_{\rm opt} r_i^2 \Delta r_i [\dots](r_i).
\end{equation}
We use the bispectrum optimization algorithm described in
\citet{smith/zaldarriaga:prep} 
to compute the optimal weights, $(w_i)_{\rm opt}$, and decide on which
quadrature points, $r_i$, to keep. We choose the number of quadrature
points such that the bispectrum computed in this way agrees with that
from more dense sampling in $r$ to $10^{-5}$,
which typically gives $\sim 5$ quadrature points for $\fnlKS$ and $\sim
15$ points for $\fnleq$.\footnote{Note that
\citet{smith/zaldarriaga:prep} use $10^{-6}$ as a criterion, which gives
more quadrature points to evaluate. We find that $10^{-5}$ is sufficient
for the size of statistical and systematic errors in the current
measurements.}  

The measurement of these estimators proceeds as follows:
\begin{enumerate}
\item Generate the simulated realizations of CMB signal maps, $T_{S}$,
      from the  
      input signal power spectrum, $C_l^{\rm cmb}$, and the beam
      transfer function, $b_l$. We have generated 300 realizations for
      the analysis given in this paper.
\item Add random noise, $T_{N}$, using the rms noise per pixel given by
      $\sigma_0/\sqrt{N_{\rm  obs}(\hat{\mathbf n})}$. 
\item Add point sources. We use a simplified treatment for the source
      simulation, 
\begin{equation}
\frac{T_{\rm src}(\hat{\mathbf n})}{2.725~{\rm K}}= 
\left[\frac{\sinh^2(x/2)}{x^4}\frac{F_{\rm src}/\Omega_{pix}}{67.55~{\rm
 MJy}}\right]\epsilon,
\end{equation}
where $\Omega_{pix}$ is the solid angle of pixel, $x=h\nu/(k_BT_{\rm
      cmb})=56.80~{\rm GHz}$ (for $T_{\rm cmb}=2.725$~K), 
      $\epsilon$ is a Poisson random variable with the mean of
      $\langle\epsilon\rangle=n_{\rm src}\Omega_{pix}$, and $n_{\rm
      src}$ is the average number of sources per steradians.  This
      simplified model assumes that there is only one population of
      sources with a fixed flux, $F_{\rm src}$, and each source's flux
      is independent of frequency. We choose $n_{\rm src}=0.85~{\rm
      sr}^{-1}$ and $F_{\rm src}=0.5~{\rm Jy}$, which yields the source
      power spectrum in Q band of $C_{ps}=8.7\times 10^{-3}~\mu{\rm
      K}^2~{\rm sr}$ and the source bispectrum in Q band of
      $\bsrc=8.7\times 10^{-5}~\mu{\rm K}^3~{\rm sr}^2$, which roughly
      reproduce the 
      measured values. 
      However, this model does not reproduce the source
      counts very well. (The source density of $n_{\rm src}=0.85~{\rm
      sr}^{-1}$ at 0.5~Jy is too low.) The main purpose of this
      phenomenological model is to reproduce 
      the power spectrum and bispectrum -- 
      we include point sources in the simulations, in
      order to take into account the potential effects of the unresolved
      sources on primordial non-Gaussianity, $\fnlKS$ and $\fnleq$.
\item Coadd them to create the simulated temperature maps,
      $T(\hat{\mathbf n})=T_{S}(\hat{\mathbf n})+T_{N}(\hat{\mathbf
      n})+T_{\rm src}(\hat{\mathbf n})$. 
 \item Mask and weight the temperature maps, $T(\hat{\mathbf
       n})\rightarrow \tilde{T}(\hat{\mathbf
       n})=W(\hat{\mathbf n})T(\hat{\mathbf n})$, where $W(\hat{\mathbf
       n})$ is given by Eq.~(\ref{eq:weightfunction}).
 \item Remove the monopole and dipole from $\tilde{T}(\hat{\mathbf
       n})$.
\item Compute the harmonic coefficients as 
\begin{equation}
 a_{lm} = \int d^2\hat{\mathbf n}~\tilde{T}(\hat{\mathbf n})Y^*_{lm}(\hat{\mathbf n}).
\end{equation}
\item Generate the filtered maps, $A_{\rm sim}$, $B_{\rm sim}$, $C_{\rm
      sim}$, $D_{\rm sim}$, and $E_{\rm sim}$, and compute the
      appropriate Monte-Carlo averages such as $\langle A_{\rm
      sim}B_{\rm sim}\rangle_{\rm
      MC}$, etc. This is the most time consuming part.
\item Compute the filtered maps from the \map\ data. When we coadd the V
      and W band data, we weight them as $T_{\rm V+W}=(T_{\rm
      V}+0.9T_{\rm W})/1.9$. The beam transfer function of the coadded
      map is given by $b_l^{\rm V+W}=(b_l^{\rm V}+0.9b_l^{\rm
      W})/1.9$, and $\sigma_0/N_{\rm obs}$ of the coadded map is given
      by
\begin{equation}
 \frac{(\sigma_0^{\rm V+W})^2}{N_{\rm obs}^{\rm V+W}(\hat{\mathbf n})}
= \frac1{1.9^2}\left[\frac{(\sigma_0^{\rm V})^2}{N_{\rm obs}^{\rm V}(\hat{\mathbf n})}+\frac{(0.9\sigma_0^{\rm W})^2}{N_{\rm obs}^{\rm W}(\hat{\mathbf n})}\right].
\end{equation}
\item Compute the skewness parameters, $S_i$, from the filtered \map\
      data, and obtain $\fnlKS$, $\fnleq$, and $\bsrc$, either jointly or
      separately. 
\item Compute these parameters from the simulated realizations as well,
      and obtain       the uncertainties.
\end{enumerate}
For the computations of $g_{Tl}(k)$ and generation of Monte Carlo
realizations, we have used the maximum-likelihood values of the \map\
3-year data (the power-law $\Lambda$CDM model fit by the \map\ data
alone with the Sunyaev--Zel'dovich effect marginalized):
$\Omega_b=0.0414$, $\Omega_{cdm}=0.1946$, $\Omega_\Lambda=0.7640$, 
$H_0=73.2~{\rm km/s/Mpc}$, $\tau=0.091$, and $n_s=0.954$
\citep{spergel/etal:2007}. These parameters yield the conformal distance
to $t=0$ as $c\tau_0=14.61$~Gpc.

\section{Axion}
\label{sec:axion}
In this Appendix we derive relations between the tensor-to-scalar ratio
$r$, the axion mass density $\Omega_ah^2$, the entropy-to-curvature
perturbation ratio $\alpha$, the phase of the Pecci-Quinn field
$\theta_a$, and the axion decay constant $f_a$.

Let us write the expectation value of the complex Pecci-Quinn field,
$\psi_{\rm PQ}$, as 
\begin{equation}
 \langle\psi_{\rm PQ}\rangle = \frac{f_a}{\sqrt{2}} e^{i\theta_a},
\end{equation}
where $f_a$ is the axion decay constant, and $\theta_a$
is the phase. Quantum fluctuations during inflation generate
fluctuations in the phase, $\delta\theta_a$, as 
\begin{equation}
 \delta\theta_a = \frac{H}{2\pi f_a}.
\end{equation}
As the number density of axions scales as the phase squared, 
$n_a\propto \theta_a^2$, 
the mass density fluctuation is given by 
\begin{equation}
\frac{\delta \rho_a}{\rho_a}=2\frac{\delta\theta_a}{\theta_a}
=\frac{H}{\pi \theta_af_a}. 
\end{equation}
As the energy density of axions was negligible during inflation, 
the axion density perturbation, $\delta\rho_a/\rho_a$, would produce the
isocurvature perturbation. While radiation (including photons) is
generated by decay of inflaton fields, (some of) dark matter is 
in the form of axions whose generation  is independent of photons;
thus, the entropy perturbation between photons and axions would be
generated. We assume that axions were not in thermal equilibrium with
photons in the subsequent evolution of the universe.

The entropy perturbations and curvature perturbations are given,
respectively, by 
\begin{eqnarray}
 \frac{k^3P_{\cal S}(k)}{2\pi^2} =
  \frac{\Omega_a^2}{\Omega_c^2}\frac{H_k^2}{\pi^2\theta_a^2f_a^2},\qquad 
 \frac{k^3P_{\cal R}(k)}{2\pi^2} = \frac{H_k^4}{4\pi^2\dot{\phi}^2_k}
\approx \frac{H_k^2}{8\pi^2M_{\rm pl}^2\epsilon}
\end{eqnarray}
where $\Omega_a\leq \Omega_c$ is the axion mass density, 
$H_k$ is the expansion rate during inflation at which the
wavenumber $k$ went outside of the horizon, $\phi_k$ is the value of 
inflation at the same time, $\epsilon\equiv-\dot{H}/H\approx (M_{\rm
pl}^2/2)(V'/V)^2$ is the usual slow-roll parameter, $V(\phi)$ is the
inflaton potential, and $M_{\rm pl}=1/\sqrt{8\pi G}$ is the reduced
Planck mass. We have used the slow-roll approximation,
$\dot{\phi}\approx -V'/(3H)$ and $H^2\approx V/(3M_{\rm pl}^2)$.

Here, let us comment on our choice of $m=1$, which makes $k^3P_{\cal
S}\propto k^{m-1}$ independent of $k$. Since $k^3P_{\cal S}\propto
H_k^2\propto k^{-2\epsilon}$, where $\epsilon=-\dot{H}/H^2$, 
this choice corresponds to having a very small slow-roll parameter,
$\epsilon\ll 1$. This is consistent with our limit on the curvature
power spectrum, $n_s=1+6\epsilon-4\eta\simeq 1-4\eta<1$, where $\eta$ is
another slow-roll parameter. 
As the current limit is $1-n_s\approx 4\eta\simeq 0.04$, our
approximation, $m=1$, is valid for $\epsilon<0.01$.
It should be straightforward to extend our
analysis to the case in which $m\neq 1$.

By dividing $P_{\cal S}(k)$ by $P_{\cal R}(k)$, we find the
entropy-to-curvature perturbation ratio for axions, $\alpha_0(k)$, as
\begin{equation}
 \frac{\alpha_0(k)}{1-\alpha_0(k)}\equiv \frac{P_{\cal S}(k)}{P_{\cal R}(k)} 
= \frac{\Omega^2_a}{\Omega^2_c}\frac{8\epsilon}{\theta^2_a(f_a/M_{\rm pl})^2}.
\label{eq:isolimit}
\end{equation}
At this point, it is clear that 
one cannot solve this constraint uniquely for any of
$\epsilon$, $f_a$, or $\theta_a$. 

In order to break the degeneracy, we use the axion mass density
\citep[][and references therein]{kawasaki/sekiguchi:prep}
\begin{equation}
 \Omega_ah^2 = 1.0\times 10^{-3}\gamma\theta_a^2\left(\frac{f_a}{10^{10}~{\rm
					 GeV}}\right)^{7/6},
\label{eq:oah2}
\end{equation}
where $\gamma$ is a dilution factor, representing 
the amount by which the axion density could have been diluted by a 
late-time entropy production between the QCD phase transition at $\sim
200$~MeV and the epoch of nucleosynthesis at $\sim 1$~Mpc.

Combining equation~(\ref{eq:isolimit}) and (\ref{eq:oah2}) to eliminate
the phase, $\theta_a$, and using the relation between the
tensor-to-scalar ratio $r$ and the slow-roll parameter $\epsilon$,
$r=16\epsilon$, we find
\begin{equation}
 r = (1.6\times
  10^{-12})\left(\frac{\Omega_ch^2}{\gamma}\right)
\left(\frac{\Omega_c}{\Omega_a}\right)\left(\frac{f_a}{10^{12}~{\rm
					 GeV}}\right)^{5/6}
\frac{\alpha_0}{1-\alpha_0}.
\label{eq:faleft}
\end{equation}
Alternatively, we can eliminate the axion decay constant, $f_a$, to obtain
\begin{equation}
 r = \frac{4.7\times 10^{-12}}{\theta_a^{10/7}}\left(\frac{\Omega_ch^2}{\gamma}\right)^{12/7}
\left(\frac{\Omega_c}{\Omega_a}\right)^{2/7}
\frac{\alpha_{0}}{1-\alpha_{0}}.
\end{equation}
This is equation~(\ref{eq:rlimitfromaxion}).

\section{Equation of State of Dark Energy: A new parametrized form}
\label{sec:eos}
\begin{figure}[ht]
\centering \noindent 
\includegraphics[width=16cm]{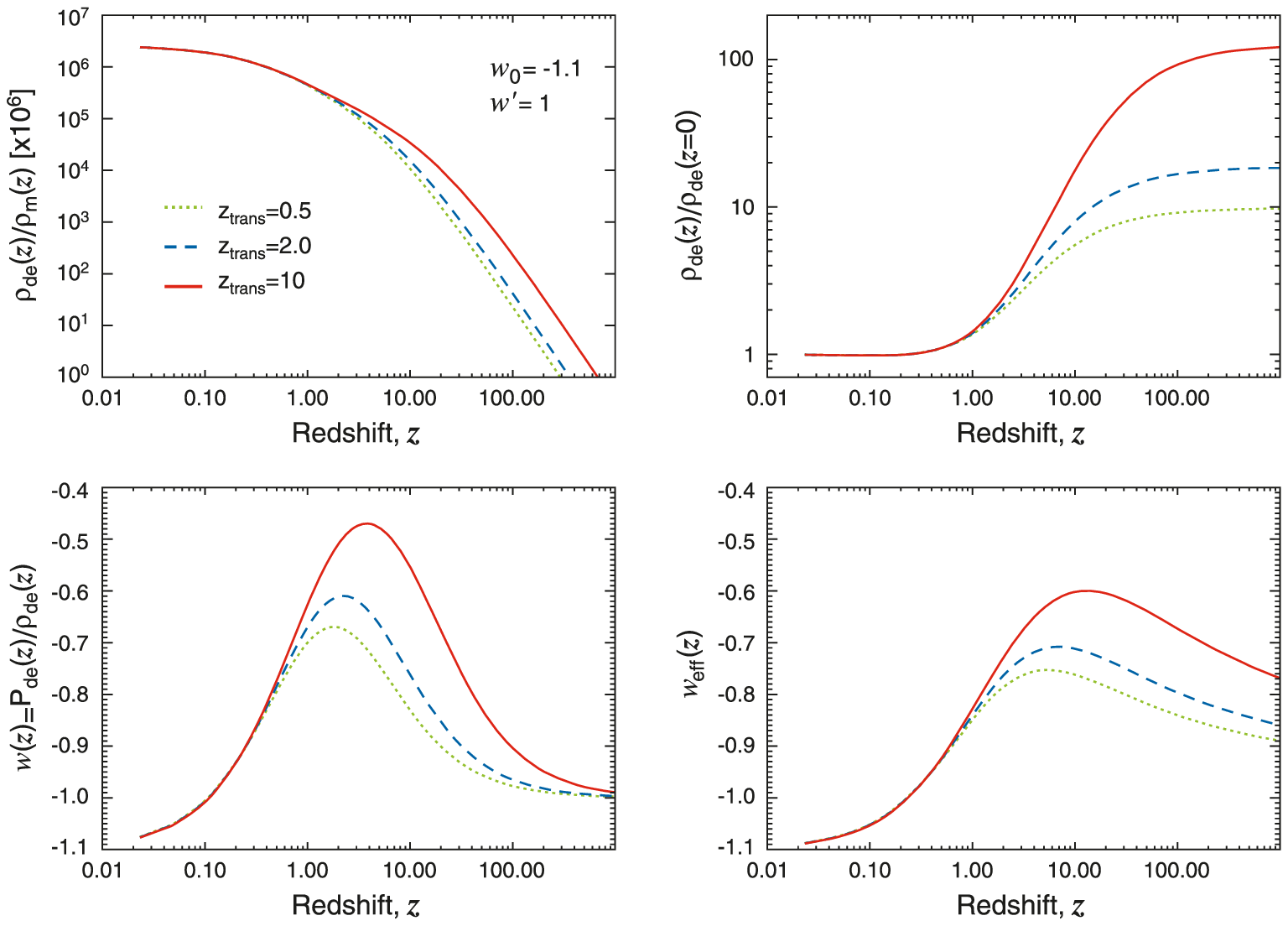}
\caption{%
 Evolution of dark energy for $w_0=-1.1$ and $w'=1$, and various
 transition redshifts, $z_{\rm trans}=0.5$, 2, and 10, above which
 $w(z)$ approaches to $-1$.
 ({\it Top Left}) Evolution of the dark energy to
 matter density ratio as a function of $z$. Note that the vertical axis
 has been multiplied by $10^6$. 
 ({\it Top Right}) Evolution of the dark energy density relative to the
 dark energy density at present. The dark energy density was nearly
 constant at high redshifts above $z_{\rm trans}$; thus, these models
 can describe the ``thawing models,'' \citep{caldwell/linder:2005} in which
 dark energy was nearly  constant at early times, and had become
 dynamical at lower redshifts. 
 ({\it Bottom Left}) Evolution of the equation of state, $w(z)=P_{\rm
 de}(z)/\rho_{\rm de}(z)$. By construction of the model, $w(z)$
 approaches to $-1$ beyond $z_{\rm trans}$. ({\it Bottom Right}) Evolution
 of the effective equation of state, $w_{\rm eff}(z)$, which determines
 the evolution of dark energy density as $\rho_{\rm de}(z)=\rho_{\rm
 de}(0)(1+z)^{3[1+w_{\rm eff}(z)]}$. 
}
\label{fig:rhode}
\end{figure}

In this Appendix, we describe the models of dark energy that we explore
in \S~\ref{sec:wmapprior}. Our goal is to obtain a sensible form of
time-dependent dark energy equation of state, $w(a)$. 
One of the most commonly used form of $w(a)$ is a linear form
\citep{chevallier/polarski:2001,linder:2003}  
\begin{equation}
 w(a) = w_0 + (1-a)w_a,
\label{eq:wlinear}
\end{equation}
where $w_0$ and $w_a$ parametrize the present-day value of $w$ and the
first derivative.
However, this form cannot be adopted as it is when one uses the CMB
data to constrain $w(a)$. Since this form is basically the leading-order term
of a Taylor series expansion, the value of $w(a)$ can become
unreasonably too large or too 
small when extrapolated to the decoupling epoch at $z_*\simeq 1090$ (or
$a_*\simeq 9.17\times 10^{-4}$), and thus one cannot extract meaningful
constraints on the 
quantities such as $w_0$ and $w_a$ that are defined at the {\it present epoch}.

To avoid this problem, yet to keep a close contact with the previous work in the
literature, we shall consider an alternative parametrized form. Our idea
is the following: we wish to keep the form given by
Eq.~(\ref{eq:wlinear}) at low redshifts, lower than some transition
redshift, $z_{\rm trans}$. However, we demand $w(a)$  approach to $-1$
at higher redshifts, $z>z_{\rm trans}$.
This form of $w(a)$ therefore has the following property: at early times, before
the transition redshift, $z_{\rm trans}$, dark energy was just like a
cosmological constant, and thus the dark energy density was nearly 
constant, i.e., $\rho_{\rm de}(z>z_{\rm trans})\approx {\rm
constant}$. Then, dark energy began to become dynamical at $z\sim z_{\rm 
trans}$, with the equation of state given by the conventional linear
form, Eq.~(\ref{eq:wlinear}). 

Some of the properties of our form of $w(a)$ are similar to those of ``thawing
models,'' \citep{caldwell/linder:2005} in which a scalar field was 
moving very slowly initially, giving $w(a)\approx -1$ at early times,
and then began to move faster towards low redshifts, causing $w(a)$ to
deviate more and more from $-1$ at low redshifts. 
Our parametrization can describe more general class of models than 
single scalar field models, as it allows for $w$ to go below $-1$. On the 
other hand, models that are based upon a single scalar field cannot have
$w<-1$ \citep[e.g.,][]{hu:2005}. 
The ``Forever regular'' parametrization explored in
\citet{wang/tegmark:2004} also approaches to a constant density at early
times, if the late-time equation of state is $w<-1$. 
The ``Kink model'' explored in 
\citet{bassett/etal:2002,bassett/corasaniti/kunz:2004,corasaniti/etal:2004}
also extrapolates a constant equation of state at early times to a different constant
equation of state at late times.
Our parametrization is more general than theirs, 
as their form allows only for a constant equation of state at late
times. 

We wish to find a smooth interpolation between $w_{\rm early}=-1$ and
$w_{\rm late}=w_0+(1-a)w_a$. 
We begin by writing 
\begin{equation}
 w(a) = \tilde{w}(a) f(a/a_{\rm trans}) + (-1)\left[1-f(a/a_{\rm
					       trans})\right], 
\end{equation}
where $a_{\rm trans}=1/(1+z_{\rm trans})$, and 
the function $f(x)$ goes to zero for $x\ll 1$, and to unity
for $x\gg 1$. Here, $\tilde{w}(a)$ is the form of $w$ at low redshifts.
Any function that has this property is adequate for $f(x)$. We
choose:
\begin{equation}
 f(x) = \frac12\left[\tanh(\ln x)+1\right] = \frac{x}{x+1},
\end{equation}
which gives the desired form of the equation of state of dark energy,
\begin{equation}
 w(a) = \frac{a\tilde{w}(a)}{a+a_{\rm trans}} - \frac{a_{\rm
  trans}}{a+a_{\rm trans}},
\label{eq:wnewform}
\end{equation}
where
\begin{equation}
 \tilde{w}(a)=\tilde{w}_0 + (1-a)\tilde{w}_a.
\end{equation}
One nice property of this form is that it allows one to obtain a closed,
analytical form of the effective equation of state, $w_{\rm eff}(a)$,
which gives the 
evolution of dark energy density, 
$\rho_{\rm de}(a)=\rho(0)a^{-3[1+w_{\rm
eff}(a)]}$:
\begin{eqnarray}
\nonumber
 w_{\rm eff}(a)&=& \frac1{\ln a}\int_0^{\ln a} d\ln a'~w(a')\\
&=& -1 + \frac{(1-a)\tilde{w}_a}{\ln a} +\frac{1+\tilde{w}_0+(1+a_{\rm
 trans})\tilde{w}_a}{\ln a}\ln\frac{a+a_{\rm trans}}{1+a_{\rm trans}}.
\label{eq:weffnewform}
\end{eqnarray}
This property allows one to compute the expansion rate, $H(a)$
 (Eq.~[\ref{eq:hubble}]), and hence the distance (Eq.~[\ref{eq:da}]),
 easily. 

Finally, we use the present-day value of $w$, $w_0\equiv w(z=0)$, and 
the first derivative, $w'\equiv \left.dw/dz\right|_{z=0}$, as free
parameters. They are related to $\tilde{w}_0$ and $\tilde{w}_a$ as
\begin{eqnarray}
 1+w_0 &=& \frac{1+\tilde{w}_0}{1+a_{\rm trans}},\\
 w' &=& \frac{\tilde{w}_a}{1+a_{\rm trans}} - \frac{a_{\rm
  trans}(1+\tilde{w}_0)}{(1+a_{\rm trans})^2}.
\end{eqnarray}
The inverse relations are
\begin{eqnarray}
 1+\tilde{w}_0 &=& (1+a_{\rm trans})(1+w_0),\\
 \tilde{w}_a &=& (1+a_{\rm trans})w' + a_{\rm trans}(1+w_0).
\end{eqnarray}
In the limit of very early transition, $a_{\rm trans}\ll 1$, one finds
$w_0\approx \tilde{w}_0$ and $w'\approx \tilde{w}_a$, as expected.
This completes the description of our form of $w(a)$.

Figure~\ref{fig:rhode} shows the evolution of dark energy density,
$\rho_{\rm de}(z)=\rho(0)(1+z)^{3[1+w_{\rm
eff}(a)]}$, the equation of state, $w(z)$, and the effective
equation of state, $w_{\rm eff}(z)$, computed from
Eq.~(\ref{eq:wnewform}) and (\ref{eq:weffnewform}).
We choose $w_0=-1.1$ and $w'=1$, which are close to the best-fitting
values that we find in \S~\ref{sec:wmapprior} (see Fig.~\ref{fig:wz}).
We show three curves for the transition redshifts of $z_{\rm trans}=0.5$, 2, and
10. We find that the form of $w(z)$ that we have derived achieves our
goal: $w(z)$ approaches to $-1$ 
and the dark energy density tends to a constant value 
at high redshifts, giving sensible results at the decoupling epoch. 
The dark energy density is totally sub-dominant compared to the matter
density at high redshifts, which is also desirable. 

The constraints that we have obtained for $w_0$ and $w'$ are not sensitive to
the exact values of $z_{\rm trans}$ (see the right panel of
Fig.~\ref{fig:wz}). This is because all of the curves
shown in Fig.~\ref{fig:rhode} are very similar at $z\lesssim 1$, where
the BAO and SN data are currently available.

\section{Comparison of Supernova Compilations and Effects of Systematic Errors}
\label{sec:sn}

\begin{deluxetable*}{llcccc}
\tablecolumns{6}
\small
\tablewidth{0pt}
\tablecaption{%
Comparison of  $\Lambda$CDM parameters from \map+BAO+SN with
various SN compilations
}
\tablehead{\colhead{Class} &
\colhead{Parameter}
&\colhead{Union\footnote{Compilation by \citet{kowalski/etal:prep}
 {\it without} the systematic errors included}}
&\colhead{Union+Sys.Err.\footnote{Compilation by
 \citet{kowalski/etal:prep} {\it with} the systematic errors included}}
&\colhead{Davis\footnote{Compilation by \citet{davis/etal:2007}}}
&\colhead{Alternative\footnote{Compilation used in the original version of this
 paper (version 1 of arXiv:0803.0547)}}
}
\startdata
Primary &
$100\Omega_bh^2$
&\ensuremath{2.267^{+ 0.058}_{- 0.059}} 
&\ensuremath{2.267\pm 0.059} 
&\ensuremath{2.270\pm 0.060} 
&\ensuremath{2.265\pm 0.059} \nl
&
$\Omega_ch^2$
&\ensuremath{0.1131\pm 0.0034} 
&\ensuremath{0.1134^{+ 0.0036}_{- 0.0037}} 
&\ensuremath{0.1121\pm 0.0035}
&\ensuremath{0.1143\pm 0.0034} \nl
&
$\Omega_\Lambda$
&\ensuremath{0.726\pm 0.015} 
&\ensuremath{0.725\pm 0.016} 
&\ensuremath{0.732^{+ 0.014}_{- 0.015}} 
&\ensuremath{0.721\pm 0.015} \nl
&
$n_s$ 
&\ensuremath{0.960\pm 0.013}
&\ensuremath{0.960\pm 0.013}
&\ensuremath{0.962\pm 0.013}
&\ensuremath{0.960^{+ 0.014}_{- 0.013}} \nl
&
$\tau$
&\ensuremath{0.084\pm 0.016}
&\ensuremath{0.085\pm 0.016}
&\ensuremath{0.085\pm 0.016}
&\ensuremath{0.084\pm 0.016} \nl
&
$\Delta^2_{\cal R}(k_0\footnote{$k_0=0.002~{\rm
 Mpc}^{-1}$. $\Delta^2_{\cal R}(k)=k^3P_{\cal R}(k)/(2\pi^2)$ (Eq.~[\ref{eq:pR}])})$
&\ensuremath{(2.445\pm 0.096)\times 10^{-9}}
&\ensuremath{(2.447^{+ 0.096}_{- 0.095})\times 10^{-9}}
&\ensuremath{(2.429^{+ 0.096}_{- 0.095})\times 10^{-9}}
&\ensuremath{(2.457^{+ 0.092}_{- 0.093})\times 10^{-9}} \nl
\hline
Derived &
$\sigma_8$
&\ensuremath{0.812\pm 0.026} 
&\ensuremath{0.813^{+ 0.026}_{- 0.027}} 
&\ensuremath{0.807\pm 0.027} 
&\ensuremath{0.817\pm 0.026} \nl
&
$H_0$
&\ensuremath{70.5\pm 1.3\ \mbox{km/s/Mpc}}
&\ensuremath{70.4\pm 1.4\ \mbox{km/s/Mpc}}
&\ensuremath{70.9\pm 1.3\ \mbox{km/s/Mpc}}
&\ensuremath{70.1\pm 1.3\ \mbox{km/s/Mpc}} \nl
&
$\Omega_b$
&\ensuremath{0.0456\pm 0.0015}
&\ensuremath{0.0458\pm 0.0016}
&\ensuremath{0.0451^{+ 0.0016}_{- 0.0015}}
&\ensuremath{0.0462\pm 0.0015} \nl
&
$\Omega_c$
&\ensuremath{0.228\pm 0.013}
&\ensuremath{0.229^{+ 0.014}_{- 0.015}}
&\ensuremath{0.223\pm 0.013}
&\ensuremath{0.233\pm 0.013} \nl
&
$\Omega_mh^2$
&\ensuremath{0.1358^{+ 0.0037}_{- 0.0036}}
&\ensuremath{0.1361^{+ 0.0038}_{- 0.0039}}
&\ensuremath{0.1348\pm 0.0038}
&\ensuremath{0.1369\pm 0.0037} \nl
&
$z_{\rm reion}$\footnote{``Redshift of reionization,'' if the universe was
 reionized instantaneously from the neutral state to the fully ionized
 state at $z_{\rm reion}$}
&\ensuremath{10.9\pm 1.4}
&\ensuremath{10.9\pm 1.4}
&\ensuremath{10.9\pm 1.4}
&\ensuremath{10.8\pm 1.4}\nl 
&
$t_0$\footnote{The present-day age of the universe}
&\ensuremath{13.72\pm 0.12\ \mbox{Gyr}} 
&\ensuremath{13.72\pm 0.12\ \mbox{Gyr}} 
&\ensuremath{13.71\pm 0.12\ \mbox{Gyr}} 
&\ensuremath{13.73\pm 0.12\ \mbox{Gyr}} 
\enddata
\label{tab:sncomp}
\end{deluxetable*}
\begin{figure*}[ht]
\centering \noindent 
\includegraphics[width=18cm]{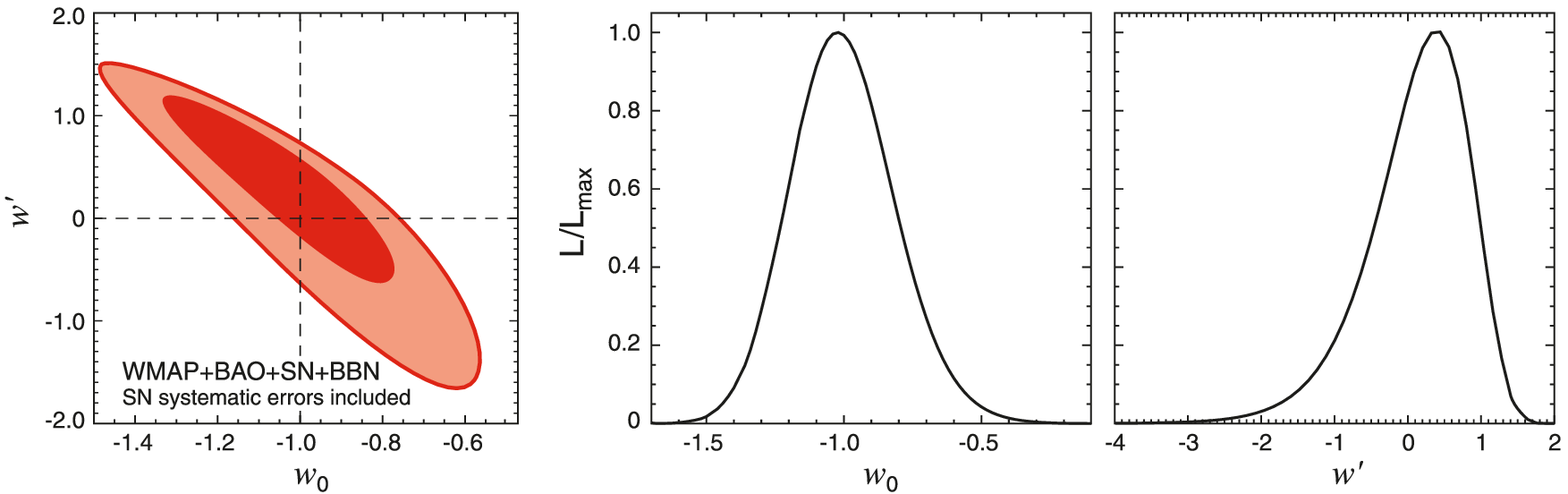}
\caption{%
The same as Fig.~\ref{fig:wright}, but the systematic errors in the Type
 Ia supernova data are included.
}
\label{fig:snsyserr}
\end{figure*}

In Table~\ref{tab:sncomp} we show the $\Lambda$CDM parameters
derived from \map+BAO+SN, where we use various SN compilations:
``Union'' for the latest union compilation \citep{kowalski/etal:prep},
``Union+Sys.Err.'' for the union compilation with systematic errors
included, ``Davis'' for the previous compilation by
\citet{davis/etal:2007}, and ``Alternative'' for the 
compilation that we used in the original version of this paper
(version 1 of arXiv:0803.0547). 

For the ``Alternative'' compilation we have combined
       measurements from the Hubble Space Telescope
       \citep{riess/etal:2004,riess/etal:2007}, the SuperNova Legacy
       Survey (SNLS) \citep{astier/etal:2006}, and the Equation of
       State: SupErNovae trace Cosmic Expansion (ESSENCE) survey
       \citep{wood-vasey/etal:2007}, as well as some nearby Type Ia
       supernovae. 
In the ``Davis'' and ``Alternative'' compilations, 
different light curve fitters were used for the supernova data taken by
       different groups, and thus these compilations were not as optimal
       as the union compilation, for which the same {\sf SALT} fitter
       was used for all the supernovae samples. Moreover, the union
       compilation is the largest of all. For this reason we have
       decided to update all the cosmological parameters using the union
       compilation.  

Nevertheless, we find that all of these compilations yield similar 
results: the mean values shift no more than $\sim 0.5\sigma$.

The effects of the systematic errors in the Type Ia supernova data on
the $\Lambda$CDM parameters are also very small for WMAP+BAO+SN; however, 
the effects on the dark energy parameters, $w_0$ and $w'$, turn out to
be  significant.
In Fig.~\ref{fig:snsyserr} we show the two-dimensional joint constraint
on $w_0$ and $w'$ from \map+BAO+SN+BBN (see also \S~\ref{sec:wz}) with
the systematic errors included. Comparing this to Fig.~\ref{fig:wright},
where the systematic errors are ignored, we find that the constraints on
$w_0$ and $w'$ weaken significantly:
we find $w_0=-1.00\pm 0.19$ and $w'=0.11\pm 0.70$ with the systematic
errors included, whereas $w_0=-1.04\pm 0.13$ and $w'=0.24\pm 0.55$
without the systematic errors.

\end{document}